\documentclass[twocolumn,showpacs,showkeys,preprintnumbers,amsmath,amssymb,aps,nofootXinbib,floatfix]{revtex4}
\pdfoutput=1
\usepackage{color}
\usepackage{indentfirst}
\usepackage{graphicx}
\usepackage{multirow}
      \def\di{\displaystyle}
      
      \def\bS{{\bf S}}
      \def\bl{{\bf l}}
      \def\bp{{\bf p}}
      
      \def\br{{\bf r}}

      \def\F{{\cal F}}
      \def\H{{\cal H}}

      \def\L{{\cal L}}
      
      \def\P{{\cal P}}
      \def\R{{\cal R}}

      \def\e{{\rm e}}

      \def\u{{{\uparrow\downarrow}}}
      \def\d{{{\downarrow\uparrow}}}

\begin{document}

\title{Triplet structure of nuclear scissors mode}

\author{ E.B. Balbutsev\email{balbuts@theor.jinr.ru}, 
I.V. Molodtsova\email{molod@theor.jinr.ru},
A.V. Sushkov\email{sushkov@theor.jinr.ru},
N.Yu. Shirikova\email{molod@theor.jinr.ru}}
\affiliation{ Joint Institute for Nuclear Research, 141980 Dubna, Moscow Region,Russia }
\author{ P. Schuck}
\affiliation{Institut de Physique Nucl\'eaire, IN2P3-CNRS, Universit\'e Paris-Sud,
F-91406 Orsay C\'edex, France;\\
Univ. Grenoble Alpes, CNRS,LPMMC, 38000 Grenoble, France} 

\begin{abstract}

The fine structure of the scissors mode is investigated within the 
Time Dependent Hartree-Fock-Bogoliubov (TDHFB) approach.
The solution of TDHFB equations by the Wigner Function Moments (WFM) method
predicts a splitting of the scissors mode into three intermingled branches. Together with the conventional scissors mode two new modes arise due to spin degrees of 
freedom. They generate significant $M1$ strength below the conventional energy range. The results of calculations 
of scissors resonances in Rare Earths and Actinides by WFM and QPNM methods
are compared with experimental data. A remarkable coherence of both methods together with experimental data is observed. 

\end{abstract}


\maketitle

\section{Introduction}

In the review \cite{Heyd} it is stated
that the scissors mode is "weakly collective, but strong 
on the single-particle scale" and further: {\it ``The weakly
collective scissors mode excitation has become an ideal test of models
-- especially microscopic models -- of nuclear vibrations. Most models
are usually calibrated to reproduce properties of strongly collective
excitations (e.g. of $J^{\pi}=2^+$ or $3^-$ states, giant resonances,
...). Weakly-collective phenomena, however, force the models to make
genuine predictions and the fact that the transitions in question are
strong on the single-particle scale makes it impossible to dismiss
failures as a mere detail, especially in the light of the overwhelming
experimental evidence for them in many nuclei \cite{Kneis,Richt}.''}

The Wigner Function Moments (WFM) or phase space moments method 
\cite{BaSc,Ann}
turns out to be very useful in this
situation. On the one hand it is a purely microscopic method, because
it is based on the Time Dependent Hartree-Fock (TDHF) equation. On the
other hand the method works with average values (moments) of operators
which have a direct relation to the considered phenomenon and, thus, make a 
natural bridge with the macroscopic description. This 
makes it an ideal instrument to describe the basic characteristics 
(energies and excitation probabilities) of collective excitations such as,
in particular, the scissors mode.

 In Ref. \cite{BaMo} the WFM method was 
applied for the first time to solve the TDHF equations including spin
dynamics. The most remarkable result was the prediction of a new type
of nuclear collective motion: rotational oscillations of "spin-up"
nucleons with respect of "spin-down" nucleons (the spin scissors mode). 

\begin{figure}[h!]
\centering
\includegraphics[width=\columnwidth]{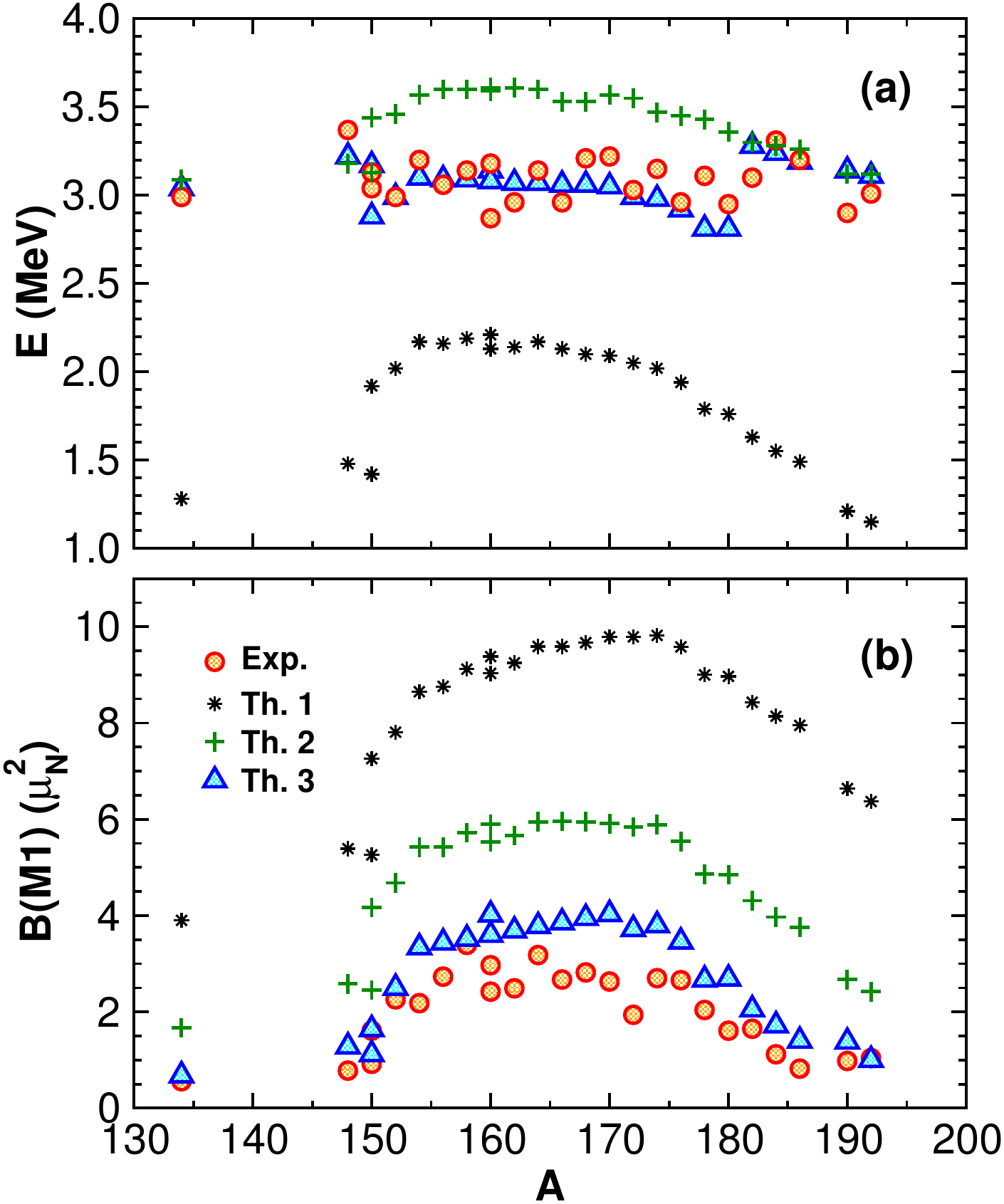}
\caption{Energy centroids $E$~(a) and summed $B(M1)$ values~(b)
of the scissors mode. 
Th. 1 -- the results of calculations without spin degrees of freedom 
and pair correlations, Th. 2 -- pairing is included, 
Th. 3 -- pairing and spin degrees of freedom are taken into account.}
\label{figMalov}\end{figure}

A generalization of the WFM method which takes into account spin degrees
of freedom and pair correlations simultaneously was outlined in 
\cite{BaMoPRC2}, where the Time Dependent Hartree-Fock-Bogoliubov (TDHFB)
equations were considered. As a result the agreement between theory and
experiment in the description of nuclear scissors modes was improved
considerably.
The evolution of our results in comparison with experimental data is
shown on Fig.~\ref{figMalov}.

By definition the scissors mode is a pure isovector mode. 
That is why we divided (approximately)
the dynamical equations describing collective  motion  
into isovector and isoscalar parts with the aim to separate the pure scissors mode. 

To study the interplay of isovector and isoscalar low-lying $1^+$ 
excitations we solved the coupled dynamical equations of our model for 
protons and neutrons exactly, without
the artificial isovector-isoscalar decoupling.
 As a result one more magnetic mode (third type of scissors) appeared.
Actually, the possible existence of three scissors states is easily explained 
by the combinatoric 
consideration -- there are only three ways to divide the four different kinds 
of objects (spin up and spin down protons and neutrons in our case) into two 
pairs. The analysis of the new situation, which appeared due to the last findings in the description of nuclear scissors, is presented in this paper.

The paper is organized as follows.
 In Sec.~II the TDHFB equations for the $2\times 2$ normal and anomalous density matrices are
formulated and their Wigner transform is found.
In Sec.~III the model Hamiltonian and the mean field are analyzed.
In Sec.~IV the collective variables are defined and the respective 
dynamical equations are derived. 
In Sec.~V the results of calculations of energies, $B(M1)$ values and
currents for nuclei of the $N=82-126$ mass region and Actinides are discussed.
The summary of main results is given in the Conclusion section. 
The mathematical details can be found in Appendix.

\section{Wigner transformation of TDHFB equations}

\hspace{5mm} The TDHFB equations in matrix formulation \cite{Solov,Ring} are
\begin{equation}
i\hbar\dot\R=[\H,\R]
\label{tHFB}
\end{equation}
with
\begin{equation}
\R={\hat\rho\qquad-\hat\kappa\choose-\hat\kappa^{\dagger}\;\;1-\hat\rho^*},
\quad\H={\hat
h\quad\;\;\hat\Delta\choose\hat\Delta^{\dagger}\quad-\hat h^*}
\end{equation}
The normal density matrix $\hat \rho$ and Hamiltonian $\hat h$ are
hermitian whereas the abnormal density $\hat\kappa$ and the pairing
gap $\hat\Delta$ are skew symmetric: $\hat\kappa^{\dagger}=-\hat\kappa^*$, 
$\hat\Delta^{\dagger}=-\hat\Delta^*$.
The detailed form of the TDHFB equations is
\begin{eqnarray}
&& i\hbar\dot{\hat\rho} =\hat h\hat\rho -\hat\rho\hat h
-\hat\Delta \hat\kappa ^{\dagger}+\hat\kappa \hat\Delta^\dagger,
\nonumber\\
&&-i\hbar\dot{\hat\rho}^*=\hat h^*\hat\rho ^*-\hat\rho ^*\hat h^*
-\hat\Delta^\dagger\hat\kappa +\hat\kappa^\dagger\hat\Delta ,
\nonumber\\
&&-i\hbar\dot{\hat\kappa} =-\hat h\hat\kappa -\hat\kappa \hat h^*+\hat\Delta
-\hat\Delta \hat\rho ^*-\hat\rho \hat\Delta ,
\nonumber\\
&&-i\hbar\dot{\hat\kappa}^\dagger=\hat h^*\hat\kappa^\dagger
+\hat\kappa^\dagger\hat h-\hat\Delta^\dagger
+\hat\Delta^\dagger\hat\rho +\hat\rho^*\hat\Delta^\dagger .
\label{HFB}
\end{eqnarray}
It is easy to see that the second and fourth equations are complex 
conjugate to the first and third ones respectively.
Let us consider their matrix form in coordinate 
space keeping all spin indices $s, s'$:
$\langle \br,s|\hat\rho|\br',s'\rangle$, 
$\langle \br,s|\hat\kappa|\br',s'\rangle$, etc.
 We do not specify the isospin indices in order to make
formulae more transparent. They will be re-introduced later. 
Let us introduce the more compact notation
$\langle \br,s|\hat X|\br',s'\rangle =X_{rr'}^{ss'}$. Then
the set of TDHFB equations (\ref{HFB}) with specified spin indices reads
\begin{widetext}
\begin{eqnarray}
&&i\hbar\dot{\rho}_{rr''}^{\uparrow\uparrow} =
\int\!d^3r'(
 h_{rr'}^{\uparrow\uparrow}\rho_{r'r''}^{\uparrow\uparrow} 
-\rho_{rr'}^{\uparrow\uparrow} h_{r'r''}^{\uparrow\uparrow}
+\hat h_{rr'}^{\uparrow\downarrow}\rho_{r'r''}^{\downarrow\uparrow} 
-\rho_{rr'}^{\uparrow\downarrow} h_{r'r''}^{\downarrow\uparrow}
-\Delta_{rr'}^{\uparrow\downarrow}{\kappa^{\dagger}}_{r'r''}^{\downarrow\uparrow}
+\kappa_{rr'}^{\uparrow\downarrow}{\Delta^{\dagger}}_{r'r''}^{\downarrow\uparrow}),
\nonumber\\
&&i\hbar\dot{\rho}_{rr''}^{\uparrow\downarrow} =
\int\!d^3r'(
 h_{rr'}^{\uparrow\uparrow}\rho_{r'r''}^{\uparrow\downarrow} 
-\rho_{rr'}^{\uparrow\uparrow} h_{r'r''}^{\uparrow\downarrow}
+\hat h_{rr'}^{\uparrow\downarrow}\rho_{r'r''}^{\downarrow\downarrow} 
-\rho_{rr'}^{\uparrow\downarrow} h_{r'r''}^{\downarrow\downarrow}),
\nonumber\\
&&i\hbar\dot{\rho}_{rr''}^{\downarrow\uparrow} =
\int\!d^3r'(
 h_{rr'}^{\downarrow\uparrow}\rho_{r'r''}^{\uparrow\uparrow} 
-\rho_{rr'}^{\downarrow\uparrow} h_{r'r''}^{\uparrow\uparrow}
+\hat h_{rr'}^{\downarrow\downarrow}\rho_{r'r''}^{\downarrow\uparrow} 
-\rho_{rr'}^{\downarrow\downarrow} h_{r'r''}^{\downarrow\uparrow}),
\nonumber\\
&&i\hbar\dot{\rho}_{rr''}^{\downarrow\downarrow} =
\int\!d^3r'(
 h_{rr'}^{\downarrow\uparrow}\rho_{r'r''}^{\uparrow\downarrow} 
-\rho_{rr'}^{\downarrow\uparrow} h_{r'r''}^{\uparrow\downarrow}
+\hat h_{rr'}^{\downarrow\downarrow}\rho_{r'r''}^{\downarrow\downarrow} 
-\rho_{rr'}^{\downarrow\downarrow} h_{r'r''}^{\downarrow\downarrow}
-\Delta_{rr'}^{\downarrow\uparrow}{\kappa^{\dagger}}_{r'r''}^{\uparrow\downarrow}
+\kappa_{rr'}^{\downarrow\uparrow}{\Delta^{\dagger}}_{r'r''}^{\uparrow\downarrow}),
\nonumber\\
&&i\hbar\dot{\kappa}_{rr''}^{\uparrow\downarrow} = -\hat\Delta_{rr''}^{\uparrow\downarrow}
+\int\!d^3r'\left(
 h_{rr'}^{\uparrow\uparrow}\kappa_{r'r''}^{\uparrow\downarrow} 
+\kappa_{rr'}^{\uparrow\downarrow} {h^*}_{r'r''}^{\downarrow\downarrow}
+\Delta_{rr'}^{\uparrow\downarrow}{\rho^*}_{r'r''}^{\downarrow\downarrow} 
+\rho_{rr'}^{\uparrow\uparrow}\Delta_{r'r''}^{\uparrow\downarrow}
\right),
\nonumber\\
&&i\hbar\dot{\kappa}_{rr''}^{\downarrow\uparrow} = -\hat\Delta_{rr''}^{\downarrow\uparrow}
+\int\!d^3r'\left(
 h_{rr'}^{\downarrow\downarrow}\kappa_{r'r''}^{\downarrow\uparrow} 
+\kappa_{rr'}^{\downarrow\uparrow} {h^*}_{r'r''}^{\uparrow\uparrow}
+\Delta_{rr'}^{\downarrow\uparrow}{\rho^*}_{r'r''}^{\uparrow\uparrow} 
+\rho_{rr'}^{\downarrow\downarrow}\Delta_{r'r''}^{\downarrow\uparrow}
\right).
\label{HFsp}
\end{eqnarray}
This set of equations must be complemented by the complex conjugated equations.
Writing these equations we neglected the diagonal in spin matrix elements
of the abnormal density:
$\kappa_{rr'}^{ss}$ and $\Delta_{rr'}^{ss}$. It was shown in~\cite{BaMoPRC2} 
that such an approximation works very well in the case of monopole pairing 
considered here.

We will work with the Wigner transform~\cite{Ring} of 
equations~(\ref{HFsp}). 
  From now on, we will not write out the coordinate 
dependence $(\br,\bp)$ of all functions in order to make the formulae 
more transparent. We have
\begin{eqnarray}
      i\hbar\dot f^{\uparrow\uparrow} &=&i\hbar\{h^{\uparrow\uparrow},f^{\uparrow\uparrow}\}
+h^{\uparrow\downarrow}f^{\downarrow\uparrow}-f^{\uparrow\downarrow}h^{\downarrow\uparrow}
+\frac{i\hbar}{2}\{h^{\uparrow\downarrow},f^{\downarrow\uparrow}\}
-\frac{i\hbar}{2}\{f^{\uparrow\downarrow},h^{\downarrow\uparrow}\}
\nonumber\\
&-&\frac{\hbar^2}{8}\{\!\{h^{\uparrow\downarrow},f^{\downarrow\uparrow}\}\!\}
+\frac{\hbar^2}{8}\{\!\{f^{\uparrow\downarrow},h^{\downarrow\uparrow}\}\!\} 
+ \kappa\Delta^* - \Delta\kappa^* 
\nonumber\\
&+&\frac{i\hbar}{2}\{\kappa,\Delta^*\}-\frac{i\hbar}{2}\{\Delta,\kappa^*\}
- \frac{\hbar^2}{8}\{\!\{\kappa,\Delta^*\}\!\} + \frac{\hbar^2}{8}\{\!\{\Delta,\kappa^*\}\!\}
+...,
\nonumber\\
      i\hbar\dot f^{\downarrow\downarrow} &=&i\hbar\{h^{\downarrow\downarrow},f^{\downarrow\downarrow}\}
+h^{\downarrow\uparrow}f^{\uparrow\downarrow}-f^{\downarrow\uparrow}h^{\uparrow\downarrow}
+\frac{i\hbar}{2}\{h^{\downarrow\uparrow},f^{\uparrow\downarrow}\}
-\frac{i\hbar}{2}\{f^{\downarrow\uparrow},h^{\uparrow\downarrow}\}
\nonumber\\
&-&\frac{\hbar^2}{8}\{\!\{h^{\downarrow\uparrow},f^{\uparrow\downarrow}\}\!\}
+\frac{\hbar^2}{8}\{\!\{f^{\downarrow\uparrow},h^{\uparrow\downarrow}\}\!\} 
+ \bar\Delta^* \bar\kappa - \bar\kappa^* \bar\Delta
\nonumber\\
&+&\frac{i\hbar}{2}\{\bar\Delta^*,\bar\kappa\}-\frac{i\hbar}{2}\{\bar\kappa^*,\bar\Delta\}
- \frac{\hbar^2}{8}\{\!\{\bar\Delta^*,\bar\kappa\}\!\} + \frac{\hbar^2}{8}\{\!\{\bar\kappa^*,\bar\Delta\}\!\}
+...,
\nonumber\\
      i\hbar\dot f^{\uparrow\downarrow} &=&
f^{\uparrow\downarrow}(h^{\uparrow\uparrow}-h^{\downarrow\downarrow})
+\frac{i\hbar}{2}\{(h^{\uparrow\uparrow}+h^{\downarrow\downarrow}),f^{\uparrow\downarrow}\}
-\frac{\hbar^2}{8}\{\!\{(h^{\uparrow\uparrow}-h^{\downarrow\downarrow}),f^{\uparrow\downarrow}\}\!\}
\nonumber\\
&-&h^{\uparrow\downarrow}(f^{\uparrow\uparrow}-f^{\downarrow\downarrow})
+\frac{i\hbar}{2}\{h^{\uparrow\downarrow},(f^{\uparrow\uparrow}+f^{\downarrow\downarrow})\}
+\frac{\hbar^2}{8}\{\!\{h^{\uparrow\downarrow},(f^{\uparrow\uparrow}-f^{\downarrow\downarrow})\}\!\}+....,
\nonumber\\
      i\hbar\dot f^{\downarrow\uparrow} &=&
f^{\downarrow\uparrow}(h^{\downarrow\downarrow}-h^{\uparrow\uparrow})
+\frac{i\hbar}{2}\{(h^{\downarrow\downarrow}+h^{\uparrow\uparrow}),f^{\downarrow\uparrow}\}
-\frac{\hbar^2}{8}\{\!\{(h^{\downarrow\downarrow}-h^{\uparrow\uparrow}),f^{\downarrow\uparrow}\}\!\}
\nonumber\\
&-&h^{\downarrow\uparrow}(f^{\downarrow\downarrow}-f^{\uparrow\uparrow})
+\frac{i\hbar}{2}\{h^{\downarrow\uparrow},(f^{\downarrow\downarrow}+f^{\uparrow\uparrow})\}
+\frac{\hbar^2}{8}\{\!\{h^{\downarrow\uparrow},(f^{\downarrow\downarrow}-f^{\uparrow\uparrow})\}\!\}+...,
\nonumber\\
      i\hbar\dot \kappa &=& \kappa\,(h^{\uparrow\uparrow}+\bar h^{\downarrow\downarrow})
  +\frac{i\hbar}{2}\{(h^{\uparrow\uparrow}-\bar h^{\downarrow\downarrow}),\kappa\}    
  -\frac{\hbar^2}{8}\{\!\{(h^{\uparrow\uparrow}+\bar h^{\downarrow\downarrow}),\kappa\}\!\}   
 \nonumber\\ 
 &+&\Delta\,(f^{\uparrow\uparrow}+\bar f^{\downarrow\downarrow})
  +\frac{i\hbar}{2}\{(f^{\uparrow\uparrow}-\bar f^{\downarrow\downarrow}),\Delta\}    
  -\frac{\hbar^2}{8}\{\!\{(f^{\uparrow\uparrow}+\bar f^{\downarrow\downarrow}),\Delta\}\!\}
  - \Delta + ...,
 \nonumber\\     
    i\hbar\dot \kappa^* &=&  -\kappa^*(h^{\uparrow\uparrow}+\bar h^{\downarrow\downarrow}) 
  +\frac{i\hbar}{2}\{(h^{\uparrow\uparrow}-\bar h^{\downarrow\downarrow}),\kappa^*\}    
  +\frac{\hbar^2}{8}\{\!\{(h^{\uparrow\uparrow}+\bar h^{\downarrow\downarrow}),\kappa^*\}\!\}   
  \nonumber\\
  &-& \Delta^*(f^{\uparrow\uparrow}+\bar f^{\downarrow\downarrow}) 
  +\frac{i\hbar}{2}\{(f^{\uparrow\uparrow}-\bar f^{\downarrow\downarrow}),\Delta^*\}    
  +\frac{\hbar^2}{8}\{\!\{(f^{\uparrow\uparrow}+\bar f^{\downarrow\downarrow}),\Delta^*\}\!\}
  + \Delta^* +..., 
\label{WHF}
\end{eqnarray} 
where the functions $h$, $f$, $\Delta$, and $\kappa$ are the Wigner
transforms of $\hat h$, $\hat\rho$, $\hat\Delta$, and $\hat\kappa$,
respectively, $\bar f(\br,\bp)=f(\br,-\bp)$, $\{f,g\}$ is the Poisson
bracket of the functions $f$ and $g$ and
$\{\{f,g\}\}$ is their double Poisson bracket, $f(\br,\bp)$ being the Wigner function.
This set of equations must be complemented by the dynamical equations for 
$\bar f^{\uparrow\uparrow}, \bar f^{\downarrow\downarrow}, \bar f^{\uparrow\downarrow}, 
\bar f^{\downarrow\uparrow},\bar\kappa,\bar\kappa^*$.
They are obtained by the change $\bp \rightarrow -\bp$ in arguments of functions and Poisson brackets. 
 We introduced the notation
$\kappa \equiv \kappa^{\uparrow\downarrow}$ and $\Delta \equiv \Delta^{\uparrow\downarrow}$.
Symmetry properties of matrices $\hat\kappa, \hat\Delta$ and the properties of their Wigner 
transforms (see \cite{BaMoPRC2}) allow us to replace the functions 
$\kappa^{\downarrow\uparrow}(\br,\bp)$ and  $\Delta^{\downarrow\uparrow}(\br,\bp)$ by the functions $\bar\kappa^{\uparrow\downarrow}(\br,\bp)$ and  $\bar\Delta^{\uparrow\downarrow}(\br,\bp)$. 
The dots stand for terms proportional to higher powers of $\hbar$ -- after 
integration over phase space these terms disappear and we arrive to the set 
of exact integral equations.
Following the paper \cite{BaMo} above equations can be rewritten in terms of spin-scalar $\quad f^+=f^{\uparrow\uparrow}+ f^{\downarrow\downarrow}
\quad$ and spin-vector
$\quad f^-=f^{\uparrow\uparrow}- f^{\downarrow\downarrow}\quad$
functions. Furthermore, it is useful to 
rewrite the obtained equations in terms of even and odd functions 
$f_{e}=\frac{1}{2}(f+\bar f)$ and $f_{o}=\frac{1}{2}(f-\bar f)$
and real and imaginary parts of $\kappa$ and $\Delta$: $\kappa^r=\frac{1}{2}(\kappa+\kappa^*),\,
\kappa^i=\frac{1}{2i}(\kappa-\kappa^*),\,\Delta^r=\frac{1}{2}(\Delta+\Delta^*),\,
\Delta^i=\frac{1}{2i}(\Delta-\Delta^*)$. These operations are straightforward
and omitted here.

\section{Model Hamiltonian}

 The microscopic Hamiltonian of the model, harmonic oscillator with 
spin orbit potential plus separable quadrupole-quadrupole and 
spin-spin residual interactions is given by
\begin{eqnarray}
\label{Ham}
 H=\sum\limits_{i=1}^A\left[\frac{\hat\bp_i^2}{2m}+\frac{1}{2}m\omega^2\br_i^2
-\eta\hat \bl_i\hat \bS_i\right]+H_{qq}+H_{ss}
\end{eqnarray}
with
\begin{eqnarray}
\label{Hqq}
&& H_{qq}=\!
\sum_{\mu=-2}^{2}(-1)^{\mu}
\left\{\bar{\kappa}
 \sum\limits_i^Z\!\sum\limits_j^N
+\frac{\kappa}{2}
\left[\sum\limits_{i,j(i\neq j)}^{Z}
+\sum\limits_{i,j(i\neq j)}^{N}
\right]
\right\}
q_{2-\mu}(\br_i)q_{2\mu}(\br_j)
,
\\
\label{Hss}
&&H_{ss}=\!
\sum_{\mu=-1}^{1}(-1)^{\mu}
\left\{\bar{\chi}
 \sum\limits_i^Z\!\sum\limits_j^N
+\frac{\chi}{2}
\left[
\sum\limits_{i,j(i\neq j)}^{Z}
+\sum\limits_{i,j(i\neq j)}^{N}
\right]
\right\}
\hat S_{-\mu}(i)\hat S_{\mu}(j)
\,\delta(\br_i-\br_j),
\end{eqnarray}
where  $q_{2\mu}=\sqrt{16\pi/5}\,r^2Y_{2\mu}=
\sqrt{6}\{r\otimes r\}_{\lambda\mu},\,
\{r\otimes r\}_{\lambda\mu}=\sum_{\sigma,\nu}
C_{1\sigma,1\nu}^{\lambda\mu}r_{\sigma}r_{\nu},\,
C_{1\sigma,1\nu}^{\lambda\mu}$ is the Clebsch-Gordan
coefficient,
cyclic coordinates $r_{-1}, r_0, r_1$ 
are defined in \cite{Var}, $N$ and $Z$ are the numbers of neutrons and 
protons. $\hat S_{\mu}$ are spin matrices \cite{Var}:
\begin{equation}
\hat S_1=-\frac{\hbar}{\sqrt2}{0\quad 1\choose 0\quad 0},\quad
\hat S_0=\frac{\hbar}{2}{1\quad\, 0\choose 0\, -\!1},\quad
\hat S_{-1}=\frac{\hbar}{\sqrt2}{0\quad 0\choose 1\quad 0}.
\label{S}
\end{equation}
\end{widetext}

\subsection{Mean Field}

Let us analyze the mean field generated by this Hamiltonian.

\subsubsection{Spin-orbit Potential}

Written in cyclic coordinates, the spin orbit part of the
Hamiltonian reads
$$\hat h_{ls}=-\eta\sum_{\mu=-1}^1(-)^{\mu}\hat l_{\mu}\hat S_{-\mu}
=-\eta{\quad\hat l_0\frac{\hbar}{2}\quad\; \hat l_{-1}\frac{\hbar}{\sqrt2} \choose 
 -\hat l_{1}\frac{\hbar}{\sqrt2}\; -\hat l_0\frac{\hbar}{2}},
$$
where \cite{Var}
\begin{equation}
\label{lqu}
\hat l_{\mu}=-\hbar\sqrt2\sum_{\nu,\alpha}C_{1\nu,1\alpha}^{1\mu}r_{\nu}\nabla_{\alpha},
\end{equation}
and
\begin{eqnarray}
&&\hat l_1=\hbar(r_0\nabla_1-r_1\nabla_0)=
-\frac{1}{\sqrt2}(\hat l_x+i\hat l_y),
\nonumber\\&&
\hat l_0=\hbar(r_{-1}\nabla_1-r_1\nabla_{-1})=\hat l_z,
\nonumber\\
&&\hat l_{-1}=\hbar(r_{-1}\nabla_0-r_0\nabla_{-1})=
\frac{1}{\sqrt2}(\hat l_x-i\hat l_y),
\nonumber\\
&&\hat l_x=-i\hbar(y\nabla_z-z\nabla_y),\quad
\hat l_y=-i\hbar(z\nabla_x-x\nabla_z),
\nonumber\\&&
\hat l_z=-i\hbar(x\nabla_y-y\nabla_x),
\label{lxyz}
\end{eqnarray}
$\eta=2\hbar\,\mathring{\omega}_0\kappa_{\rm Nils}$.
 The matrix elements of $\hat h_{ls}$ in coordinate space can be written \cite{BaMo} as
\begin{eqnarray}
\langle \br_1,s_1|\hat h_{ls}|\br_2,s_2\rangle 
=-\frac{\hbar}{2}\eta\left[\hat l_{0}(\br_1)(\delta_{s_1\uparrow}\delta_{s_2\uparrow}
-\delta_{s_1\downarrow}\delta_{s_2\downarrow})\right.
\nonumber\\
+\left.\sqrt2\, \hat l_{-1}(\br_1)\delta_{s_1\uparrow}\delta_{s_2\downarrow}
-\sqrt2\, \hat l_{1}(\br_1)\delta_{s_1\downarrow}\delta_{s_2\uparrow}\right]\delta(\br_1-\br_2).
\nonumber
\end{eqnarray}
Their Wigner transform reads \cite{BaMo}:
\begin{eqnarray}
 h_{ls}^{s_1s_2}(\br,\bp)
=-\frac{\hbar}{2}\eta\left[l_{0}(\br,\bp)(\delta_{s_1\uparrow}\delta_{s_2\uparrow}
-\delta_{s_1\downarrow}\delta_{s_2\downarrow})\right.
\nonumber\\
+\left.\sqrt2 l_{-1}(\br,\bp)\delta_{s_1\uparrow}\delta_{s_2\downarrow}
-\sqrt2 l_{1}(\br,\bp)\delta_{s_1\downarrow}\delta_{s_2\uparrow}\right],\ 
\label{Hrp}
\end{eqnarray}
with
$l_{\mu}(\br,\bp)=
-i\sqrt2\sum_{\nu,\alpha}C_{1\nu,1\alpha}^{1\mu}r_{\nu}p_{\alpha}$.

\subsubsection{Quadrupole-quadrupole interaction}

 The contribution of $H_{qq}$ to the mean field potential is easily
found by replacing one of the $q_{2\mu}$ operators by its average value.
We have
\begin{equation}
\label{potenirr}
V^{\tau}_{qq}=\sqrt6\sum_{\mu}(-1)^{\mu}Z_{2-\mu}^{\tau +}q_{2\mu}.
\end{equation}
 Here
\begin{eqnarray}
\label{Z2mu}
&&Z_{2\mu}^{{\rm n}+}=\kappa R_{2\mu}^{{\rm n}+}
+\bar{\kappa}R_{2\mu}^{{\rm p}+},\quad
Z_{2\mu}^{{\rm p}+}=\kappa R_{2\mu}^{{\rm p}+}
+\bar{\kappa}R_{2\mu}^{{\rm n}+},\nonumber\\&&  R_{2\mu}^{\tau+}(t)=
\frac{1}{\sqrt6}\int d(\bp,\br)
q_{2\mu}(\br)f^{\tau+}(\br,\bp,t)
\end{eqnarray}
 with
 $\int\! d(\bp,\br)\equiv
(2\pi\hbar)^{-3}\int\! d^3p\,\int\! d^3r$ and $\tau$ being the isospin index.

\subsubsection{Spin-spin interaction}

The analogous expression for $H_{ss}$ is found in a
standard way \cite{BaMoPRC} with the following result for
the Wigner transform of the proton mean field:
\begin{eqnarray}
\label{Vp}
V_{\rm p}^{s s'}(\br,t)=
3\chi\frac{\hbar^2}{8}
\left[
\delta_{s\downarrow}\delta_{s'\uparrow}n_{\rm p}^{\downarrow\uparrow}+
\delta_{s\uparrow}\delta_{s'\downarrow}n_{\rm p}^{\uparrow\downarrow}
\right.\nonumber\\ \left.
-\delta_{s\downarrow}\delta_{s'\downarrow}n_{\rm p}^{\uparrow\uparrow}
-\delta_{s\uparrow}\delta_{s'\uparrow}n_{\rm p}^{\downarrow\downarrow}
\right]
\nonumber\\
+\bar\chi\frac{\hbar^2}{8}
\left[
2\delta_{s\downarrow}\delta_{s'\uparrow}n_{\rm n}^{\downarrow\uparrow}+
2\delta_{s\uparrow}\delta_{s'\downarrow}n_{\rm n}^{\uparrow\downarrow}
\right.\nonumber\\ \left.
+(\delta_{s\uparrow}\delta_{s'\uparrow}-
\delta_{s\downarrow}\delta_{s'\downarrow})(n_{\rm n}^{\uparrow\uparrow}-
n_{\rm n}^{\downarrow\downarrow})
\right],
\end{eqnarray}
where 
${\di n_{\tau}^{ss'}(\br,t)=\int\frac{d\bp}{(2\pi\hbar)^3}f^{ss'}_{\tau}(\br,\bp,t)}$.
The Wigner transform of the neutron mean field $V_{\rm n}^{ss'}$ is 
obtained from (\ref{Vp}) by the obvious change of indices ${\rm p}\leftrightarrow {\rm n}$.

\subsection{Pair potential}

The Wigner transform of the pair potential (pairing gap) $\Delta^\tau(\br,\bp)$ is related to 
the Wigner transform of the anomalous density by \cite{Ring}
\begin{equation}
\Delta^\tau(\br,\bp)=-\int\! \frac{d\bp'}{(2\pi\hbar)^3}
v(|\bp-\bp'|)\kappa^\tau(\br,\bp'),
\label{DK}
\end{equation}
where $v(p)$ is a Fourier transform of the two-body interaction.
We take for the pairing interaction a simple Gaussian of strength $V_0$ 
and range $r_p$  \cite{Ring}
\begin{equation}
v(p)=\beta {\rm e}^{-\alpha p^2}\!,
\label{v_p}
\end{equation}
with $\beta=-|V_0|(r_p\sqrt{\pi})^3$ and $\alpha=r_p^2/4\hbar^2$. 
The following values of parameters were used in calculations:  
$V_0^{\rm p}=27$~MeV, $V_0^{\rm n}=23$~MeV, $r_p^{\rm p}=1.50$~fm, $r_p^{\rm n}=1.85$~fm for nuclei with $A=150-186$
and  $V_0^{\rm p}=25.5$ MeV, $V_0^{\rm n}=21.5$ MeV, $r_p^{\rm p}=1.5$ fm, $r_p^{\rm n}=1.80$ fm for Actinides.
Exceptions are listed in the captions to the corresponding Tables.

\section{Equations of motion}

Equations (\ref{WHF}) will be solved by the method of moments in a small 
amplitude approximation. To this end 
all functions $f(\br,\bp,t)$ and $\kappa(\br,\bp,t)$ are divided into an equilibrium part 
and a variation: $f(\br,\bp,t)=f(\br,\bp)_{\rm eq}+\delta f(\br,\bp,t)$, 
$\kappa(\br,\bp,t)=\kappa(\br,\bp)_{\rm eq}+\delta \kappa(\br,\bp,t)$.
Then the equations are linearized  neglecting  terms quadratic in the variations.

After linearization, the phase of $\Delta$
(and of $\kappa$) is expressed by $\delta\Delta^i$ (and $\delta\kappa^i$), while
$\delta\Delta^r$ (and $\delta\kappa^r$) describes oscillations of the
magnitude of $\Delta$ (and of~$\kappa$). 
  From general arguments one can expect that the phase of $\Delta$ (and
of $\kappa$, since both are linked, according to equation (\ref{DK}))
is much more relevant than its magnitude, since the former determines
the superfluid velocity. Let us therefore assume that
\begin{equation}
\delta\kappa^r(\br,\bp)\ll\delta\kappa^i(\br,\bp).
\label{approx1}
\end{equation}
This assumption was explicitly confirmed in \cite{M.Urban} for the case
of superfluid trapped fermionic atoms, where it was shown that
$\delta\Delta^r$ is suppressed with respect to $\delta\Delta^i$ by one
order of $\Delta/E_{\rm F}$, where $E_{\rm F}$ denotes the Fermi energy.
The assumption~(\ref{approx1}) allows one to neglect all terms containing the variations
$\delta \kappa^r$ and $\delta \Delta^r$ in the equations~(\ref{WHF}) after their linearization.

 Integrating these equations over phase space 
with the weights 
\begin{equation}
W =\{r\otimes p\}_{\lambda\mu},\,\{r\otimes r\}_{\lambda\mu},\,
\{p\otimes p\}_{\lambda\mu}, \mbox{ and } 1
\nonumber
\end{equation}
one gets dynamic equations for 
the following collective variables:
\begin{eqnarray}
&&\L^{\tau\varsigma}_{\lambda\mu}(t)=\int\! d(\bp,\br) \{r\otimes p\}_{\lambda\mu}
\delta f^{\tau\varsigma}_o(\br,\bp,t),
\nonumber\\&&
\R^{\tau\varsigma}_{\lambda\mu}(t)=\int\! d(\bp,\br) \{r\otimes r\}_{\lambda\mu}
\delta f^{\tau\varsigma}_e(\br,\bp,t),
\nonumber\\
&&\P^{\tau\varsigma}_{\lambda\mu}(t)=\int\! d(\bp,\br) \{p\otimes p\}_{\lambda\mu}
\delta f^{\tau\varsigma}_e(\br,\bp,t),
\nonumber\\&&
\F^{\tau\varsigma}(t)=\int\! d(\bp,\br)
\delta f^{\tau\varsigma}_e(\br,\bp,t),
\nonumber\\
&&\tilde{\L}^{\tau}_{\lambda\mu}(t)=\int\! d(\bp,\br) \{r\otimes p\}_{\lambda\mu}
\delta \kappa^{\tau i}_o(\br,\bp,t),
\nonumber\\&&
\tilde{\R}^{\tau}_{\lambda\mu}(t)=\int\! d(\bp,\br) \{r\otimes r\}_{\lambda\mu}
\delta \kappa^{\tau i}_e(\br,\bp,t),
\nonumber\\
&&\tilde{\P}^{\tau}_{\lambda\mu}(t)=\int\! d(\bp,\br) \{p\otimes p\}_{\lambda\mu}
\delta \kappa^{\tau i}_e(\br,\bp,t),
\nonumber
\end{eqnarray}
 where 
$\varsigma\!=+,\,-,\,\uparrow\downarrow,\,\downarrow\uparrow,$

The required expressions for 
$h^{\pm}$, $h^{\uparrow\downarrow}$ and $h^{\downarrow\uparrow}$ are
\begin{eqnarray}
h_{\tau}^{+}&=&\frac{p^2}{m}+m\,\omega^2r^2
+12\sum_{\mu}(-1)^{\mu}Z_{2\mu}^{\tau+}(t)\{r\otimes r\}_{2-\mu}\qquad
\nonumber\\
&+&V_{\tau}^+(\br,t)-\mu^{\tau}, 
\nonumber
\end{eqnarray}
$\mu^{\tau}$ being the chemical potential of protons ($\tau={\rm p}$) or neutrons ($\tau={\rm n}$),
\begin{eqnarray}
&&h_{\tau}^-=-\hbar\eta l_0+V_{\tau}^-(\br,t), 
\nonumber\\
&&h_{\tau}^{\uparrow\downarrow}=-\frac{\hbar}{\sqrt2}\eta l_{-1}+V_{\tau}^{\uparrow\downarrow}(\br,t),
\nonumber\\ 
&&h_{\tau}^{\downarrow\uparrow}=\frac{\hbar}{\sqrt2}\eta l_{1}+V_{\tau}^{\downarrow\uparrow}(\br,t),
\nonumber
\end{eqnarray}
where according to (\ref{Vp})
\begin{eqnarray}
\label{Vss}
&&V_{\rm p}^+(\br,t)=-3\frac{\hbar^2}{8}\chi n_{\rm p}^+(\br,t),
\nonumber\\
&&V_{\rm p}^-(\br,t)=3\frac{\hbar^2}{8}\chi n_{\rm p}^-(\br,t)+\frac{\hbar^2}{4}\bar\chi n_{\rm n}^-(\br,t),
\nonumber\\
&&V_{\rm p}^{\uparrow\downarrow}(\br,t)=3\frac{\hbar^2}{8}\chi n_{\rm p}^{\uparrow\downarrow}(\br,t)
+\frac{\hbar^2}{4}\bar\chi n_{\rm n}^{\uparrow\downarrow}(\br,t),
\nonumber\\
&&V_{\rm p}^{\downarrow\uparrow}(\br,t)=3\frac{\hbar^2}{8}\chi n_{\rm p}^{\downarrow\uparrow}(\br,t)
+\frac{\hbar^2}{4}\bar\chi n_{\rm n}^{\downarrow\uparrow}(\br,t)
\end{eqnarray}
and the neutron potentials $V_{\rm n}^{\varsigma}$ are
obtained by the obvious change of indices ${\rm p}\leftrightarrow{\rm n}$.
Variations of these mean fields read:

$$\delta h_{\tau}^{+}=12\sum_{\mu}(-1)^{\mu}\delta Z_{2\mu}^{\tau+}(t)\{r\otimes r\}_{2-\mu}
+\delta V_{\tau}^+(\br,t),$$
where
$\delta Z_{2\mu}^{{\rm p}+}=\kappa \delta R_{2\mu}^{{\rm p}+}
+\bar{\kappa}\delta R_{2\mu}^{{\rm n}+}$,  $\delta R_{\lambda\mu}^{\tau+}(t)\equiv
\R_{\lambda\mu}^{\tau+}(t)$ and
\begin{eqnarray}
&&\delta V_{\rm p}^+(\br,t)=-3\frac{\hbar^2}{8}\chi \delta n_{\rm p}^+(\br,t),
\nonumber\\ \nonumber
&&\delta n_{\rm p}^{+}(\br,t)=\int\frac{d^3p}{(2\pi\hbar)^3}\delta f^{+}_{\rm p}(\br,\bp,t).
\end{eqnarray}
Variations of
$h^{-}$, $h^{\uparrow\downarrow}$ and $h^{\downarrow\uparrow}$ are obtained in a similar way. The detailed procedure of the density variation is 
described in Appendix B of \cite{BaMoPRC}. 
Variation of the pair potential is
\begin{equation}
\delta \Delta^\tau(\br,\bp,t)=-\int\! \frac{d\bp'}{(2\pi\hbar)^3}
v(|\bp-\bp'|)\delta \kappa^\tau(\br,\bp',t).
\label{DKvar}
\end{equation}

We are interested in the scissors mode with quantum number $K^{\pi}=1^+$. 
Therefore, we only need the part of dynamic equations  with $\mu=1$. 

  It is convenient to rewrite the dynamical equations for neutron and 
proton variables in terms of isoscalar and isovector variables
\begin{eqnarray}
\label{Isovs}
&&\R_{\lambda\mu}=\R_{\lambda\mu}^{\rm n}+\R_{\lambda\mu}^{\rm p},\quad
\bar \R_{\lambda\mu}=\R_{\lambda\mu}^{\rm n}-\R_{\lambda\mu}^{\rm p},
\nonumber\\ 
&&\P_{\lambda\mu}=\P_{\lambda\mu}^{\rm n}+\P_{\lambda\mu}^{\rm p},\quad
\bar \P_{\lambda\mu}=\P_{\lambda\mu}^{\rm n}-\P_{\lambda\mu}^{\rm p},
\nonumber\\
&&\L_{\lambda\mu}=\L_{\lambda\mu}^{\rm n}+\L_{\lambda\mu}^{\rm p},\quad
\bar \L_{\lambda\mu}=\L_{\lambda\mu}^{\rm n}-\L_{\lambda\mu}^{\rm p}.
\nonumber\\
&&\F=\F^{\rm n}+\F^{\rm p},\qquad \bar \F=\F^{\rm n}-\F^{\rm p}.
\end{eqnarray}
We also define isovector and isoscalar strength constants
$\kappa_1=\frac{1}{2}(\kappa-\bar\kappa)$ and
$\kappa_0=\frac{1}{2}(\kappa+\bar\kappa)$ connected by the relation
$\kappa_1=\alpha\kappa_0$ with $\alpha=-2$ \cite{BaSc}.

The integration yields the following set of equations for isovector variables:

\begin{widetext}

\begin{eqnarray}
     \dot {\bar\L}^{+}_{21}&=&
\frac{1}{m}\bar\P_{21}^{+}
-\left[m\,\omega^2+\kappa_0 \left(4\alpha Q_{00}+(1+\alpha)Q_{20}\right)\right]\bar\R^{+}_{21}
-i\hbar\frac{\eta}{2}\left[\bar\L_{21}^-
+2\bar\L^\u_{22}+
\sqrt6\bar\L^\d_{20}\right]
\nonumber\\&&-\underbrace{\kappa_0\left(4\bar Q_{00}+(1+\alpha) \bar Q_{20}\right)
\R^{+}_{21}}_{\text{coupling term}},
\nonumber\\
     \dot {\bar\L}^{-}_{21}&=&
\frac{1}{m}\bar\P_{21}^{-}
-\left[m\,\omega^2+\kappa_0 Q_{20}
-\frac{\hbar^2}{15} \left( 3\chi-\bar\chi \right) \frac{I_1}{A_1A_2} \left(Q_{00}+Q_{20}/4\right)
\right]\bar\R^{-}_{21}-i\hbar\frac{\eta}{2}\bar\L_{21}^+
 +\frac{4}{\hbar} I_{rp}^{\kappa\Delta}(r') {\bar{\tilde\L}}_{21}
\nonumber\\
&&
-\underbrace{\left[\alpha\kappa_0 \bar Q_{20}
- \frac{\hbar^2}{15}
\left( 3\chi+\bar\chi \right) \frac{I_1}{A_1A_2} \left(\bar Q_{00}+\bar Q_{20}/4\right)\right]\R^{-}_{21}
+\frac{4}{\hbar}\bar I_{rp}^{\kappa\Delta}(r') {{\tilde\L}}_{21}
},
\nonumber\\
     \dot {\bar\L}^\u_{22}&=&
\frac{1}{m}\bar\P_{22}^\u-
\left[m\,\omega^2-2\kappa_0 Q_{20}
-4\hbar^2 
\left( 3\chi-\bar\chi \right)\frac{I_1}{A_1A_2}\left(Q_{20}+Q_{00}\right)
\right]\bar\R^\u_{22}
-i\hbar\frac{\eta}{2}\bar\L_{21}^+
\nonumber\\
&&+\underbrace{\left[2\alpha\kappa_0 \bar Q_{20} + 4\hbar^2 
\left( 3\chi+\bar\chi \right)\frac{I_1}{A_1A_2}\left(\bar Q_{20}+\bar Q_{00}\right)
\right]\R^\u_{22}},
\nonumber\\
\dot {\bar\L}^\d_{20}&=&
\frac{1}{m}\bar\P_{20}^\d-
\left[m\,\omega^2
+2\kappa_0 Q_{20}\right]\bar\R^\d_{20}
+2\sqrt2\kappa_0 Q_{20}\,\bar\R^\d_{00}
-i\hbar\frac{\eta}{2}\sqrt{\frac{3}{2}}\bar\L_{21}^+ 
\nonumber\\
&&+\frac{\hbar^2}{15} 
\left( 3\chi-\bar\chi \right)\frac{I_1}{A_1A_2} \,
\left[
Q_{00}\bar\R_{20}^\d+
Q_{20}\bar\R_{00}^\d/\sqrt2
\right],
\nonumber\\
&&\underbrace{-2\alpha\kappa_0 \bar Q_{20}\left[ \R_{20}^\d+\sqrt2\R_{00}^\d\right]
+ \frac{\hbar^2}{15} 
\left( 3\chi+\bar\chi \right)\frac{I_1}{A_1A_2} \,
\left[
\bar Q_{00}\R_{20}^\d+
\bar Q_{20}\R_{00}^\d/\sqrt2
\right]
},
\nonumber\\
\dot {\bar\L}^{+}_{11}&=&
-3(1-\alpha)\kappa_0 Q_{20}\,\bar\R^{+}_{21}
-i\hbar\frac{\eta}{2}\left[\bar\L_{11}^- 
+\sqrt2\bar\L^\d_{10}\right]
+\underbrace{3(1-\alpha)\kappa_0 \bar Q_{20}\,\R^{+}_{21}},
\nonumber\\
     \dot {\bar\L}^{-}_{11}&=&
-\left[3\kappa_0 Q_{20}
+\frac{\hbar^2}{20} 
\left( 3\chi-\bar\chi \right)\frac{I_1}{A_1A_2} Q_{20}
\right]\bar\R^{-}_{21}
 +\frac{4}{\hbar} I_{rp}^{\kappa\Delta}(r') {\bar{\tilde\L}}_{11}
 -\hbar\frac{\eta}{2}\left[i\bar\L_{11}^+
+\hbar \bar\F^\d\right]
\nonumber\\
&&
-\underbrace{\left[3\alpha\kappa_0 \bar Q_{20}
+\frac{\hbar^2}{20}  
\left( 3\chi+\bar\chi \right)\frac{I_1}{A_1A_2}\bar Q_{20}\right]\R^{-}_{21}
+\frac{4}{\hbar}\bar I_{rp}^{\kappa\Delta}(r') {{\tilde\L}}_{11}
},
\nonumber\\
     \dot{\bar\L}^\d_{10}&=&
-\hbar\frac{\eta}{2\sqrt2}\left[i\bar\L_{11}^+
+\hbar \bar\F^\d\right],
\nonumber\\
     \dot{\bar\F}^\d&=&
-\eta\left[\bar\L_{11}^- +\sqrt2\bar\L^\d_{10}\right],
\nonumber
\end{eqnarray}\begin{eqnarray}
\dot {\bar\R}^{+}_{21}&=&
\frac{2}{m}\bar\L_{21}^{+}
-i\hbar\frac{\eta}{2}\left[\bar\R_{21}^-
+2\bar\R^\u_{22}+
\sqrt6\bar\R^\d_{20}\right],
\nonumber\\
     \dot {\bar\R}^{-}_{21}&=&
\frac{2}{m}\bar\L_{21}^{-}
-i\hbar\frac{\eta}{2}\bar\R_{21}^+,
\nonumber\\
     \dot {\bar\R}^\u_{22}&=&
\frac{2}{m}\bar\L_{22}^\u
-i\hbar\frac{\eta}{2}\bar\R_{21}^+,
\nonumber\\
     \dot {\bar\R}^\d_{20}&=&
\frac{2}{m}\bar\L_{20}^\d
-i\hbar\frac{\eta}{2}\sqrt{\frac{3}{2}}\bar\R_{21}^+,
\nonumber\\
     \dot {\bar\P}^{+}_{21}&=&
-2\left[m\,\omega^2+\kappa_0 Q_{20}\right]\bar\L^{+}_{21}
+6\kappa_0 Q_{20}\bar\L^{+}_{11}
-i\hbar\frac{\eta}{2}\left[\bar\P_{21}^- 
+2\bar\P^\u_{22}+\sqrt6\bar\P^\d_{20}\right]
\nonumber\\
&&+\frac{3}{8} \hbar^2\chi\frac{I_2}{A_1A_2}\left[\left(Q_{20}+4 Q_{00}\right)\bar\L_{21}^{+} +3 Q_{20} \bar\L_{11}^{+}\right]
+\frac{4}{\hbar}|V_0| I_{pp}^{\kappa\Delta}(r') {\bar{\tilde\P}}_{21}
\nonumber\\
&&+\underbrace{2\alpha\kappa_0 \bar Q_{20}\left(3\L^{+}_{11}-\L^{+}_{21}\right)
+\frac{3}{8} \hbar^2\chi\frac{I_2}{A_1A_2}\left[\left(\bar Q_{20}+4\bar Q_{00}\right)\L_{21}^{+} +3\bar Q_{20} \L_{11}^{+}\right]
+\frac{4}{\hbar}|V_0| \bar I_{pp}^{\kappa\Delta}(r') {{\tilde\P}}_{21}
},
\nonumber\\
     \dot {\bar\P}^{-}_{21}&=&
-2\left[m\,\omega^2+\kappa_0 Q_{20}\right]\bar\L^{-}_{21}
+6\kappa_0 Q_{20}\bar\L^{-}_{11}
-6\sqrt2\alpha\kappa_0 L_{10}^-({\rm eq})\bar\R^{+}_{21}
-i\hbar\frac{\eta}{2}\bar\P_{21}^{+}
\nonumber\\
&&+\frac{3}{8} \hbar^2\chi\frac{I_2}{A_1A_2}\left[\left(Q_{20}+4 Q_{00}\right)\bar\L_{21}^{-} +3 Q_{20} \bar\L_{11}^{-}\right]
\nonumber\\
&&+\underbrace{2\alpha\kappa_0 \bar Q_{20}\left(3\L^{-}_{11}-\L^{-}_{21}\right)
+\frac{3}{8} \hbar^2\chi\frac{I_2}{A_1A_2}\left[\left(\bar Q_{20}+4\bar Q_{00}\right)\L_{21}^{-} +3\bar Q_{20} \L_{11}^{-}\right]
-6\sqrt2\kappa_0 \bar L_{10}^-({\rm eq})\R^{+}_{21}
},
\nonumber\\
     \dot {\bar\P}^\u_{22}&=&
-2\left[m\,\omega^2-2\kappa_0 Q_{20}\right]\bar\L^\u_{22}
-i\hbar\frac{\eta}{2}\bar\P_{21}^{+}
+\frac{3}{2}\hbar^2 \chi \frac{I_2}{A_1A_2}\left(Q_{20}+Q_{00}\right)
\bar\L^\u_{22}
\nonumber\\
&&+\underbrace{4\alpha\kappa_0 \bar Q_{20} \L^\u_{22}
+\frac{3}{2}\hbar^2 \chi \frac{I_2}{A_1A_2}\left(\bar Q_{20}+\bar Q_{00}\right)\L^\u_{22}
},
\nonumber\\
     \dot {\bar\P}^\d_{20}&=&
-2\left[m\,\omega^2+2\kappa_0 Q_{20}\right]\bar\L^\d_{20}
+4\sqrt2\kappa_0 Q_{20}\bar\L^\d_{00}
-i\hbar\frac{\eta}{2}\sqrt{\frac{3}{2}}\bar\P_{21}^{+}
+\frac{3}{2}\hbar^2 \chi \frac{I_2}{A_1A_2}\left[Q_{00}\bar\L_{20}^\d+Q_{20} \bar\L_{00}^\d/ \sqrt2\right]
\nonumber\\
&&+\underbrace{4\alpha\kappa_0\bar Q_{20}\left(\sqrt2\L^\d_{00}-\L^\d_{20}\right)
+\frac{3}{2}\hbar^2 \chi \frac{I_2}{A_1A_2}\left[\bar Q_{00}\L_{20}^\d+\bar Q_{20} \L_{00}^\d/ \sqrt2\right]
}
\nonumber\\
     \dot {\bar\L}^\d_{00}&=&
\frac{1}{m}\bar\P_{00}^\d-m\,\omega^2\bar\R^\d_{00}
+2\sqrt2\kappa_0 Q_{20}\bar\R^\d_{20}
+\frac{\hbar^2}{4\,A_1A_2} 
\left[\left( \chi-\frac{\bar\chi}{3} \right)I_1-\frac{9}{4}\chi I_2\right]
\left[\left(2 Q_{00}+Q_{20}\right)\bar\R_{00}^\d+
\sqrt2 Q_{20}\bar\R_{20}^\d
\right]
\nonumber\\
&&+\underbrace{2\sqrt2\alpha\kappa_0\bar Q_{20} \R^\d_{20}
+\frac{\hbar^2}{4\,A_1A_2} 
\left[\left( \chi+\frac{\bar\chi}{3} \right)I_1-\frac{9}{4}\chi I_2\right]
\left[\left(2 \bar Q_{00}+\bar Q_{20}\right)\R_{00}^\d+
\sqrt2\bar Q_{20}\R_{20}^\d
\right]
},
\nonumber\\
     \dot {\bar\R}^\d_{00}&=&
\frac{2}{m}\bar\L_{00}^\d,
\nonumber\\
     \dot {\bar\P}^\d_{00}&=&
-2m\,\omega^2\bar\L^\d_{00}
+4\sqrt2\kappa_0 Q_{20}\bar\L^\d_{20}
+\frac{3}{4}\hbar^2 
\chi \frac{I_2}{A_1A_2}
\left[\left(2 Q_{00}+Q_{20}\right)\bar \L_{00}^\d+
\sqrt2 Q_{20}\bar \L_{20}^\d
\right]
 \nonumber\\
&&+\underbrace{4\sqrt2\alpha\kappa_0\bar Q_{20}\L^\d_{20}
+\frac{3}{4}\hbar^2 
\chi \frac{I_2}{A_1A_2}
\left[\left(2 \bar Q_{00}+\bar Q_{20}\right)\L_{00}^\d+
\sqrt2\bar Q_{20}\L_{20}^\d
\right]
},
\nonumber\\ 
     \dot{\bar{\tilde\P}}_{21} &=& -\frac{1}{\hbar}\Delta(r') \bar\P^+_{21} + 
 6 \hbar\alpha\kappa_0 K_0{\bar\R}^+_{21}
\underbrace{-\frac{1}{\hbar}\bar \Delta(r') \P^+_{21} + 
 6 \hbar\kappa_0\bar K_0{\R}^+_{21}
}, 
\nonumber\\
     \dot{\bar{\tilde\L}}_{21} &=& -\frac{1}{\hbar}\Delta(r') \bar\L^-_{21}-\underbrace{\frac{1}{\hbar}\bar \Delta(r') \L^-_{21}},
\nonumber\\
     \dot{\bar{\tilde\L}}_{11} &=& -\frac{1}{\hbar}\Delta(r') \bar\L^-_{11}-\underbrace{\frac{1}{\hbar}\bar \Delta(r') \L^-_{11}},
\label{iv}\end{eqnarray}

The set of equations for isoscalar variables reads:

\begin{eqnarray}
     \dot {\L}^{+}_{21}&=&
\frac{1}{m}\P_{21}^{+}
-\left[m\,\omega^2+2\kappa_0 \left(2 Q_{00}+Q_{20}\right)\right]\R^{+}_{21}
-i\hbar\frac{\eta}{2}\left[\L_{21}^-
+2\L^\u_{22}+
\sqrt6\L^\d_{20}\right]
\nonumber\\&&-\underbrace{\alpha\kappa_0\left(2\bar Q_{00}+ \bar Q_{20}\right)
\bar\R^{+}_{21}}_{\text{coupling term}},
\nonumber\\
     \dot {\L}^{-}_{21}&=&
\frac{1}{m}\P_{21}^{-}
-\left[m\,\omega^2+\kappa_0 Q_{20}
-\frac{\hbar^2}{15} \left( 3\chi+\bar\chi \right) \frac{I_1}{A_1A_2} \left(Q_{00}+Q_{20}/4\right)
\right]\R^{-}_{21}-i\hbar\frac{\eta}{2}\L_{21}^+
 +\frac{4}{\hbar} I_{rp}^{\kappa\Delta}(r') {{\tilde\L}}_{21}
\nonumber\\
&&
-\underbrace{\left[\alpha\kappa_0 \bar Q_{20}
- \frac{\hbar^2}{15}
\left( 3\chi-\bar\chi \right) \frac{I_1}{A_1A_2} \left(\bar Q_{00}+\bar Q_{20}/4\right)\right]\bar\R^{-}_{21}
+\frac{4}{\hbar}\bar I_{rp}^{\kappa\Delta}(r') {\bar{\tilde\L}}_{21}
},
\nonumber\\
     \dot {\L}^\u_{22}&=&
\frac{1}{m}\P_{22}^\u-
\left[m\,\omega^2-2\kappa_0 Q_{20}
-4\hbar^2 
\left( 3\chi+\bar\chi \right)\frac{I_1}{A_1A_2}\left(Q_{20}+Q_{00}\right)
\right]\R^\u_{22}
-i\hbar\frac{\eta}{2}\L_{21}^+
\nonumber\\
&&+\underbrace{\left[2\alpha\kappa_0 \bar Q_{20} + 4\hbar^2 
\left( 3\chi-\bar\chi \right)\frac{I_1}{A_1A_2}\left(\bar Q_{20}+\bar Q_{00}\right)
\right]\bar\R^\u_{22}},
\nonumber\\
\dot {\L}^\d_{20}&=&
\frac{1}{m}\P_{20}^\d-
\left[m\,\omega^2
+2\kappa_0 Q_{20}\right]\R^\d_{20}
+2\sqrt2\kappa_0 Q_{20}\,\R^\d_{00}
-i\hbar\frac{\eta}{2}\sqrt{\frac{3}{2}}\L_{21}^+ 
\nonumber\\
&&+\frac{\hbar^2}{15} 
\left( 3\chi+\bar\chi \right)\frac{I_1}{A_1A_2} \,
\left[
Q_{00}\R_{20}^\d+
Q_{20}\R_{00}^\d/\sqrt2
\right],
\nonumber\\
&&\underbrace{-2\alpha\kappa_0 \bar Q_{20}\left[ \bar\R_{20}^\d+\sqrt2\bar\R_{00}^\d\right]
+ \frac{\hbar^2}{15} 
\left( 3\chi-\bar\chi \right)\frac{I_1}{A_1A_2} \,
\left[
\bar Q_{00}\bar\R_{20}^\d+
\bar Q_{20}\bar\R_{00}^\d/\sqrt2
\right]
},
\nonumber\\
\dot {\L}^{+}_{11}&=&
-i\hbar\frac{\eta}{2}\left[\L_{11}^- +\sqrt2\L^\d_{10}\right],
\nonumber\\
     \dot {\L}^{-}_{11}&=&
-\left[3\kappa_0 Q_{20}
+\frac{\hbar^2}{20} 
\left( 3\chi+\bar\chi \right)\frac{I_1}{A_1A_2} Q_{20}
\right]\R^{-}_{21}
 +\frac{4}{\hbar} I_{rp}^{\kappa\Delta}(r') {{\tilde\L}}_{11}
 -\hbar\frac{\eta}{2}\left[i\L_{11}^+
+\hbar \F^\d\right]
\nonumber\\
&&
-\underbrace{\left[3\alpha\kappa_0 \bar Q_{20}
+\frac{\hbar^2}{20}  
\left( 3\chi-\bar\chi \right)\frac{I_1}{A_1A_2}\bar Q_{20}\right]\bar\R^{-}_{21}
+\frac{4}{\hbar}\bar I_{rp}^{\kappa\Delta}(r') {\bar{\tilde\L}}_{11}
},
\nonumber\\
     \dot{\L}^\d_{10}&=&
-\hbar\frac{\eta}{2\sqrt2}\left[i\L_{11}^+
+\hbar \F^\d\right],
\nonumber\\
     \dot{\F}^\d&=&
-\eta\left[\L_{11}^- +\sqrt2\L^\d_{10}\right],
\nonumber\\
\dot {\R}^{+}_{21}&=&
\frac{2}{m}\L_{21}^{+}
-i\hbar\frac{\eta}{2}\left[\R_{21}^-
+2\R^\u_{22}+
\sqrt6\R^\d_{20}\right],
\nonumber\\
     \dot {\R}^{-}_{21}&=&
\frac{2}{m}\L_{21}^{-}
-i\hbar\frac{\eta}{2}\R_{21}^+,
\nonumber\\
     \dot {\R}^\u_{22}&=&
\frac{2}{m}\L_{22}^\u
-i\hbar\frac{\eta}{2}\R_{21}^+,
\nonumber\\
     \dot {\R}^\d_{20}&=&
\frac{2}{m}\L_{20}^\d
-i\hbar\frac{\eta}{2}\sqrt{\frac{3}{2}}\R_{21}^+,
\nonumber\\
     \dot {\P}^{+}_{21}&=&
-2\left[m\,\omega^2+\kappa_0 Q_{20}\right]\L^{+}_{21}
+6\kappa_0 Q_{20}\L^{+}_{11}
-i\hbar\frac{\eta}{2}\left[\P_{21}^- 
+2\P^\u_{22}+\sqrt6\P^\d_{20}\right]
\nonumber\\
&&+\frac{3}{8} \hbar^2\chi\frac{I_2}{A_1A_2}\left[\left(Q_{20}+4 Q_{00}\right)\L_{21}^{+} +3 Q_{20} \L_{11}^{+}\right]
+\frac{4}{\hbar}|V_0| I_{pp}^{\kappa\Delta}(r') {{\tilde\P}}_{21}
\nonumber\\
&&+\underbrace{2\alpha\kappa_0 \bar Q_{20}\left(3\bar\L^{+}_{11}-\bar\L^{+}_{21}\right)
+\frac{3}{8} \hbar^2\chi\frac{I_2}{A_1A_2}\left[\left(\bar Q_{20}+4\bar Q_{00}\right)\bar\L_{21}^{+} +3\bar Q_{20} \bar\L_{11}^{+}\right]
+\frac{4}{\hbar}|V_0| \bar I_{pp}^{\kappa\Delta}(r') {\bar{\tilde\P}}_{21}
},
\nonumber\\
     \dot {\P}^{-}_{21}&=&
-2\left[m\,\omega^2+\kappa_0 Q_{20}\right]\L^{-}_{21}
+6\kappa_0 Q_{20}\L^{-}_{11}
-6\sqrt2\kappa_0 L_{10}^-({\rm eq})\R^{+}_{21}
-i\hbar\frac{\eta}{2}\P_{21}^{+}
\nonumber\\
&&+\frac{3}{8} \hbar^2\chi\frac{I_2}{A_1A_2}\left[\left(Q_{20}+4 Q_{00}\right)\L_{21}^{-} +3 Q_{20} \L_{11}^{-}\right]
\nonumber\\
&&+\underbrace{2\alpha\kappa_0 \bar Q_{20}\left(3\bar\L^{-}_{11}-\bar\L^{-}_{21}\right)
+\frac{3}{8} \hbar^2\chi\frac{I_2}{A_1A_2}\left[\left(\bar Q_{20}+4\bar Q_{00}\right)\bar\L_{21}^{-} +3\bar Q_{20} \bar\L_{11}^{-}\right]
-6\sqrt2\alpha\kappa_0 \bar L_{10}^-({\rm eq})\bar\R^{+}_{21}
},
\nonumber
\end{eqnarray}\begin{eqnarray}
     \dot {\P}^\u_{22}&=&
-2\left[m\,\omega^2-2\kappa_0 Q_{20}\right]\L^\u_{22}
-i\hbar\frac{\eta}{2}\P_{21}^{+}
+\frac{3}{2}\hbar^2 \chi \frac{I_2}{A_1A_2}\left(Q_{20}+Q_{00}\right)
\L^\u_{22}
\nonumber\\
&&+\underbrace{4\alpha\kappa_0 \bar Q_{20} \bar\L^\u_{22}
+\frac{3}{2}\hbar^2 \chi \frac{I_2}{A_1A_2}\left(\bar Q_{20}+\bar Q_{00}\right)\bar\L^\u_{22}
},
\nonumber\\
     \dot{\P}^\d_{20}&=&
-2\left[m\,\omega^2+2\kappa_0 Q_{20}\right]\L^\d_{20}
+4\sqrt2\kappa_0 Q_{20}\L^\d_{00}
-i\hbar\frac{\eta}{2}\sqrt{\frac{3}{2}}\P_{21}^{+}
+\frac{3}{2}\hbar^2 \chi \frac{I_2}{A_1A_2}\left[Q_{00}\L_{20}^\d+Q_{20} \L_{00}^\d/ \sqrt2\right]
\nonumber\\
&&+\underbrace{4\alpha\kappa_0\bar Q_{20}\left(\sqrt2\bar\L^\d_{00}-\bar\L^\d_{20}\right)
+\frac{3}{2}\hbar^2 \chi \frac{I_2}{A_1A_2}\left[\bar Q_{00}\L_{20}^\d+\bar Q_{20} \bar\L_{00}^\d/ \sqrt2\right]
}
\nonumber\\
     \dot {\L}^\d_{00}&=&
\frac{1}{m}\P_{00}^\d-m\,\omega^2\R^\d_{00}
+2\sqrt2\kappa_0 Q_{20}\R^\d_{20}
+\frac{\hbar^2}{4\,A_1A_2} 
\left[\left( \chi+\frac{\bar\chi}{3} \right)I_1-\frac{9}{4}\chi I_2\right]
\left[\left(2 Q_{00}+Q_{20}\right)\R_{00}^\d+
\sqrt2 Q_{20}\R_{20}^\d
\right]
\nonumber\\
&&+\underbrace{2\sqrt2\alpha\kappa_0\bar Q_{20} \bar\R^\d_{20}
+\frac{\hbar^2}{4\,A_1A_2} 
\left[\left( \chi-\frac{\bar\chi}{3} \right)I_1-\frac{9}{4}\chi I_2\right]
\left[\left(2 \bar Q_{00}+\bar Q_{20}\right)\bar\R_{00}^\d+
\sqrt2\bar Q_{20}\bar\R_{20}^\d
\right]
},
\nonumber\\
     \dot {\R}^\d_{00}&=&
\frac{2}{m}\L_{00}^\d,
\nonumber\\
     \dot {\P}^\d_{00}&=&
-2m\,\omega^2\L^\d_{00}
+4\sqrt2\kappa_0 Q_{20}\L^\d_{20}
+\frac{3}{4}\hbar^2 
\chi \frac{I_2}{A_1A_2}
\left[\left(2 Q_{00}+Q_{20}\right)\bar \L_{00}^\d+
\sqrt2 Q_{20}\bar \L_{20}^\d
\right]
 \nonumber\\
&&+\underbrace{4\sqrt2\alpha\kappa_0\bar Q_{20}\bar\L^\d_{20}
+\frac{3}{4}\hbar^2 
\chi \frac{I_2}{A_1A_2}
\left[\left(2 \bar Q_{00}+\bar Q_{20}\right)\bar\L_{00}^\d+
\sqrt2\bar Q_{20}\bar\L_{20}^\d
\right]
},
\nonumber\\ 
     \dot{{\tilde\P}}_{21} &=& -\frac{1}{\hbar}\Delta(r') \P^+_{21} + 
 6 \hbar\kappa_0 K_0\R^+_{21}
\underbrace{-\frac{1}{\hbar}\bar \Delta(r') \bar\P^+_{21} + 
 6 \hbar\alpha\kappa_0\bar K_0\bar\R^+_{21}
}, 
\nonumber\\
     \dot{{\tilde\L}}_{21} &=& -\frac{1}{\hbar}\Delta(r') \L^-_{21}-\underbrace{\frac{1}{\hbar}\bar \Delta(r') \bar\L^-_{21}},
\nonumber\\
     \dot{{\tilde\L}}_{11} &=& -\frac{1}{\hbar}\Delta(r') \L^-_{11}-\underbrace{\frac{1}{\hbar}\bar \Delta(r') \bar\L^-_{11}},
\label{is}\end{eqnarray}

\end{widetext}
where the terms coupling isovector and isoscalar sets of equations are 
underlined by the braces and
\begin{eqnarray}
\label{Ai}
A_1&=&\sqrt2\, R_{20}^{\rm eq}-R_{00}^{\rm eq}=\frac{Q_{00}}{\sqrt3}\left(1+\frac{4}{3}\delta\right),
\nonumber\\
A_2&=& R_{20}^{\rm eq}/\sqrt2+R_{00}^{\rm eq}=-\frac{Q_{00}}{\sqrt3}\left(1-\frac{2}{3}\delta\right),\quad
\end{eqnarray}
$Q_{00}=\frac{3}{5}AR^2/
[(1+\frac{4}{3}\delta)^{1/3}(1-\frac{2}{3}\delta)^{2/3}],$\\
$\delta$ -- deformation parameter,\quad $Q_{20}=\frac{4}{3}\delta Q_{00},$\\
$\bar Q_{00}=Q_{00}^{\rm n}-Q_{00}^{\rm p},$ \,
$\bar Q_{20}=Q_{20}^{\rm n}-Q_{20}^{\rm p},$ \\
$\bar \Delta=\Delta^{\rm n}-\Delta^{\rm p}, \,\,R=r_0A^{1/3},$ \,
$r_0=1.2$~fm,\, 
$$I_1=\frac{\pi}{4}\int\limits_{0}^{\infty}dr\, r^4\left(\frac{\partial n(r)}{\partial r}\right)^2,\
I_2=\frac{\pi}{4}\int\limits_{0}^{\infty}dr\, r^2 n(r)^2,$$
$n(r)= n_0\left(1+{\rm e}^{\frac{r-R}{a}}\right)^{-1}$ -- nuclear density, $a=0.53$,
$$K_0=\int d(\br,\bp) \kappa_0(\br,\bp),\quad
\bar K_0=K_0^{\rm n}-K_0^{\rm p}.$$
The functions $\Delta(r')$, $I_{rp}^{\kappa\Delta}(r')$ and 
$I_{pp}^{\kappa\Delta}(r')$
are outlined in Appendix.
Deriving these equations we neglected double Poisson brackets containing $\kappa$~or~$\Delta$,
which are the quantum corrections to pair correlations.

\section{Results of calculations and discussion}

The calculations were performed for nuclei of the $N=82-126$ mass region and
for Actinides. The calculations procedure and parameters are mostly the same
as in our previous paper \cite{BaMoPRC1}.

\subsection{Nuclei of the $N=82-126$ mass region}

\subsubsection{WFM}

\begin{figure*}
\centering{
\includegraphics[width=0.27\textwidth]{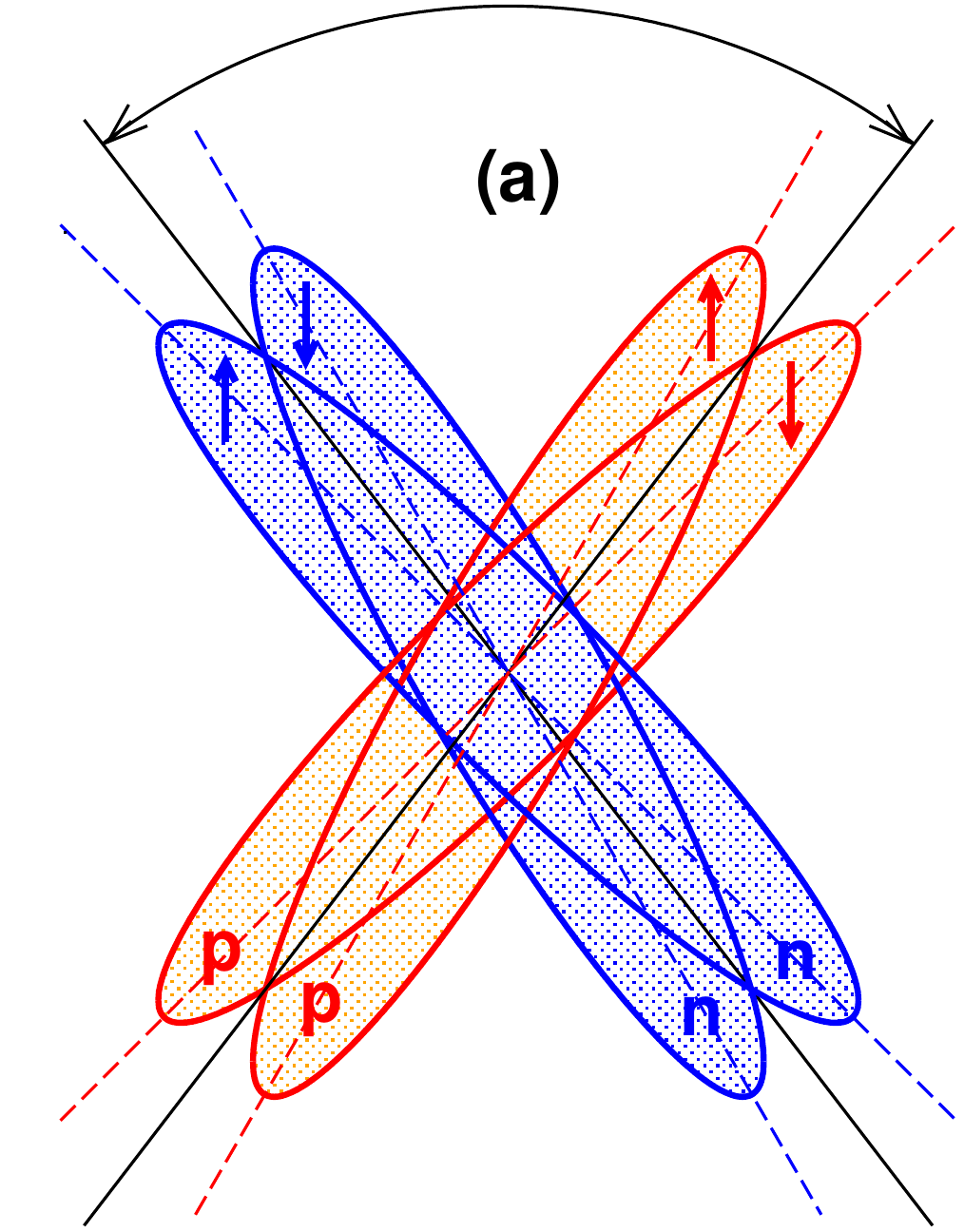}\hspace*{5mm}
\includegraphics[width=0.27\textwidth]{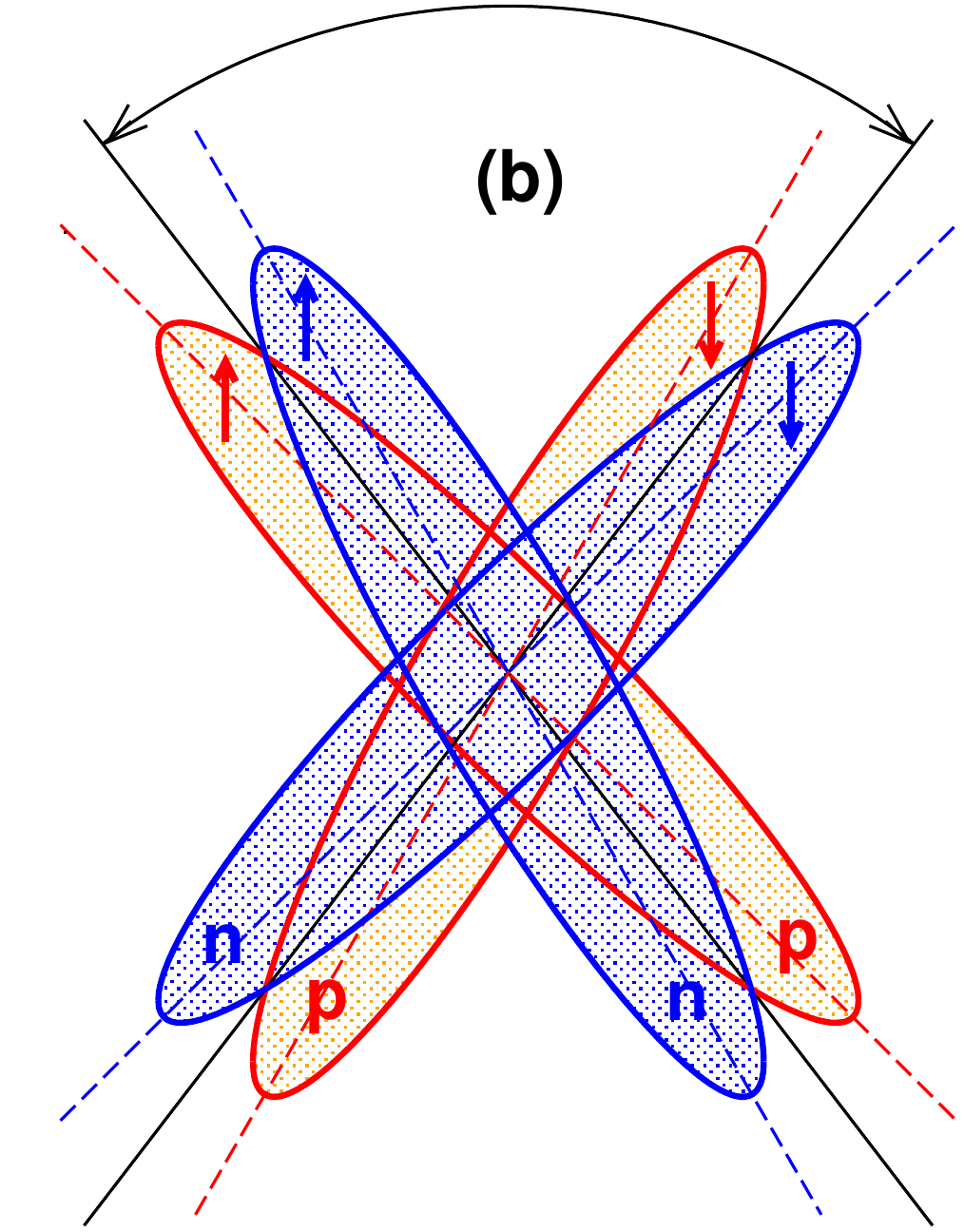}\hspace*{5mm}
\includegraphics[width=0.27\textwidth]{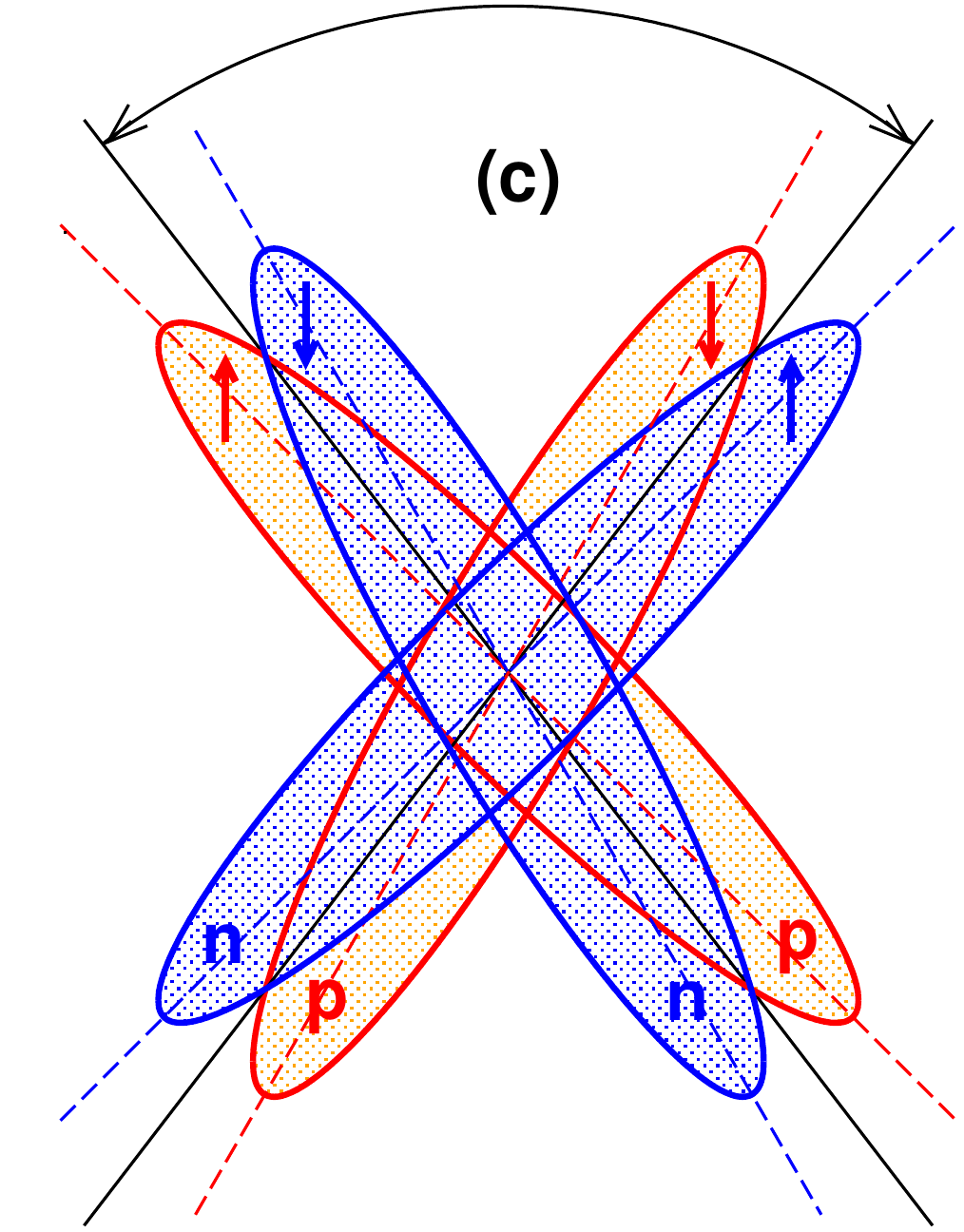}}
\caption{Schematic representation of three scissors modes:
(a)~spin-scalar isovector (conventional, orbital scissors), (b)~spin-vector isoscalar (spin scissors), (c)~spin-vector isovector (spin scissors).
Arrows show the direction of spin projections;
p -- protons, n -- neutrons. The small angle spread between the various distributions is only for presentation purposes. 
In reality the distributions are perfectly overlapping. 
}\label{fig1}
\end{figure*}

Let us analyze in detail the results of systematic calculations for nuclei of 
this mass region considering the example of Dy isotopes. The most interesting of them is $^{164}$Dy, where a  rather exceptional experimental situation 
with the low-lying $1^+$ excitations exists. The results of the solution of 
equations (\ref{iv}, \ref{is}) for this nucleus are presented in the 
Table~\ref{tab1}, where the  energies of $1^+$ levels with their magnetic dipole 
and electric quadrupole strengths are shown.
Left panel -- the solutions of decoupled equations, right -- isoscalar-isovector coupling is taken into account.
\begin{table}[h!]
\caption{The results of WFM calculations for $^{164}$Dy for energies $E$ (MeV), magnetic dipole $B(M1)$ ($\mu_N^2$) 
and electric quadrupole $B(E2)$ (W. u.) strengths of $1^+$ excitations. 
The marks IS -- isoscalar and IV -- isovector are valid only
for the decoupled case.}
\begin{ruledtabular}\begin{tabular}{lccccccc}
         \multicolumn{4}{c}{Decoupled equations}  &                       
        &\multicolumn{3}{c}{Coupled equations  }\\  
 \cline{1-4}   \cline{6-8}
     & $E$   & $B(M1)$    & $B(E2)$  & &
       $E$   & $B(M1)$    & $B(E2)$  \\    
 \cline{2-4}   \cline{6-8} 
  IS    &       1.29  &      0.01  & 53.25 & &       1.47  &      0.17  & 25.44  \\
  IV    &  {\bf 2.44} & {\bf 2.03} &  0.34 & &  {\bf 2.20} & {\bf 1.76} &  3.30  \\
  IS    &  { 2.62} & { 0.09} &  2.91 & &  {\bf 2.87} & {\bf 2.24} &  0.34  \\ 
  IV    &  {\bf 3.35} & {\bf 1.36} &  1.62 & &  {\bf 3.59} & {\bf 1.56} &  4.37  \\*[.5mm] 
  IS    &      10.94  & 0.00 & 55.12 &  & 10.92 & 0.04 & 50.37  \\ 
  IV    &      14.04  & 0.00 &  2.78 &  & 13.10 & 0.00 &  2.85  \\
  IS    &      14.60  & 0.06 &  0.48 &  & 15.42 & 0.07 &  0.57  \\
  IV    &      15.88  & 0.00 &  0.55 &  & 15.55 & 0.00 &  1.12  \\ 
  IS    &      16.46  & 0.07 &  0.36 &  & 16.78 & 0.06 &  0.53  \\ 
  IV    &      17.69  & 0.00 &  0.45 &  & 17.69 & 0.01 &  0.68  \\
  IS    &      17.90  & 0.00 &  0.51 &  & 17.91 & 0.00 &  0.53  \\ 
  IV    &      18.22  & 0.18 &  1.85 &  & 18.22 & 0.13 &  0.89  \\ 
  IS    &      19.32  & 0.10 &  0.97 &  & 19.32 & 0.08 &  0.61  \\ 
  IV    &      21.29  & 2.47 & 31.38 &  & 21.26 & 2.03 & 21.60  \\            
\end{tabular}\end{ruledtabular}
\label{tab1}
\end{table} 
The first observation is that the high-lying levels are less sensitive to decoupling.
Among the high-lying states, $\mu=1$ branches of isoscalar (at the energy of $10.92$ MeV) and isovector ($E=21.26$ MeV) Giant Quadrupole Resonances
are distinguished by the large $B(E2)$ values. 
The rest of high-lying states have quite small excitation probabilities and we omit them from further discussion.

Comparing the left and right panels, we see that the most remarkable change happens with the third low-lying level, an isoscalar one without coupling,
-- it acquires a rather big magnetic strength.
The "jump" from 0.09 $\mu_N^2$ to 2.24 $\mu_N^2$ looks a little bit 
surprising. However it is explained quite naturally by the structure of the
matrix element of the excitation operator (see Appendix C). According to 
formula (\ref{psiO}) the contribution of isoscalar variables occurs  with the
factor $[g_s^{\rm n}+g_s^{\rm p}-g_l^{\rm p}]$. Its numerical value 
(including the quenching factor $q=0.7$) is $0.23$. The contribition of 
isovector variables goes with the factors $\frac{1}{2}(g_s^{\rm p}-g_s^{\rm n})=3.29$ and
$g_l^{\rm p}=1$, i.e. $\sim 20$ times bigger than isoscalar one. In the decoupled 
case the third level, being the isoscalar one,
has the contribution only from isoscalar variables, which is 
obviously small. In the case with coupling it gets the additional
contribution from isovector variables, which is an order of magnitude bigger.
This  explains the effect of the big increase of the $B(M1)$ value.

So, in the decoupled case there are 2 isoscalar electric and 2 isovector magnetic low-lying levels, and
in the coupled case there are 1 electric and 3 magnetic
levels of  mixed isovector-isoscalar nature.
The interpretation of the lowest electric level is not clear at this moment and requires the separate investigation. 
 It shows practically zero $B(M1)$ strength and, thus, is not of direct interest to this work. Its nature shall be studied in future work. 


The three magnetic states correspond to three physically possible 
types of scissors modes already mentioned in the introduction. 
Roughly speaking the state at the energy $3.59$ MeV in $^{164}$Dy is predominately the conventional "orbital" scissors mode, the last two states at the energies $2.20$ MeV and $2.87$ MeV 
are predominately the "spin" scissors modes. 
The detailed analysis of these three states is given in section C, 
"Currents".

The Fig.~\ref{fig1} 
shows a schematic representation of these modes: 
the orbital scissors (neutrons versus protons)
and two spin scissors (spin-up nucleons versus spin-down nucleons and more complicated -- 
spin-up protons together with spin-down neutrons versus spin-down protons with spin-up neutrons).
 Both spin scissors exist only due to spin degrees of freedom.
 If we remove the arrows from the picture, nothing will change for the conventional scissors~(a).
However figures~(b) and~(c) in this case become identical and senseless,
because the division of identical particles (neutrons or protons) in two parts 
becomes irrelevant.

The natural question arises here: what is the origin of forces who coerce the
spin-up and spin-down particles to move out of phase? There is no analogous
problem with the conventional scissors, because the Hamiltonian (\ref{Ham}) 
includes the neutron-proton quadrupole-quadrupole (q-q) interaction 
(\ref{Hqq}), that makes protons and neutrons move out of phase. But what 
generates a similar motion of spin-up and spin-down particles? It turns out that again the main working element is the nucleon-nucleon q-q interaction.
However, this time it works together with the spin-orbital part of the mean 
field. 

Let us consider in detail the "life" of, for example, 
the system of spin-up protons and spin-down neutrons within the mean 
field. Due to the neutron-proton q-q interaction protons push neutrons and force them to move generating in such a way for example the scissors modes
(Fig. \ref{fig1}b)). Neutrons have spins, so due to the spin-orbital term 
their motion will depend on their spin projection. That means that the result of pushing will depend on the spin projections of the pushed neutrons. In addition,the pushing protons also have spins, therefore the result of pushing 
will depend on their spin projection too. Moreover, due to 
proton-proton q-q interaction spin-up protons will push spin-down protons 
and again the result of their interaction will be influenced by the 
spin-orbital potential. 
As we see, there is no necessity to introduce the special kind of interaction
to activate the spin degrees of freedom and to generate in such a way the
spin dependent excitation.
It is done quite 
naturally by the usual q-q interaction, the result of the activation being 
dependent on spin projections due to the spin-orbital potential, which can lead to the appearance of three different types of scissors.

\begin{figure}[t!]
\centering{\includegraphics[width=\columnwidth]{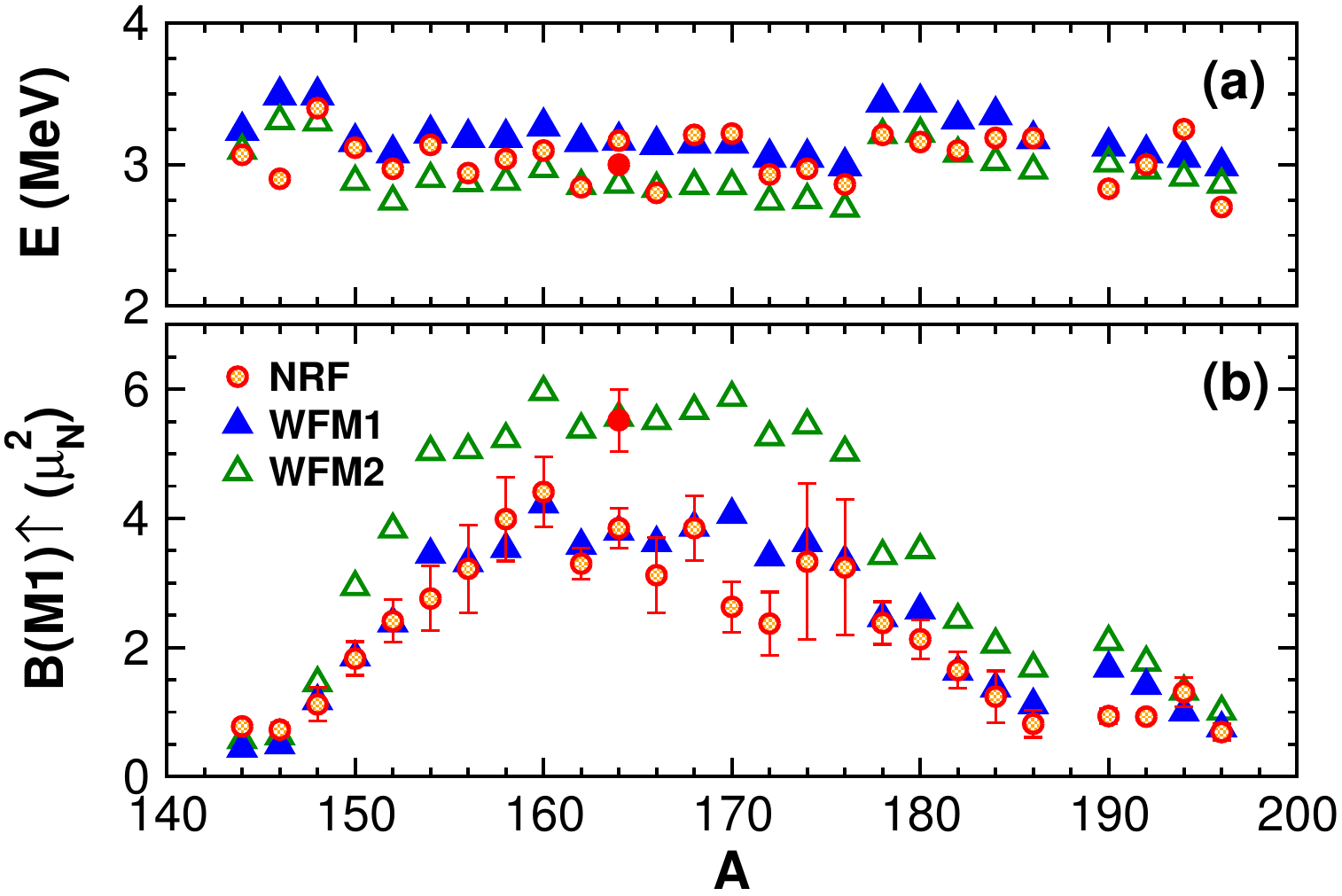}}
\caption{Calculated (WFM) and experimental (NRF) mean exitation energies~(a) 
and summed $M1$ strengths~(b) of the scissors mode. 
WFM1 -- the sum of two highest scissors, WFM2 -- the sum of three scissors. 
Experimental data are taken from the papers listed in the Table~1 of~\cite{Rich}. 
The solid circle marks the experimental result for $^{164}$Dy when summed in the energy range from 2 to 4 MeV.}
\label{fig2}
\end{figure}

So, the calculations without an artificial decoupling produce three low-lying 
magnetic states (instead of two without coupling).
The energies and $B(M1)$ values of all three types of scissors calculated for
other nuclei are shown in Tables \ref{tab11}, \ref{tab13s}.

\begin{table*}
\caption{The nuclear scissors mode fine structure. The results of calculations by the WFM method: energies $E_i$ with correspondings $B_i(M1)$-values.
Energy centroids $\bar E$ and summed $M1$ strength are also presented.
Parameters of pair correlations: $V_0^{\rm p}=27$~MeV, $V_0^{\rm n}=23$~MeV, $r_p^{\rm p}=1.50$~fm, $r_p^{\rm n}=1.85$~fm for nuclei with $A=150-186$.
The spin-orbit strength constant $\kappa_{\rm Nils}=0.0637$, quenching factor $q=0.7$. 
Parameters that differ from those indicated: $V_0^{\rm p}=26.5$~MeV, $V_0^{\rm n}=22.6$~MeV for Hf and W isotops,
$q=0.57$ for $^{150}$Sm,
$r_p^{\rm p}=1.57$~fm,  $q=0.78$ for $^{150}$Nd,
$V_0^{\rm p}=23$~MeV, $V_0^{\rm n}=20$~MeV, $r_p^{\rm p}=2.0$~fm, $r_p^{\rm n}=2.4$~fm,  $q=0.5$ for $^{148}$Sm and $^{148}$Nd,
$V_0^{\rm p}=23$~MeV, $V_0^{\rm n}=20$~MeV, $r_p^{\rm p}=2.05$~fm, $r_p^{\rm n}=2.5$~fm,  $\kappa_{\rm Nils}=0.05$, $q=0.5$ for $^{146}$Nd.
}
\begin{ruledtabular}\begin{tabular}{ccccccccccc}
 \multirow{2}{*}{Nuclei}&\multirow{2}{*}{$\delta$}  & \multirow{2}{*}{$i$} & $E_i$ (MeV)  & $B_i(M1)$ ($\mu_N^2$) &  \multicolumn{2}{c}{$\bar E_{[2-3]}$(MeV)} &  
 \multicolumn{2}{c}{$\sum\limits_{i=2}^{3}\!\!B_i(M1)$ ($\mu_N^2$)} &  $\bar E_{[1-3]}$ (MeV)  &  $\sum\limits_{i=1}^{3}\!\!B_i(M1)$ ($\mu_N^2$)\\   
   &  &   &\multicolumn{2}{c}{WFM}   & WFM & NRF &  WFM & NRF & \multicolumn{2}{c}{WFM}    \\
\hline\hline
             &      & 1 &  2.66 & 0.13  &                       &                       &                       &                           &      &       \\          
 $^{146}$Nd  & 0.13 & 2 &  3.41 & 0.38  & \multirow{2}{*}{3.49} & \multirow{2}{*}{2.90} & \multirow{2}{*}{0.49} & \multirow{2}{*}{0.73(10)} & 3.31 & 0.63  \\ 
             &      & 3 &  3.74 & 0.12  &                       &                       &                       &                           &      &       \\ 
\hline                                                                                                                
             &      & 1 &  2.51 & 0.28  &                       &                       &                       &                           &      &       \\
 $^{148}$Nd  & 0.17 & 2 &  3.36 & 0.80  & \multirow{2}{*}{3.49} & \multirow{2}{*}{3.40} & \multirow{2}{*}{1.17} & \multirow{2}{*}{1.12(26)} & 3.30 & 1.45  \\ 
             &      & 3 &  3.78 & 0.37  &                       &                       &                       &                           &      &       \\ 
 \hline                                                                                                               
             &      & 1 &  2.39 & 1.08  &                       &                       &                       &                           &      &       \\
 $^{150}$Nd  & 0.23 & 2 &  3.07 & 1.57  & \multirow{2}{*}{3.16} & \multirow{2}{*}{3.12} & \multirow{2}{*}{1.85} & \multirow{2}{*}{1.83(27)} & 2.88 & 2.94  \\ 
             &      & 3 &  3.70 & 0.28  &                       &                       &                       &                           &      &       \\ 
 \hline\hline
             &      & 1 &  2.48 & 0.08  &                       &                       &                       &                           &      &       \\          
 $^{148}$Sm  & 0.12 & 2 &  3.16 & 0.19  & \multirow{2}{*}{3.21} & \multirow{2}{*}{3.07} & \multirow{2}{*}{0.23} & \multirow{2}{*}{0.51(12)} & 3.02 & 0.31  \\ 
             &      & 3 &  3.48 & 0.04  &                       &                       &                       &                           &      &       \\ 
\hline 
             &      & 1 &  2.27 & 0.49  &                       &                       &                       &                           &      &       \\          
 $^{150}$Sm  & 0.16 & 2 &  2.68 & 0.23  & \multirow{2}{*}{3.00} & \multirow{2}{*}{3.18} & \multirow{2}{*}{0.61} & \multirow{2}{*}{0.97(17)} & 2.67 & 1.10  \\ 
             &      & 3 &  3.18 & 0.38  &                       &                       &                       &                           &      &       \\ 
\hline                                                                                                                
             &      & 1 &  2.18 & 1.45  &                       &                       &                       &                           &      &       \\
 $^{152}$Sm  & 0.24 & 2 &  2.75 & 1.31  & \multirow{2}{*}{3.08} & \multirow{2}{*}{2.97} & \multirow{2}{*}{2.38} & \multirow{2}{*}{2.41(33)} & 2.74 & 3.83  \\ 
             &      & 3 &  3.48 & 1.07  &                       &                       &                       &                           &      &       \\ 
 \hline                                                                                                               
             &      & 1 &  2.22 & 1.59  &                       &                       &                       &                           &      &       \\
 $^{154}$Sm  & 0.26 & 2 &  2.91 & 1.98  & \multirow{2}{*}{3.22} & \multirow{2}{*}{3.14} & \multirow{2}{*}{3.44} & \multirow{2}{*}{2.76(50)} & 2.90 & 5.03  \\ 
             &      & 3 &  3.64 & 1.46  &                       &                       &                       &                           &      &       \\ 
 \hline\hline                                                                                                  
             &      & 1 &  2.24 & 1.70  &                       &                       &                       &                           &      &       \\
 $^{154}$Gd  & 0.25 & 2 &  2.78 & 1.40  & \multirow{2}{*}{3.12} & \multirow{2}{*}{3.00} & \multirow{2}{*}{2.65} & \multirow{2}{*}{2.60(5)} & 2.78 & 4.35  \\ 
             &      & 3 &  3.51 & 1.25  &                       &                       &                       &                           &      &       \\ 
 \hline                                                                                                               
             &      & 1 &  2.25 & 1.75  &                       &                       &                       &                           &      &       \\
 $^{156}$Gd  & 0.26 & 2 &  2.86 & 1.84  & \multirow{2}{*}{3.19} & \multirow{2}{*}{2.94} & \multirow{2}{*}{3.31} & \multirow{2}{*}{3.22(68)} & 2.87 & 5.06  \\ 
             &      & 3 &  3.60 & 1.46  &                       &                       &                       &                           &      &       \\ 
\hline                                                                                                                
             &      & 1 &  2.22 & 1.70  &                       &                       &                       &                           &      &       \\
 $^{158}$Gd  & 0.26 & 2 &  2.88 & 2.04  & \multirow{2}{*}{3.19} & \multirow{2}{*}{3.04} & \multirow{2}{*}{3.53} & \multirow{2}{*}{3.99(65)} & 2.88 & 5.23  \\ 
             &      & 3 &  3.61 & 1.49  &                       &                       &                       &                           &      &       \\ 
\hline                                                                                                              
             &      & 1 &  2.23 & 1.74  &                       &                       &                       &                           &      &       \\
 $^{160}$Gd  & 0.27 & 2 &  2.97 & 2.50  & \multirow{2}{*}{3.27} & \multirow{2}{*}{3.10} & \multirow{2}{*}{4.22} & \multirow{2}{*}{4.41(54)} & 2.97 & 5.96  \\ 
             &      & 3 &  3.70 & 1.72  &                       &                       &                       &                           &      &       \\ 
\hline\hline                                                                                                              
             &      & 1 &  2.25 & 1.84  &                       &                       &                       &                           &      &       \\
 $^{160}$Dy  & 0.26 & 2 &  2.84 & 1.85  & \multirow{2}{*}{3.17} & \multirow{2}{*}{2.87} & \multirow{2}{*}{3.35} & \multirow{2}{*}{2.42(30)} & 2.84 & 5.19  \\ 
             &      & 3 &  3.56 & 1.50  &                       &                       &                       &                           &      &       \\ 
\hline                                                                                                                
             &      & 1 &  2.22 & 1.80  &                       &                       &                       &                           &      &       \\
 $^{162}$Dy  & 0.26 & 2 &  2.86 & 2.04  & \multirow{2}{*}{3.16} & \multirow{2}{*}{2.84} & \multirow{2}{*}{3.58} & \multirow{2}{*}{3.30(24)} & 2.85 & 5.38  \\ 
             &      & 3 &  3.57 & 1.53  &                       &                       &                       &                           &      &       \\ 
\hline                                                                                                              
             &      & 1 &  2.20 & 1.76  &                       &                       &                       &                           &      &       \\
 $^{164}$Dy  & 0.26 & 2 &  2.87 & 2.24  & \multirow{2}{*}{3.17} & \multirow{2}{*}{3.17} & \multirow{2}{*}{3.80} & \multirow{2}{*}{3.85(31)} & 2.86 & 5.56  \\ 
             &      & 3 &  3.59 & 1.56  &                       &                       &                       &                           &      &       \\ 
\end{tabular}\end{ruledtabular}\label{tab11}
\end{table*}

\begin{table*}
\caption{Continuation of Table II.}
\begin{ruledtabular}\begin{tabular}{ccccccccccc}
 \multirow{2}{*}{Nuclei}&\multirow{2}{*}{$\delta$}  & \multirow{2}{*}{$i$} & $E_i$ (MeV)  & $B_i(M1)$ ($\mu_N^2$) &  \multicolumn{2}{c}{$\bar E_{[2-3]}$(MeV)} &  
 \multicolumn{2}{c}{$\sum\limits_{i=2}^{3}\!\!B_i(M1)$ ($\mu_N^2$)} &  $\bar E_{[1-3]}$ (MeV)  &  $\sum\limits_{i=1}^{3}\!\!B_i(M1)$ ($\mu_N^2$)\\    
   &  &   &\multicolumn{2}{c}{WFM}   & WFM & NRF &  WFM & NRF & \multicolumn{2}{c}{WFM}    \\
\hline\hline

             &      & 1 &  2.23 & 1.89  &                       &                       &                       &                           &      &       \\
 $^{166}$Er  & 0.26 & 2 &  2.83 & 2.05  & \multirow{2}{*}{3.14} & \multirow{2}{*}{2.79} & \multirow{2}{*}{3.62} & \multirow{2}{*}{3.12(58)} & 2.83 & 5.51  \\ 
             &      & 3 &  3.54 & 1.57  &                       &                       &                       &                           &      &       \\ 
\hline            
             &      & 1 &  2.20 & 1.81  &                       &                       &                       &                           &      &       \\
 $^{168}$Er  & 0.26 & 2 &  2.87 & 2.28  & \multirow{2}{*}{3.15} & \multirow{2}{*}{3.21} & \multirow{2}{*}{3.86} & \multirow{2}{*}{3.85(50)} & 2.85 & 5.67  \\ 
             &      & 3 &  3.57 & 1.58  &                       &                       &                       &                           &      &       \\ 
 \hline           
             &      & 1 &  2.18 & 1.81  &                       &                       &                       &                           &      &       \\
 $^{170}$Er  & 0.26 & 2 &  2.86 & 2.43  & \multirow{2}{*}{3.15} & \multirow{2}{*}{3.22} & \multirow{2}{*}{4.06} & \multirow{2}{*}{2.63(39)} & 2.85 & 5.87  \\ 
             &      & 3 &  3.57 & 1.63  &                       &                       &                       &                           &      &       \\ 
\hline\hline                                                                                                       
             &      & 1 &  2.18 & 1.86  &                       &                       &                       &                           &      &       \\
 $^{172}$Yb  & 0.25 & 2 &  2.76 & 1.97  & \multirow{2}{*}{3.05} & \multirow{2}{*}{2.93} & \multirow{2}{*}{3.40} & \multirow{2}{*}{2.37(49)} & 2.74 & 5.26  \\ 
             &      & 3 &  3.45 & 1.43  &                       &                       &                       &                           &      &       \\ 
\hline                                                                                                                
             &      & 1 &  2.16 & 1.82  &                       &                       &                       &                           &      &       \\
 $^{174}$Yb  & 0.25 & 2 &  2.78 & 2.16  & \multirow{2}{*}{3.05} & \multirow{2}{*}{2.96} & \multirow{2}{*}{3.62} &\multirow{2}{*}{3.33(1.21)}& 2.75 & 5.44  \\ 
             &      & 3 &  3.46 & 1.46  &                       &                       &                       &                           &      &       \\ 
\hline                                                                                                                
             &      & 1 &  2.11 & 1.69  &                       &                       &                       &                           &      &       \\
 $^{176}$Yb  & 0.24 & 2 &  2.73 & 2.04  & \multirow{2}{*}{2.99} & \multirow{2}{*}{2.86} & \multirow{2}{*}{3.33} &\multirow{2}{*}{3.24(1.05)}& 2.69 & 5.02  \\ 
             &      & 3 &  3.40 & 1.29  &                       &                       &                       &                           &      &       \\              
\hline
%
%
             &      & 1 &  2.66 & 1.08  &                       &                       &                       &                           &      &       \\
 $^{176}$Hf  & 0.23 & 2 &  3.36 & 1.96  & \multirow{2}{*}{3.50} & \multirow{2}{*}{3.22} & \multirow{2}{*}{2.71} & \multirow{2}{*}{3.32(28)} & 3.26 & 3.79  \\ 
             &      & 3 &  3.86 & 0.76  &                       &                       &                       &                           &      &       \\ 
 \hline                                                                                                               
             &      & 1 &  2.62 & 0.96  &                       &                       &                       &                           &      &       \\
 $^{178}$Hf  & 0.22 & 2 &  3.33 & 1.89  & \multirow{2}{*}{3.44} & \multirow{2}{*}{3.21} & \multirow{2}{*}{2.46} & \multirow{2}{*}{2.38(33)} & 3.21 & 3.42  \\ 
             &      & 3 &  3.82 & 0.58  &                       &                       &                       &                           &      &       \\ 
\hline                                                                                                                
             &      & 1 &  2.60 & 0.93  &                       &                       &                       &                           &      &       \\
 $^{180}$Hf  & 0.22 & 2 &  3.34 & 2.04  & \multirow{2}{*}{3.44} & \multirow{2}{*}{3.16} & \multirow{2}{*}{2.58} & \multirow{2}{*}{2.13(30)} & 3.22 & 3.51  \\ 
             &      & 3 &  3.83 & 0.54  &                       &                       &                       &                           &      &       \\ 
\hline\hline                                                    
             &      & 1 &  2.59 & 0.80  &                       &                       &                       &                           &      &       \\          
 $^{182}$W   & 0.20 & 2 &  3.23 & 1.30  & \multirow{2}{*}{3.32} & \multirow{2}{*}{3.10} & \multirow{2}{*}{1.63} & \multirow{2}{*}{1.65(28)} & 3.08 & 2.43  \\ 
             &      & 3 &  3.68 & 0.33  &                       &                       &                       &                           &      &       \\ 
\hline                                                                                                               
             &      & 1 &  2.56 & 0.68  &                       &                       &                       &                           &      &       \\
 $^{184}$W   & 0.19 & 2 &  3.20 & 1.20  & \multirow{2}{*}{3.35} & \multirow{2}{*}{3.19} & \multirow{2}{*}{1.37} & \multirow{2}{*}{1.24(37)} & 3.02 & 2.05  \\ 
             &      & 3 &  3.63 & 0.17  &                       &                       &                       &                           &      &       \\ 
 \hline                                                                                                               
             &      & 1 &  2.53 & 0.57  &                       &                       &                       &                           &      &       \\
 $^{186}$W   & 0.18 & 2 &  3.17 & 1.09  & \multirow{2}{*}{3.18} & \multirow{2}{*}{3.19} & \multirow{2}{*}{1.11} & \multirow{2}{*}{0.82(21)} & 2.96 & 1.68  \\ 
             &      & 3 &  3.59 & 0.02  &                       &                       &                       &                           &      &       \\ 
\end{tabular}\end{ruledtabular}\label{tab13s}
\end{table*}

In our example of $^{164}$Dy the summarized magnetic strength 
$\sum B(M1)=5.56\ \mu_N^2$ of three scissors is
remarkably stronger than the analogous value $\sum B(M1)=3.39\ \mu_N^2$ of two 
magnetic states in the case of decoupling (see Table~\ref{tab1}). One may say 
that it is also stronger
than the respective experimental value. However, one must be careful here.

Trying to compare the theoretical results with the existing experimental data for the scissors mode, we encounter different summing interval conventions.
It is assumed that the scissors mode includes only the states in a certain energy 
range. As a rule, the following two conventions are chosen, 
which lead to slightly different results for the summed $M1$ strength:\\
$2.7 < E < 3.7$ MeV for $Z<68$ and\\
$2.4 < E < 3.7$ MeV for $Z\geq 68$~\cite{Pietr95},\\
$2.5 < E < 4.0$ MeV for $82\leq N \leq 126$~\cite{Rich}.

Obviously, only the two highest scissors fall into both of these intervals. It turns out that their summed  $B(M1)$  agrees rather well with 
the majority of experimental values found by NRF 
(Nuclear Resonance Fluorescence)
experiments~\cite {Pietr95,Pietr98,Rich} for nuclei of $N=82-126$ mass region (see Fig.~\ref{fig2} and columns 6 and 13 of Table VI).
The situation with the lowest scissors is very interesting. It helps to explain the long-standing problem of the $1^+$ spectrum of $^{164}$Dy.
%

\begin{figure}[h!]
\centering
\includegraphics[width=\columnwidth]{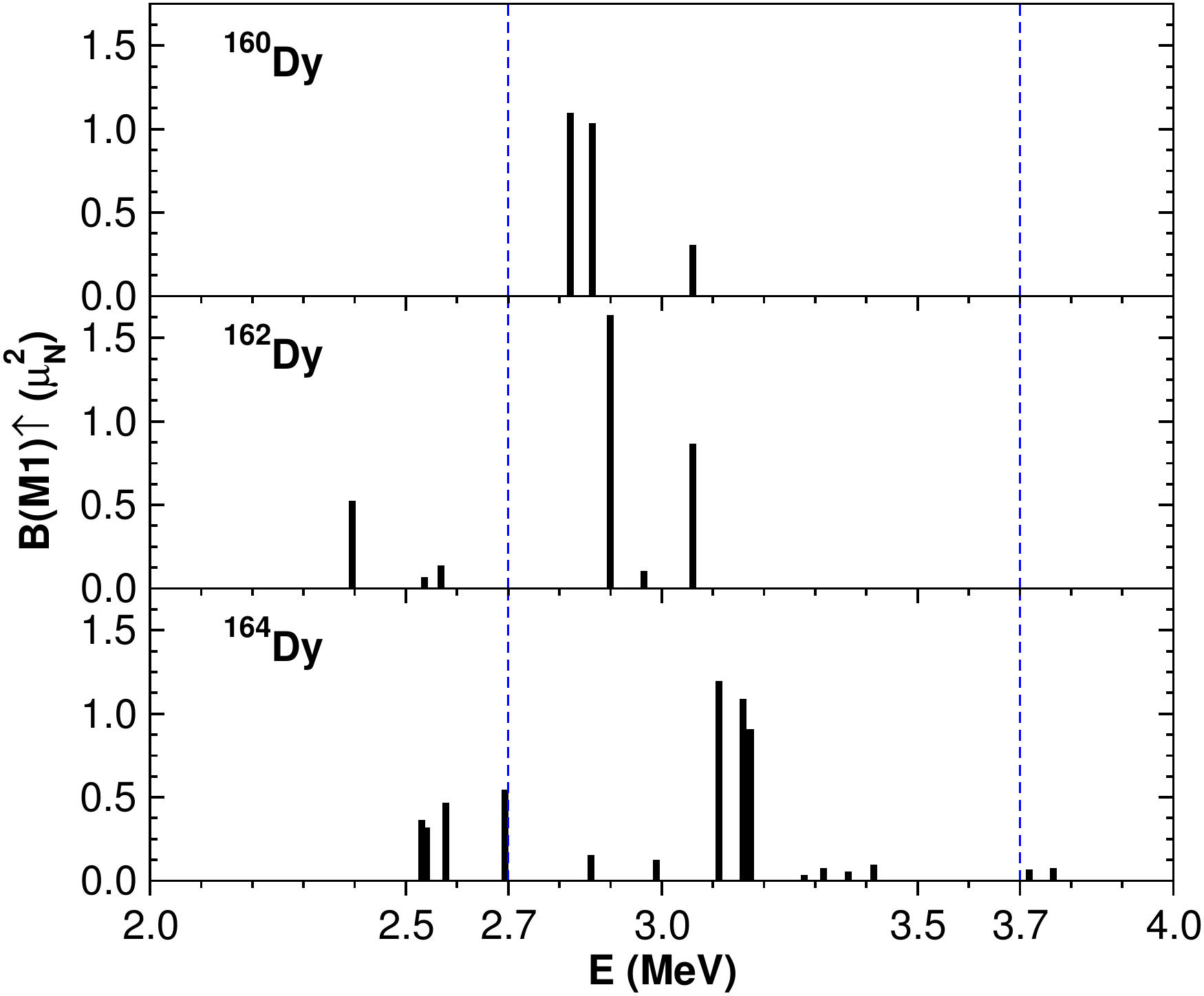}
\caption{Excitation energies $E$ with the corresponding $B(M1)$ values, obtained by the NRF experiment~\cite{Margraf}.
The dashed lines mark the boundaries of the conventional interval from~\cite{Pietr95}.}
\label{fig3}\end{figure}
The Fig.~\ref{fig3} demonstrates experimental $M1$ strength distributions in $^{160,162,164}$Dy in the energy range between 2 MeV and 4 MeV, 
reported by Margraf~{\it et~al.}~\cite{Margraf}.
Obviously, there are two groups of strong  $M1$ excitations in $^{164}$Dy around $2.6$ and $3.1$~MeV.
However, only the upper group was attributed to the scissors mode, and
the group around $2.6$ MeV was not included because it has a rather big spin
contribution and one level has pure two-quasiparticle nature and the summed 
$M1$ strength of both groups strongly deviates from the scissors mode  
systematics in the Rare Earth nuclei~\cite{Rich}.
The results of our calculations allow one to clarify the origin of both groups.
Table~\ref{tab2} demonstrates that the energy centroid and summed 
$B(M1)$-value of the observed lower group 
agree very well with the calculated energy $E$ and $B(M1)$ value of the lowest scissors. 
The respective values of the observed higher group are in
excellent agreement with the calculated energy centroid and summed $B(M1)$ of 
two remaining (higher in energy) scissors.

\begin{table}[h!]
\caption{The calculated energies $E$ (MeV) and excitation probabilities 
$B(M1)$ ($\mu_N^2$) of three scissors 
are compared with experimental values $\bar E$ and $\sum B(M1)$ of two groups of $1^+$ levels in $^{164}$Dy~\cite{Margraf}. }
\begin{ruledtabular}\begin{tabular}{ccccccc}
 \multicolumn{4}{c}{Theory (WFM)} & & \multicolumn{2}{c}{Experiment (NRF)} \\
\cline{1-4} \cline{6-7}
 $E$  & $B(M1)$  & $\bar E$  & $\sum B(M1)$  & & $\bar E$  & $\sum B(M1)$   \\
\cline{1-4} \cline{6-7}
 2.20 & 1.76 & 2.20 & 1.76 &  &   2.60 & 1.67(14)  \\ 
    2.87 & 2.24 &\multirow{2}{*}{3.17}&\multirow{2}{*}{3.80}&\multirow{2}{*}{ }&\multirow{2}{*}{3.17}&\multirow{2}{*}{3.85(31)} \\ 
    3.59 & 1.56 &                     &                     &                  &                     &           \\
\end{tabular}\end{ruledtabular}\label{tab2}
\end{table}

So, according to our calculations, the low-energy group of states in $^{164}$Dy is also a branch of the scissors mode (spin-vector isovector
scissors) and the calculated summed magnetic strength 5.56~$\mu_N^2$ is in excellent agreement with the experimental value 5.52~$\mu_N^2$. 
Analogous values for two other Dy isotopes, $^{160}$Dy and $^{162}$Dy, are predicted to 
be 5.19~$\mu_N^2$ and 5.38~$\mu_N^2$ (see Table~\ref{tab3}). 
  From a first glance on Fig.~\ref{fig3} it becomes clear 
that in those nuclei nothing similar to the $^{164}$Dy case was observed by NRF experiments. 
Nevertheless our prediction was confirmed recently by another experiment --
photo-neutron measurements performed by the Oslo group.
In Ref.~\cite{Renstrom} the authors revised  their previous data on the Scissors Resonance (SR) in $^{160-164}$Dy obtained by the Oslo method.
The essence of their findings is formulated in the following quotation from~\cite{Renstrom}:
{\it ``...If we integrate Eq. (19) over all transition energies, we find
a total, summed SR strength of $4.6(12)-5.8(26) \mu_N^2$...
The present fit strategy gives about $40\%$ higher summed SR strengths than the reported  NRF results. However, if we apply the NRF energy limits to
Eq. (19), we obtain excellent agreement with the NRF results...
It is interesting to note that $\approx 40-60\%$ of our measured SR strength lies in the energy region below $2.7$ MeV. In traditional NRF experiments 
using bremsstrahlung, the transitions in this energy range are
quite difficult to separate from the sizable atomic background.''}

This is exactly the point! The statements about "$40\%$ higher" and 
"$\approx 40-60\%$... below 2.7 MeV" are in qualitative and often in very good 
agreement with our findings!

\begin{figure}[h!]
\centering{\includegraphics[width=\columnwidth]{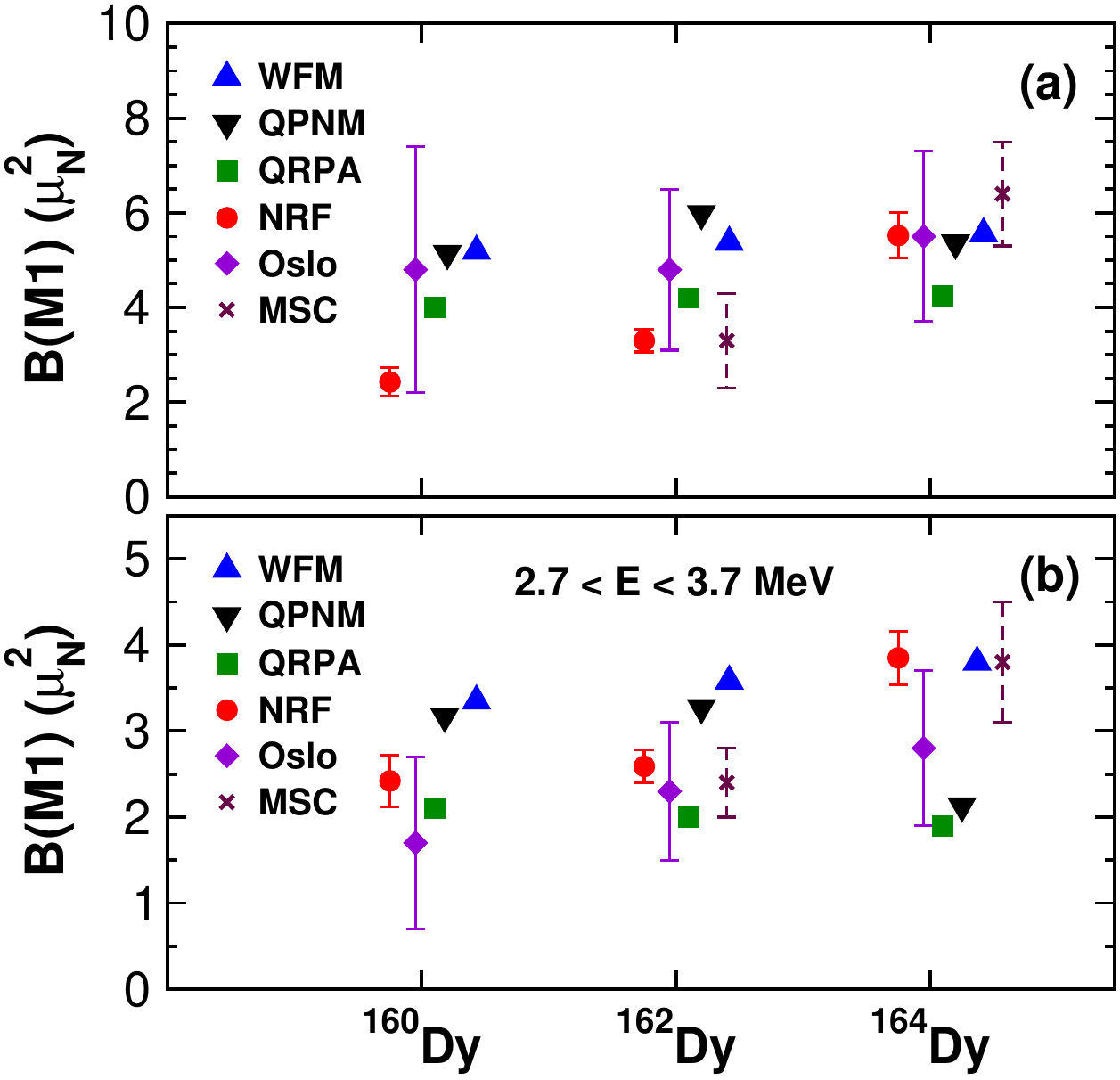}}
\caption{Comparison of the summed  $B(M1)$  values for SR in $^{160,162,164}$Dy from the present WFM theory,
the QPNM~\cite{VGS1,VGS2} and Gogny QRPA~\cite{Renstrom} calculations
with the experimental values
from the NRF~\cite{Margraf}, photo-neutron measurements (Oslo)~\cite{Renstrom} and from multistep-cascade (MSC) measurements
of $\gamma$ decay following neutron capture~\cite{Valenta}. 
Panel (a) --  averaging energy intervals are $2-4$~MeV for WFM, QPNM and NRF; $0-3.5$~MeV for QRPA; $0-10$~MeV for Oslo and MSC,
(b) --  averaging interval is $2.7-3.7$~MeV.}
\label{fig4}\end{figure}

\subsubsection{WFM versus QPNM}

\begin{table*}[t!]
\caption{The energy centroids $\bar E$  and corresponding summed 
$B(M1)$  values
given by WFM theory and QPNM calculations
are compared with experimental results by the NRF~\cite{Margraf} 
and photo-neutron measurements (Oslo)~\cite{Renstrom} for $^{160,162,164}$Dy.
Comparison is presented for various energy intervals.}
\begin{ruledtabular}\begin{tabular}{cccccccccc}
 & \multicolumn{4}{c}{Theory} & &\multicolumn{4}{c}{Experiment} \\
\cline{2-5}\cline{7-10}
$^{\rm A}$Dy & \multicolumn{2}{c}{WFM} &  \multicolumn{2}{c}{QPNM} & &
 \multicolumn{2}{c}{NRF} & \multicolumn{2}{c}{Oslo} \\  \cline{2-5}\cline{7-10}
 &  $\bar E$ (MeV) & $B(M1)$ ($\mu_N^2$) & $\bar E$ (MeV) & $B(M1)$ ($\mu_N^2$) &  &
    $\bar E$ (MeV) & $B(M1)$ ($\mu_N^2$) & $\bar E$ (MeV) & $B(M1)$ ($\mu_N^2$) \\   
\cline{2-5}\cline{7-10} 
 & \multicolumn{4}{c}{\small $2.7<E<3.7$ MeV }  & & \multicolumn{4}{c}{\small $2.7<E<3.7$ MeV } \\
 $^{160}$Dy & 3.17 & 3.35 & 3.05 &  3.17~ & & 2.87 & 2.42(30) & 2.66(12) & 1.7(10) \\ 
 $^{162}$Dy & 3.16 & 3.58 & 3.08 &  3.27~ & & 2.96 & 2.59(19) & 2.81(8)\ \ & 2.3(8)\ \  \\ 
 $^{164}$Dy & 3.17 & 3.80 & 3.26 &  2.13~ & & 3.17 & 3.85(31) & 2.83(8)\ \  & 2.8(9)\ \  \\ 
\cline{2-5}\cline{7-10}
 &  \multicolumn{4}{c}{\small $2.0<E<4.0$ MeV} & & \multicolumn{2}{c}{\small $2.0<E<4.0$ MeV} &  \multicolumn{2}{c}{\small $0<E<10$ MeV}  \\
 $^{160}$Dy & 2.84 & 5.19 & 3.05  &  5.14~ & & 2.87 & 2.42(30) & 2.66(12) & 4.8(26) \\ 
 $^{162}$Dy & 2.85 & 5.38 & 3.10  &  5.98~ & & 2.84 & 3.30(24) & 2.81(8)\ \  & 4.8(17) \\ 
 $^{164}$Dy & 2.86 & 5.56 & 2.87  &  5.36~ & & 3.00 & 5.52(48) & 2.83(8)\ \  & 5.5(18) \\ 
\end{tabular}\end{ruledtabular}\label{tab3}
\end{table*}
\begin{figure}[b!]
\centering
\includegraphics[width=\columnwidth]{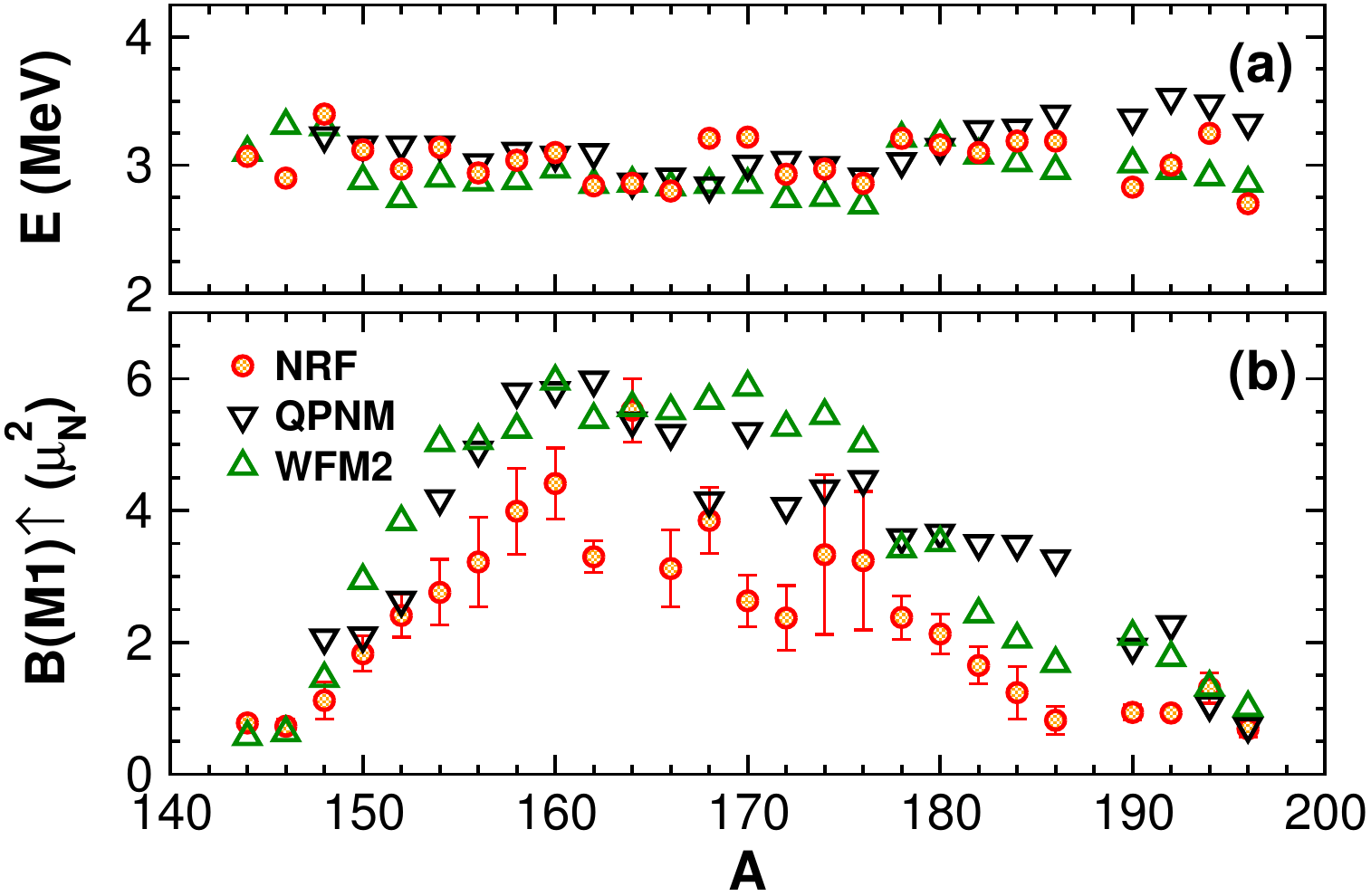}
\caption{WFM2 -- energy centroid of three scisors and the respective $B(M1)$ value given by WFM method, QPNM -- analogous values calculated in the frame 
of QPNM in the energy range $2-4$ MeV, NRF -- experimental data.}
\label{fig9}\end{figure}

In the rest nuclei of $N=82-126$ mass region an equally  significant low 
energy $M1$ strength was not detected in the NRF experiments.
However, our calculations predict the existence of comparable magnetic strength in all well-deformed nuclei of this  mass region (see WFM2 in Fig.~\ref{fig2}). This prediction is supported by calculations in the frame 
of Quasiparticle-Phonon Nuclear Model (QPNM), which also predicts remarkable 
$M1$ strength below the conventional energy interval.
A short outline of QPNM together with calculation details can be found in
the papers \cite{VGS2,VGS1}.
\begin{table*}[ht!]
\vspace*{-5mm}\caption{\footnotesize The analysis of the nuclear scissors mode structure for specific energy regions.
The results of calculations by the WFM method and QPNM: energies $E_i$ with correspondings 
$B_i(M1)$. The energy ranges for QPNM are:\\
I\ : $1.8-4.0$ MeV ($i=1:\ 1.8-2.5$ MeV, $i=2:\ 2.5-4.0$ MeV);\\ II: $1.8-3.7$ MeV ($i=1:\ 1.8-2.7$ MeV, $i=2:\ 2.7-3.7$ MeV).\\
Energy centroids $\bar E$ and summed $M1$ strength from WFM, QPNM calculations and NRF experiments are presented for specified energy ranges. 
The parameter values for WFM calculations are indicatedfig9 in the caption to Table~\ref{tab11}.}
\begin{ruledtabular}\begin{tabular}{cccccccccccclccc}
 \multirow{3}{*}{Nuclei}  &  & 
 \multicolumn{3}{c}{$E_i$~(MeV)}  & \multicolumn{3}{c}{$B_i(M1)$~($\mu_N^2$)} &   
 \multicolumn{4}{c}{$\bar E$~(MeV)}  &  \multicolumn{4}{c}{$\sum B(M1)$~($\mu_N^2$)}\\   
   &  & WFM & \multicolumn{2}{c}{QPNM} & WFM & \multicolumn{2}{c}{QPNM} & NRF & WFM & \multicolumn{2}{c}{QPNM} & \ \ NRF & WFM & \multicolumn{2}{c}{QPNM} \\
   & $i$ & & I & II &  & I & II & {\footnotesize $2.0-4.0$} &  & {\footnotesize $1.8-4.0$} & {\footnotesize $1.8-4.2$} & {\footnotesize $2.0-4.0$} &  & {\footnotesize $1.8-4.0$} & {\footnotesize $1.8-4.2$} \\
\hline\hline
\multirow{2}{*}{$^{148}$Nd} & 1 & 2.51 & 2.47 & 2.47 & 0.28 & 0.41 & 0.41 &\multirow{2}{*}{3.40}&\multirow{2}{*}{3.30}&\multirow{2}{*}{3.23}&\multirow{2}{*}{3.25}&\multirow{2}{*}{ 1.12(26) }&\multirow{2}{*}{1.45}&\multirow{2}{*}{2.07}&\multirow{2}{*}{2.11}\\
                            & 2 & 3.49 & 3.42 & 3.18 & 1.17 & 1.65 & 1.15 &                     &                     &                     &                     &                           &                     &                     &                     \\ 
\hline                                                                                                                                                     
\multirow{2}{*}{$^{150}$Nd} & 1 & 2.39 & 2.46 & 2.46 & 1.08 & 0.52 & 0.52 &\multirow{2}{*}{3.12}&\multirow{2}{*}{2.88}&\multirow{2}{*}{3.16}&\multirow{2}{*}{3.17}&\multirow{2}{*}{ 1.83(27) }&\multirow{2}{*}{2.94}&\multirow{2}{*}{2.10}&\multirow{2}{*}{2.13}\\
                            & 2 & 3.16 & 3.39 & 3.18 & 1.85 & 1.59 & 1.13 &                     &                     &                     &                     &                           &                     &                     &                     \\ 
\hline\hline
\multirow{2}{*}{$^{148}$Sm} & 1 & 2.48 & 2.43 & 2.55 & 0.08 & 0.22 & 0.74 &\multirow{2}{*}{3.07}&\multirow{2}{*}{3.02}&\multirow{2}{*}{2.88}&\multirow{2}{*}{3.25}&\multirow{2}{*}{ 0.51(12) }&\multirow{2}{*}{0.31}&\multirow{2}{*}{1.57}&\multirow{2}{*}{2.24}\\
                            & 2 & 3.21 & 2.95 & 3.17 & 0.23 & 1.36 & 0.83 &                     &                     &                     &                     &                           &                     &                     &                     \\ 
\hline 
\multirow{2}{*}{$^{150}$Sm} & 1 & 2.27 &  --  & 2.53 & 0.49 &  --  & 0.40 &\multirow{2}{*}{3.18}&\multirow{2}{*}{2.67}&\multirow{2}{*}{3.11}&\multirow{2}{*}{3.38}&\multirow{2}{*}{ 0.97(17) }&\multirow{2}{*}{1.10}&\multirow{2}{*}{1.59}&\multirow{2}{*}{2.17}\\
                            & 2 & 3.00 & 3.11 & 3.30 & 0.61 & 1.59 & 1.17 &                     &                     &                     &                     &                           &                     &                     &                     \\ 
\hline                                                                                                                                                     
\multirow{2}{*}{$^{152}$Sm} & 1 & 2.18 & 2.31 & 2.31 & 1.45 & 0.13 & 0.13 &\multirow{2}{*}{2.97}&\multirow{2}{*}{2.74}&\multirow{2}{*}{3.16}&\multirow{2}{*}{3.40}&\multirow{2}{*}{ 2.41(33) }&\multirow{2}{*}{3.83}&\multirow{2}{*}{2.64}&\multirow{2}{*}{3.50}\\
                            & 2 & 3.08 & 3.21 & 3.21 & 2.38 & 2.51 & 2.50 &                     &                     &                     &                     &                           &                     &                     &                     \\ 
\hline                                                                                                                                                     
\multirow{2}{*}{$^{154}$Sm} & 1 & 2.22 & 2.19 & 2.19 & 1.59 & 0.83 & 0.83 &\multirow{2}{*}{3.14}&\multirow{2}{*}{2.90}&\multirow{2}{*}{3.16}&\multirow{2}{*}{3.22}&\multirow{2}{*}{ 2.76(50) }&\multirow{2}{*}{5.03}&\multirow{2}{*}{4.18}&\multirow{2}{*}{4.45}\\
                            & 2 & 3.22 & 3.41 & 3.28 & 3.44 & 3.34 & 2.67 &                     &                     &                     &                     &                           &                     &                     &                     \\                                                                          
\hline\hline
\multirow{2}{*}{$^{156}$Gd} & 1 & 2.25 & 2.04 & 2.04 & 1.75 & 0.79 & 0.79 &\multirow{2}{*}{2.94}&\multirow{2}{*}{2.87}&\multirow{2}{*}{3.02}&\multirow{2}{*}{3.10}&\multirow{2}{*}{ 3.22(68) }&\multirow{2}{*}{5.06}&\multirow{2}{*}{4.92}&\multirow{2}{*}{5.30}\\
                            & 2 & 3.19 & 3.20 & 3.15 & 3.31 & 4.13 & 3.81 &                     &                     &                     &                     &                           &                     &                     &                     \\ 
\hline                                                                                                                                                     
\multirow{2}{*}{$^{158}$Gd} & 1 & 2.22 & 2.34 & 2.34 & 1.70 & 0.48 & 0.48 &\multirow{2}{*}{3.04}&\multirow{2}{*}{2.88}&\multirow{2}{*}{3.11}&\multirow{2}{*}{3.13}&\multirow{2}{*}{ 3.99(65) }&\multirow{2}{*}{5.23}&\multirow{2}{*}{5.80}&\multirow{2}{*}{5.87}\\
                            & 2 & 3.19 & 3.18 & 3.08 & 3.53 & 5.32 & 4.64 &                     &                     &                     &                     &                           &                     &                     &                     \\ 
\hline                                                                                                  
\multirow{2}{*}{$^{160}$Gd} & 1 & 2.23 & 2.47 & 2.56 & 1.74 & 0.63 & 1.28 &\multirow{2}{*}{3.10}&\multirow{2}{*}{2.97}&\multirow{2}{*}{3.08}&\multirow{2}{*}{3.13}&\multirow{2}{*}{ 4.41(54) }&\multirow{2}{*}{5.96}&\multirow{2}{*}{5.82}&\multirow{2}{*}{6.14}\\
                            & 2 & 3.27 & 3.15 & 3.10 & 4.22 & 5.18 & 3.79 &                     &                     &                     &                     &                           &                     &                     &                     \\ 
\hline\hline                                                                                            
\multirow{2}{*}{$^{160}$Dy} & 1 & 2.25 & 2.43 & 2.43 & 1.84 & 1.07 & 1.08 &\multirow{2}{*}{2.87}&\multirow{2}{*}{2.84}&\multirow{2}{*}{3.05}&\multirow{2}{*}{3.14}&\multirow{2}{*}{ 2.42(30) }&\multirow{2}{*}{5.19}&\multirow{2}{*}{5.14}&\multirow{2}{*}{5.62}\\
                            & 2 & 3.17 & 3.22 & 3.05 & 3.35 & 4.07 & 3.17 &                                           &                     &                     &                           &                     &                     &                     \\ 
\hline                                                                                                  
\multirow{2}{*}{$^{162}$Dy} & 1 & 2.22 & 2.46 & 2.47 & 1.80 & 1.40 & 1.41 &\multirow{2}{*}{2.84}&\multirow{2}{*}{2.85}&\multirow{2}{*}{3.10}&\multirow{2}{*}{3.11}&\multirow{2}{*}{ 3.30(24) }&\multirow{2}{*}{5.38}&\multirow{2}{*}{5.98}&\multirow{2}{*}{6.05}\\
                            & 2 & 3.16 & 3.30 & 3.08 & 3.58 & 4.58 & 3.27 &                     &                     &                     &                     &                           &                     &                     &                     \\  
\hline                                                                                                                                                
\multirow{2}{*}{$^{164}$Dy} & 1 & 2.20 & 2.08 & 2.35 & 1.76 & 1.26 & 2.69 &\multirow{2}{*}{3.00}&\multirow{2}{*}{2.86}&\multirow{2}{*}{2.87}&\multirow{2}{*}{2.89}&\multirow{2}{*}{ 5.52(48) }&\multirow{2}{*}{5.56}&\multirow{2}{*}{5.36}&\multirow{2}{*}{5.45}\\
                            & 2 & 3.17 & 3.11 & 3.26 & 3.80 & 4.10 & 2.13 &                     &                     &                     &                     &                           &                     &                     &                     \\  
\hline\hline                                                                                            
\multirow{2}{*}{$^{166}$Er} & 1 & 2.23 & 2.04 & 2.29 & 1.89 & 1.35 & 2.30 &\multirow{2}{*}{2.79}&\multirow{2}{*}{2.83}&\multirow{2}{*}{2.91}&\multirow{2}{*}{2.95}&\multirow{2}{*}{ 3.12(58) }&\multirow{2}{*}{5.51}&\multirow{2}{*}{5.17}&\multirow{2}{*}{5.36}\\
                            & 2 & 3.14 & 3.21 & 3.24 & 3.62 & 3.82 & 2.13 &                     &                     &                     &                     &                           &                     &                     &                     \\                                                                                             
\hline
\multirow{2}{*}{$^{168}$Er} & 1 & 2.20 & 2.32 & 2.48 & 1.81 & 1.14 & 2.20 &\multirow{2}{*}{3.21}&\multirow{2}{*}{2.85}&\multirow{2}{*}{2.84}&\multirow{2}{*}{2.85}&\multirow{2}{*}{ 3.85(50) }&\multirow{2}{*}{5.67}&\multirow{2}{*}{4.15}&\multirow{2}{*}{4.21}\\
                            & 2 & 3.15 & 3.03 & 3.19 & 3.86 & 3.01 & 1.81 &                     &                     &                     &                     &                           &                     &                     &                     \\  
\hline\hline                                                                                            
\multirow{2}{*}{$^{172}$Yb} & 1 & 2.18 & 2.16 & 2.52 & 1.86 & 0.53 & 1.96 &\multirow{2}{*}{2.93}&\multirow{2}{*}{2.74}&\multirow{2}{*}{3.04}&\multirow{2}{*}{3.07}&\multirow{2}{*}{ 2.37(49) }&\multirow{2}{*}{5.26}&\multirow{2}{*}{4.06}&\multirow{2}{*}{4.19}\\
                            & 2 & 3.05 & 3.17 & 3.27 & 3.40 & 3.53 & 1.22 &                     &                     &                     &                     &                           &                     &                     &                     \\ 
\hline                                                                                                  
\multirow{2}{*}{$^{174}$Yb} & 1 & 2.16 & 2.10 & 2.40 & 1.82 & 0.86 & 1.89 &\multirow{2}{*}{2.96}&\multirow{2}{*}{2.75}&\multirow{2}{*}{3.00}&\multirow{2}{*}{3.00}&\multirow{2}{*}{3.33(1.21)}&\multirow{2}{*}{5.44}&\multirow{2}{*}{4.33}&\multirow{2}{*}{4.34}\\
                            & 2 & 3.05 & 3.22 & 3.33 & 3.62 & 3.48 & 1.83 &                     &                     &                     &                     &                           &                     &                     &                     \\ 
\hline                                                                                                  
\multirow{2}{*}{$^{176}$Yb} & 1 & 2.11 & 1.88 & 2.26 & 1.69 & 1.11 & 2.20 &\multirow{2}{*}{2.86}&\multirow{2}{*}{2.69}&\multirow{2}{*}{2.91}&\multirow{2}{*}{2.91}&\multirow{2}{*}{3.24(1.05)}&\multirow{2}{*}{5.02}&\multirow{2}{*}{4.47}&\multirow{2}{*}{4.47}\\
                            & 2 & 2.99 & 3.25 & 3.54 & 3.33 & 3.36 & 2.27 &                     &                     &                     &                     &                           &                     &                     &                     \\ 
\hline\hline                                                                                            
\multirow{2}{*}{$^{176}$Hf} & 1 & 2.66 & 2.21 & 2.21 & 1.08 & 1.04 & 1.04 &\multirow{2}{*}{3.22}&\multirow{2}{*}{3.26}&\multirow{2}{*}{3.12}&\multirow{2}{*}{3.12}&\multirow{2}{*}{ 3.32(28) }&\multirow{2}{*}{3.79}&\multirow{2}{*}{3.93}&\multirow{2}{*}{3.93}\\
                            & 2 & 3.50 & 3.45 & 3.39 & 2.71 & 2.89 & 2.57 &                     &                     &                     &                     &                           &                     &                     &                     \\ 
\hline                                                                                                  
\multirow{2}{*}{$^{178}$Hf} & 1 & 2.62 & 2.22 & 2.27 & 0.96 & 0.92 & 1.04 &\multirow{2}{*}{3.21}&\multirow{2}{*}{3.21}&\multirow{2}{*}{3.03}&\multirow{2}{*}{3.03}&\multirow{2}{*}{2.38(33)}  &\multirow{2}{*}{3.42}&\multirow{2}{*}{3.59}&\multirow{2}{*}{3.59}\\
                            & 2 & 3.44 & 3.31 & 3.34 & 2.46 & 2.66 & 2.54 &                     &                     &                     &                     &                           &                     &                     &                     \\ 
\hline                                                                                                  
\multirow{2}{*}{$^{180}$Hf} & 1 & 2.60 & 2.29 & 2.29 & 0.93 & 0.79 & 0.79 &\multirow{2}{*}{3.16}&\multirow{2}{*}{3.22}&\multirow{2}{*}{3.14}&\multirow{2}{*}{3.14}&\multirow{2}{*}{2.13(30)}  &\multirow{2}{*}{3.51}&\multirow{2}{*}{3.66}&\multirow{2}{*}{3.66}\\
                            & 2 & 3.44 & 3.37 & 3.18 & 2.58 & 2.86 & 2.01 &                     &                     &                     &                     &                           &                     &                     &                     \\ 
\hline\hline                                                                                            
\multirow{2}{*}{$^{182}$W}  & 1 & 2.59 & 2.29 & 2.52 & 0.80 & 0.01 & 0.13 &\multirow{2}{*}{3.10}&\multirow{2}{*}{3.08}&\multirow{2}{*}{3.28}&\multirow{2}{*}{3.28}&\multirow{2}{*}{ 1.65(28) }&\multirow{2}{*}{2.43}&\multirow{2}{*}{3.50}&\multirow{2}{*}{3.51}\\
                            & 2 & 3.32 & 3.28 & 3.19 & 1.63 & 3.49 & 2.78 &                     &                     &                     &                     &                           &                     &                     &                     \\ 
\hline                                                                                                  
\multirow{2}{*}{$^{184}$W}  & 1 & 2.56 & 2.39 & 2.39 & 0.68 & 0.54 & 0.54 &\multirow{2}{*}{3.19}&\multirow{2}{*}{3.02}&\multirow{2}{*}{3.29}&\multirow{2}{*}{3.29}&\multirow{2}{*}{1.24(37)}  &\multirow{2}{*}{2.05}&\multirow{2}{*}{3.49}&\multirow{2}{*}{3.49}\\
                            & 2 & 3.35 & 3.45 & 3.32 & 1.37 & 2.96 & 2.30 &                     &                     &                     &                     &                           &                     &                     &                     \\ 
\hline                                                                                                  
\multirow{2}{*}{$^{186}$W}  & 1 & 2.53 & 2.40 & 2.61 & 0.57 & 0.01 & 0.68 &\multirow{2}{*}{3.19}&\multirow{2}{*}{2.96}&\multirow{2}{*}{3.40}&\multirow{2}{*}{3.40}&\multirow{2}{*}{0.82(21)}  &\multirow{2}{*}{1.68}&\multirow{2}{*}{3.27}&\multirow{2}{*}{3.28}\\
                            & 2 & 3.18 & 3.40 & 3.50 & 1.11 & 3.25 & 1.98 &                     &                     &                     &                     &                           &                     &                     &                     \\ 
\end{tabular}\end{ruledtabular}\label{tab6}
\end{table*}
\begin{table*}[ht!]
\caption{
The continuation of Table~\ref{tab6}. Parameters for WFM calculations: 
$V_0^{\rm p}=26.5$~MeV, $V_0^{\rm n}=22.6$~MeV, $r_p^{\rm p}=1.7$ fm, $r_p^{\rm n}=2.1$ fm, 
$\kappa_{\rm Nils}=0.05$, $q=0.57$.
}
\begin{ruledtabular}\begin{tabular}{cccccccccccclccc}
 \multirow{3}{*}{Nuclei}  &  & 
 \multicolumn{3}{c}{$E_i$ (MeV)}  & \multicolumn{3}{c}{$B_i(M1)$~($\mu_N^2$)} &   
 \multicolumn{4}{c}{$\bar E$ (MeV)}  &  \multicolumn{4}{c}{$\sum B(M1)$~($\mu_N^2$)}\\   
   &  & WFM & \multicolumn{2}{c}{QPNM} & WFM & \multicolumn{2}{c}{QPNM} & NRF & WFM & \multicolumn{2}{c}{QPNM} & \ \ NRF & WFM & \multicolumn{2}{c}{QPNM} \\
   & $i$ & & I & II &  & I & II & {\footnotesize $2.0-4.0$} &  & {\footnotesize $1.8-4.0$} & {\footnotesize $1.8-4.2$} & {\footnotesize $2.0-4.0$} &  & {\footnotesize $1.8-4.0$} & {\footnotesize $1.8-4.2$} \\
\hline\hline
\multirow{2}{*}{$^{190}$Os} & 1 & 2.51 &  --  &  --  & 0.42 &  --  &  --  &\multirow{2}{*}{2.83}&\multirow{2}{*}{3.01}&\multirow{2}{*}{3.37}&\multirow{2}{*}{3.54}&\multirow{2}{*}{ 0.94(12) }&\multirow{2}{*}{2.09}&\multirow{2}{*}{1.93}&\multirow{2}{*}{2.55}\\
                            & 2 & 3.13 & 3.37 & 3.36 & 1.68 & 1.93 & 1.91 &                     &                     &                     &                     &                           &                     &                     &                     \\ 
\hline                                                                                                                                                     
\multirow{2}{*}{$^{192}$Os} & 1 & 2.49 &  --  &  --  & 0.35 &  --  &  --  &\multirow{2}{*}{3.00}&\multirow{2}{*}{2.96}&\multirow{2}{*}{3.53}&\multirow{2}{*}{3.61}&\multirow{2}{*}{ 0.93(06) }&\multirow{2}{*}{1.77}&\multirow{2}{*}{2.27}&\multirow{2}{*}{2.65}\\
                            & 2 & 3.08 & 3.53 & 3.44 & 1.41 & 2.27 & 1.73 &                     &                     &                     &                     &                           &                     &                     &                     \\ 
\hline\hline
\multirow{2}{*}{$^{194}$Pt} & 1 & 2.50 & 2.23 & 2.38 & 0.32 & 0.06 & 0.11 &\multirow{2}{*}{3.25}&\multirow{2}{*}{2.91}&\multirow{2}{*}{3.48}&\multirow{2}{*}{3.51}&\multirow{2}{*}{ 1.31(23) }&\multirow{2}{*}{1.32}&\multirow{2}{*}{1.06}&\multirow{2}{*}{1.11}\\
                            & 2 & 3.05 & 3.56 & 3.34 & 1.00 & 1.00 & 0.48 &                     &                     &                     &                     &                           &                     &                     &                     \\ 
\hline 
\multirow{2}{*}{$^{196}$Pt} & 1 & 2.47 & 2.27 & 2.35 & 0.26 & 0.02 & 0.03 &\multirow{2}{*}{2.70}&\multirow{2}{*}{2.86}&\multirow{2}{*}{3.33}&\multirow{2}{*}{3.59}&\multirow{2}{*}{ 0.69(13) }&\multirow{2}{*}{1.01}&\multirow{2}{*}{0.73}&\multirow{2}{*}{1.14}\\
                            & 2 & 2.99 & 3.36 & 3.24 & 0.75 & 0.71 & 0.55 &                     &                     &                     &                     &                           &                     &                     &                     \\ 
\end{tabular}\end{ruledtabular}\label{tab6a}
\end{table*}

The energy centroids and corresponding summed  $B(M1)$  given by
the WFM theory and by the QPNM calculations for Dy isotopes are compared with
experimental results from the NRF and from photo-neutron measurements (Oslo)~\cite{Renstrom} in Table~\ref{tab3}.
The results are shown for various energy averaging intervals.
As it is seen, the theoretical results and experimental data 
of Oslo group are in very good overall agreement for all three Dy isotopes. 
It is worthwhile to remark the excellent agreement between all theoretical and
experimental results for $^{164}$Dy.
One may be worried by the comparatively big interval $0<E<10$ MeV employed by the Oslo group~\cite{Renstrom} for centroids energies and $B(M1)$. 
However, in their paper they say that 
all spin-flip excitations are eliminated in their averages and, thus, their averaging becomes equal to the theoretical one, 
since it is generally believed that all what is above 4 MeV excitation energy does not belong to SR excitations. 
In Fig.~\ref{fig4} the summed $B(M1)$ values are also shown for $^{160,162,164}$Dy including this time the results from Gogny QRPA calculations
and the experimental results obtained by the radiative capture of resonance neutrons~\cite{Valenta}. 
It is remarkable to which extent theory and experiment agree taking the NRF as well as the Oslo averaging intervals. 
This yields strong support to our interpretation that there are in fact not one but three intermingled scissors modes at play: 
the conventional one and two spin scissors which may be predominately  isovector spin-vector and isoscalar spin-vector in nature. 
As mentioned, this is just the natural triplet of scissors modes which one obtains from pure combinatorics.

The energy centroids and summed $B(M1)$ values of two highest scissors are 
compared with the experimental NRF  
data in the columns 6, 7 and 8, 9 of Tables \ref{tab11}, \ref{tab13s}. 
This comparison is demonstrated also in Fig.~\ref{fig2}.
It is seen that the overall agreement between the theoretical and experimental 
results is very good -- there are only 3 (of 27) remarkable differences for $B(M1)$ ($^{160}$Dy, $^{170}$Er, $^{172}$Yb) and 
four ones for energies ($^{146}$Nd, $^{160,162}$Dy, $^{166}$Er).
The energy centroids and summed $B(M1)$ values of all three scissors are
shown in columns 10 and 11. These data are also shown on Fig.~\ref{fig2}. Only 
for $^{164}$Dy there is the excellent agreement of the theory and experiment. 
For all the rest nuclei the significant additional $M1$ strength, given by the 
WFM method, is a prediction. This prediction is supported by QPNM 
calculations (see Tables~\ref{tab6},~\ref{tab6a} and Fig.~\ref{fig9}).

\subsection{Actinides}

\begin{figure}[b!]
\centering{\includegraphics[width=\columnwidth]{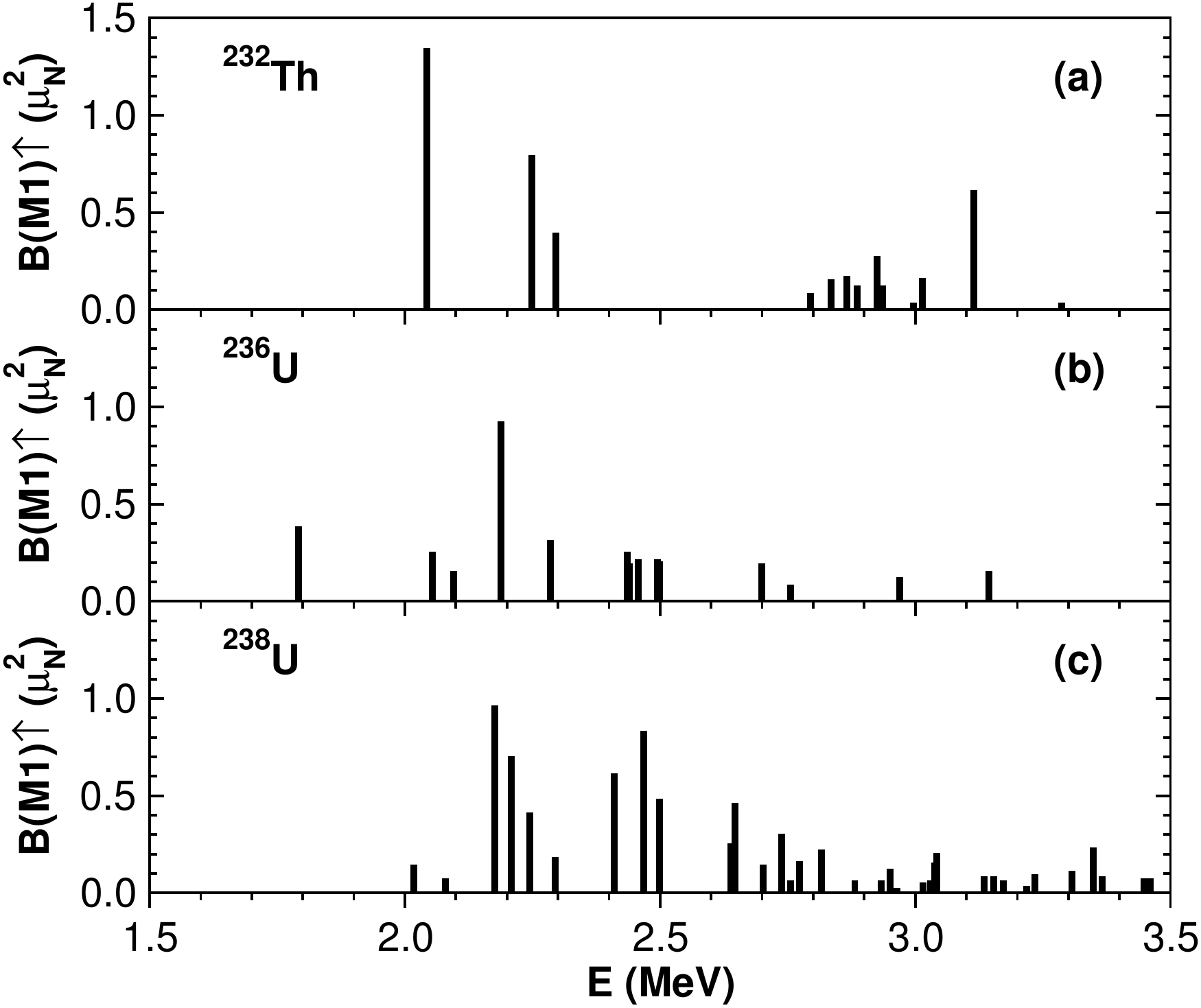}}
\caption{The experimentally observed spectra of $1^+$ excitations in 
(a) $^{232}$Th -- \cite{Adekola}, (b) $^{236}$U -- \cite{U236} and (c) 
$^{238}$U -- \cite{Hammo}.}\label{Acti}
\end{figure}

\begin{table*}[t!]
\caption{The nuclear scissors mode fine structure. The results of calculations by the WFM method: energies $E_i$ with correspondings $B_i(M1)$-values.
Energy centroids $\bar E$ and summed $M1$ strength are also presented.
Parameters of pair correlations: $V_0^{\rm p}=25.5$ MeV, $V_0^{\rm n}=21.5$ MeV, $r_p^{\rm p}=1.5$ fm, $r_p^{\rm n}=1.80$ fm; $\kappa_{\rm Nils}=0.06$, $q=0.7$.}
\begin{ruledtabular}\begin{tabular}{ccccccccccc}
 \multirow{2}{*}{Nuclei}&\multirow{2}{*}{$\delta$}  & \multirow{2}{*}{$i$} & $E_i$ (MeV)  & $B_i(M1)$ ($\mu_N^2$) &  \multicolumn{2}{c}{$\bar E_{[2-3]}$ (MeV)} &  
 \multicolumn{2}{c}{$\sum\limits_{i=2}^{3}\!\!B_i(M1)\ [\mu_N^2]$} &  $\bar E_{[1-3]}$ (MeV)  &  $\sum\limits_{i=1}^{3}\!\!B_i(M1)$ ($\mu_N^2$)\\   
   &  &   &\multicolumn{2}{c}{WFM}   & WFM & NRF &  WFM & NRF & \multicolumn{2}{c}{WFM}    \\
\hline\hline                                                                                                              
             &      & 1 &  1.53 & 1.70  &                       &                       &                       &                           &      &       \\
 $^{232}$Th  &0.216 & 2 &  2.21 & 2.55  & \multirow{2}{*}{2.43} & \multirow{2}{*}{2.49} & \multirow{2}{*}{4.07} & \multirow{2}{*}{4.26(64)} & 2.16 & 5.77  \\ 
             &      & 3 &  2.81 & 1.51  &                       &                       &                       &                           &      &       \\ 
 \hline                                                                                                               
             &      & 1 &  1.54 & 1.91  &                       &                       &                       &                           &      &       \\
 $^{236}$U~  &0.220 & 2 &  2.22 & 2.87  & \multirow{2}{*}{2.44} & \multirow{2}{*}{2.35} & \multirow{2}{*}{4.51} & \multirow{2}{*}{4.06(61)} & 2.17 & 6.41  \\ 
             &      & 3 &  2.82 & 1.64  &                       &                       &                       &                           &      &       \\ 
\hline                                                                                                                
             &      & 1 &  1.57 & 2.12  &                       &                       &                       &                           &      &       \\
 $^{238}$U~  &0.234 & 2 &  2.32 & 3.69  & \multirow{2}{*}{2.54} & \multirow{2}{*}{2.58} & \multirow{2}{*}{5.80} &\multirow{2}{*}{7.59(1.2)} & 2.28 & 7.92  \\ 
             &      & 3 &  2.93 & 2.10  &                       &                       &                       &                           &      &       \\ 
\end{tabular}\end{ruledtabular}\label{Actin}
\end{table*}

\begin{figure}[h!]
\centering{\includegraphics[width=\columnwidth]{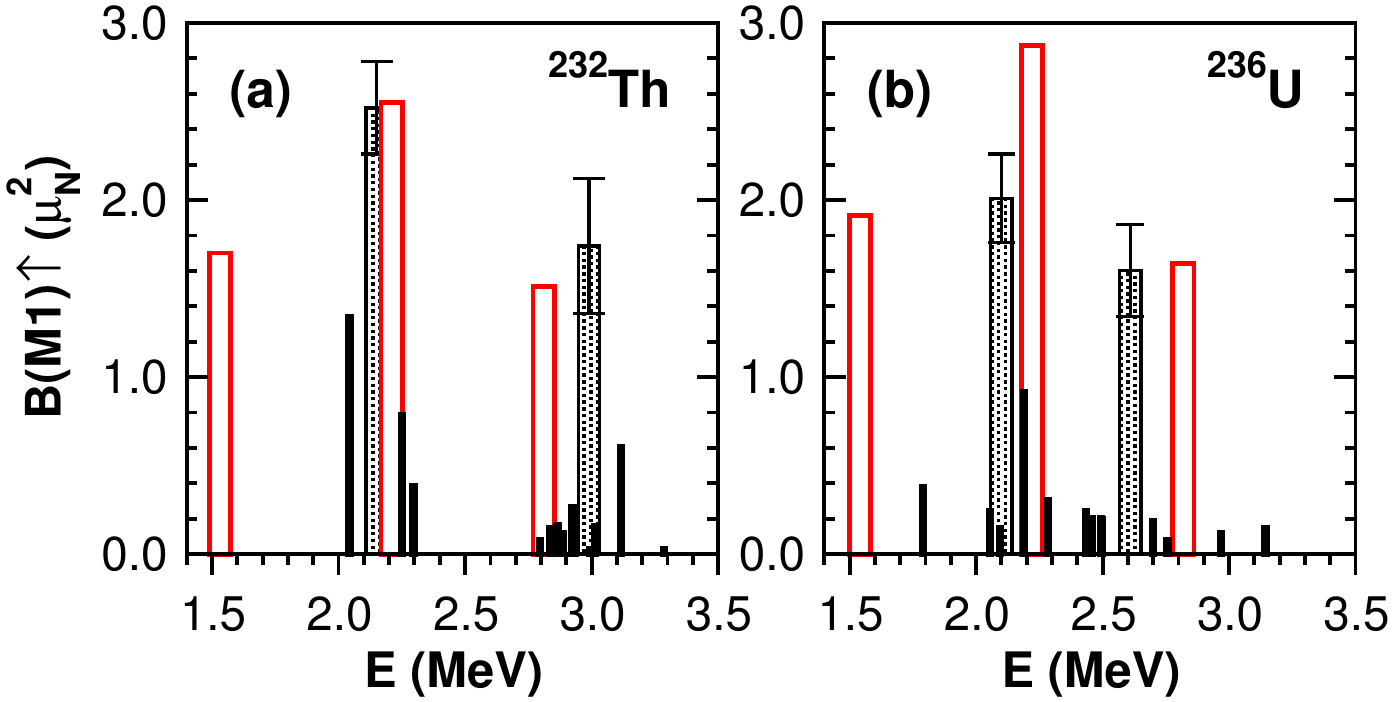}}
\caption{The centroids of experimentally observed spectra of $1^+$ 
excitations in $^{232}$Th (a) and $^{236}$U (b) (black rectangles with error 
bars) are compared with the results of WFM calculations (red 
rectangles).}\label{f3}
\end{figure}

The case of Actinides is similar to the Rare Earth region, see 
Table~\ref{Actin} and Figs.~\ref{Acti},~\ref{f3}. The calculated energy 
centroids and summed $B(M1)$ values of two highest scissors
in $^{232}$Th are in excellent agreement with experimental NRF data. The agreement between the analogous values in $^{236}$U can be characterized as
acceptable. In addition, Fig.~\ref{f3} demonstrates that the average energy and the
summed magnetic strength of the lower group of levels in $^{232}$Th
practically exactly coincides with the energy and $B(M1)$ value of the middle
($E=2.2$~MeV) calculated scissors mode and the analogous values of the higher group of levels are in very good agreement with the energy and $B(M1)$ value of
the highest ($E=2.81$~MeV) scissors mode given by the theory. A similar 
picture can be obtained for $^{236}$U if to divide its spectrum in two groups,
the boundary between them being chosen in the energy window 
2.3 MeV $<E<$ 2.4 MeV (see Fig.~\ref{Acti}).
It is worth to note an interesting detail of $^{236}$U spectrum. Its lowest level is disposed remarkably lower 2 MeV and is separated by the remarkable
energy gap from the higher levels. This makes it possible to interpret this 
level as a small fraction of the lowest scissors predicted by the theory.

One observes an unexpectedly large value of 
the summed $B(M1)$ for $^{238}$U in  comparison with that of $^{236}$U and 
$^{232}$Th and with the theoretical result. 
The possible reason of this discrepancy was indicated by the authors of
\cite{Hammo}: {\it ``$M1$ excitations are observed at approximately $2.0$ MeV $<E_{\gamma}< 3.5$ MeV with a strong concentration of $M1$ states around 2.5 MeV. ... 
The observed $M1$ strength may include states from both the scissors 
mode and the spin-flip mode, which are indistinguishable from each other based
exclusively on the use of the NRF technique.''} 
The most reasonable (and quite natural) place for the boundary between the scissors mode and the 
spin-flip resonance is located in the spectrum gap between 2.5 MeV and 2.62 MeV (see Fig.~\ref{Acti}). The summed $M1$ strength 
of scissors in this case becomes $B(M1)=4.38\pm 0.5\ \mu_N^2$ in rather good agreement with  $^{236}$U and $^{232}$Th. However one can not be satisfied 
by this agreement, because this value turns out a little bit too small in
comparison with the theoretical result 5.8 $\mu_N^2$. 
In addition, having remarkably bigger
deformation, $^{238}$U is expected to have bigger $M1$ strength than $^{236}$U
and $^{232}$Th according to the experimentally established rule 
$B(M1)\sim \delta^2$. In this connection it makes sense to consider another
possible place for the required boundary. If one puts it into the less
pronounced spectrum gap between 2.82 MeV and 2.88 MeV, then the summed $B(M1)$
of scissors becomes 5.97 $\mu_N^2$ which agrees rather well with the theoretical value.

\subsection{Currents}\label{Currents}

Figure ~\ref{fig1} gives a schematic view of all possible nuclear
scissors. To obtain a more objective picture of the phenomenon it is necessary
to study the distribution of neutron and proton currents 
$J_i^{\varsigma}(\br)$.
By definition the current is obtained by the odd in {\bf p} part of the phase space distribution
\begin{eqnarray}
J^{\varsigma}_i(\br,t)=\int\!\frac{d^3p}{(2\pi\hbar)^3} p_i
f^{\varsigma}_o(\br,\bp,t).
\label{Cur}
\end{eqnarray}
An isospin index is omitted for simplicity.
In Ref. \cite{BaSc}, where the simple model of a harmonic oscillator 
with separable QQ interaction was considered, the analytical formula for 
the nucleons' flows was derived. In the case with spin degrees of freedom and pair 
correlations the currents can be constructed only numerically. According to 
the approximation suggested in \cite{Balb,BaMoPRC} the current variation is expanded in the following series:
\begin{eqnarray}
\delta J^{\varsigma}_{i}(\br,t)
=n^+(\br)\left[K^{\varsigma}_{i}(t)
+\sum_{j}(-1)^{j}K^{\varsigma}_{i,-j}(t)r_{j}\right.
\nonumber\\
\left. +\sum_{\lambda',\mu'}(-1)^{\mu'}K^{\varsigma}_{i,\lambda'-\mu'}(t)
\{r\otimes r\}_{\lambda'\mu'}+...\right].\;\,
\label{VarJ}
\end{eqnarray}
All terms containing expansion coefficients $K$ with odd numbers of indexes
disappear due to axial symmetry. Furthermore,
we truncate this series omitting all terms generating higher than 
second order moments. So, finally the following expression is used:
\begin{eqnarray}
\label{VarJtr}
\delta J^{\varsigma}_{i}(\br,t)
=n^+(\br)\sum_{j}(-1)^{j}K^{\varsigma}_{i,-j}(t)r_{j}.
\end{eqnarray}
The detailed expressions are:
\begin{eqnarray}
\label{VarJdet}
&&\delta J^{\varsigma}_1
=n^+\left(K^{\varsigma}_{1,0}r_{0}-K^{\varsigma}_{1,-1}r_{1}-K^{\varsigma}_{1,1}r_{-1}\right),
\nonumber\\
&&\delta J^{\varsigma}_0
=n^+\left(K^{\varsigma}_{0,0}r_{0}-K^{\varsigma}_{0,-1}r_{1}-K^{\varsigma}_{0,1}r_{-1}\right),
\nonumber\\
&&\delta J^{\varsigma}_{-1}
=n^+\left(K^{\varsigma}_{-1,0}r_{0}-K^{\varsigma}_{-1,-1}r_{1}-K^{\varsigma}_{-1,1}r_{-1}\right).
\nonumber
\end{eqnarray}
The coefficients $K^{\varsigma}_{i,-j}(t)$ are connected by linear relations 
with the collective variables $\L_{\lambda\mu}^{\varsigma}(t)$ 
(see Appendix~\ref{AppB}). 
Taking into account, that in the frame of the problem 
considered here $\L_{\lambda 0}^{\varsigma}=\L_{\lambda 2}^{\varsigma}=0$ 
for $\varsigma=+,-$, 
we find 
\begin{eqnarray}
\label{VarJL}
&&\delta J^{\varsigma}_1 = n^+\alpha_1 
\left(\L^{\varsigma}_{21}-\L^{\varsigma}_{11}\right)r_{0},
\nonumber\\
&&\delta J^{\varsigma}_0 = n^+\alpha_2
\left[ \left(\L^{\varsigma}_{2-1}-\L^{\varsigma}_{1-1}\right)r_{1}+\left(\L^{\varsigma}_{21}+\L^{\varsigma}_{11}\right)r_{-1} \right],
\nonumber\\
&&\delta J^{\varsigma}_{-1} = n^+\alpha_1
\left(\L^{\varsigma}_{2-1}+\L^{\varsigma}_{1-1}\right)r_{0},
\nonumber
\end{eqnarray}
where $\alpha_i=\sqrt3 /(\sqrt2A_i)$ and $A_i$ are defined by~(\ref{Ai}).
The expressions for currents in Cartesian coordinates are written:
\begin{eqnarray}
&&\delta J^{\varsigma}_x=(\delta J^{\varsigma}_{-1}-\delta J^{\varsigma}_{1})/\sqrt2 
\nonumber\\
&&\qquad=\frac{1}{\sqrt2}n^+\alpha_1\left(\L^{\varsigma}_{2-1}-\L^{\varsigma}_{21}+\L^{\varsigma}_{1-1}+\L^{\varsigma}_{11}\right)z,
\nonumber\\
&&\delta J^{\varsigma}_y=i(\delta J^{\varsigma}_{-1}+\delta J^{\varsigma}_{1})/\sqrt2
\nonumber\\
&&\qquad=\frac{i}{\sqrt2}n^+\alpha_1\left(\L^{\varsigma}_{2-1}+\L^{\varsigma}_{21}+\L^{\varsigma}_{1-1}-\L^{\varsigma}_{11}\right)z,
\nonumber\\
&&\delta J^{\varsigma}_z=\delta J^{\varsigma}_0
=n^+\alpha_2
\bigg[
\left(\L^{\varsigma}_{21}-\L^{\varsigma}_{2-1}+\L^{\varsigma}_{11}+\L^{\varsigma}_{1-1}\right)x
\nonumber\\
&&\qquad-\frac{i}{\sqrt2}\left(\L^{\varsigma}_{21}+\L^{\varsigma}_{2-1}+\L^{\varsigma}_{11}-\L^{\varsigma}_{1-1}\right)y
\bigg].
\end{eqnarray}
\begin{table}[t!] 
\caption{Strengths (amplitudes) of currents in $^{164}$Dy.
$\beta=-B/A.$}
\begin{ruledtabular}\begin{tabular}{cccccc}
 $E$ (MeV)            & (i) &  $B\ (10^{-2})$ & $A\ (10^{-2})$ & \% & $\beta$ \\ 
\hline
                      & (a) &  0.75 & -0.47 & \multirow{2}{*}{~1.75} &  1.60 \\
                      & (b) &  0.51 & -0.18 &                        &  2.79 \\
\multirow{2}{*}{2.20} & (c) & -1.46 &  2.77 & \multirow{2}{*}{47.29} &  0.53 \\
                      & (d) &  2.72 & -3.42 &                        &  0.79 \\
                      & (e) &  2.87 & -3.50 & \multirow{2}{*}{50.95} &  0.82 \\
                      & (f) & -1.61 &  2.85 &                        &  0.57 \\  
\hline
                      & (a) &  1.99 & -2.44 & \multirow{2}{*}{31.90} &  0.82 \\
                      & (b) & -2.94 &  4.00 &                        &  0.74 \\
\multirow{2}{*}{2.87} & (c) &  2.90 & -3.32 & \multirow{2}{*}{53.71} &  0.87 \\
                      & (d) & -3.85 &  4.89 &                        &  0.79 \\
                      & (e) &  1.22 & -1.24 & \multirow{2}{*}{14.39} &  0.99 \\
                      & (f) & -2.17 &  2.80 &                        &  0.78 \\ 
\hline
                      & (a) & 11.57 &-12.14 & \multirow{2}{*}{61.55} &  0.95 \\
                      & (b) & -8.17 & 15.05 &                        &  0.54 \\
\multirow{2}{*}{3.59} & (c) & -1.87 &  5.75 & \multirow{2}{*}{~7.76} &  0.33 \\
                      & (d) &  5.27 & -2.84 &                        &  1.86 \\
                      & (e) & -5.95 & 10.39 & \multirow{2}{*}{30.69} &  0.57 \\
                      & (f) &  9.35 & -7.48 &                        &  1.25 \\
\end{tabular}\end{ruledtabular}\label{tab-cur}
\end{table}\begin{figure}[h!]
\includegraphics[width=0.5\columnwidth]{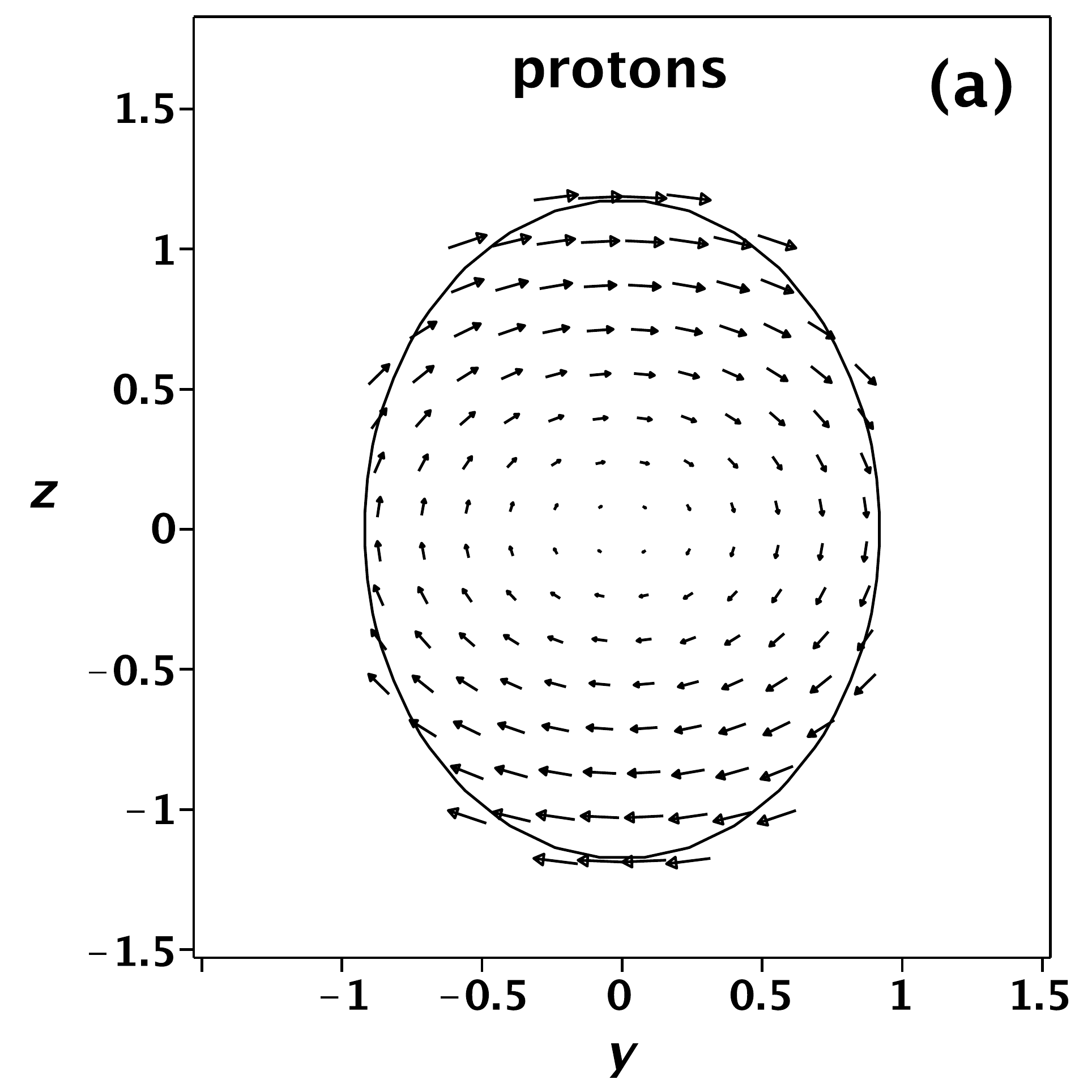}\includegraphics[width=0.5\columnwidth]{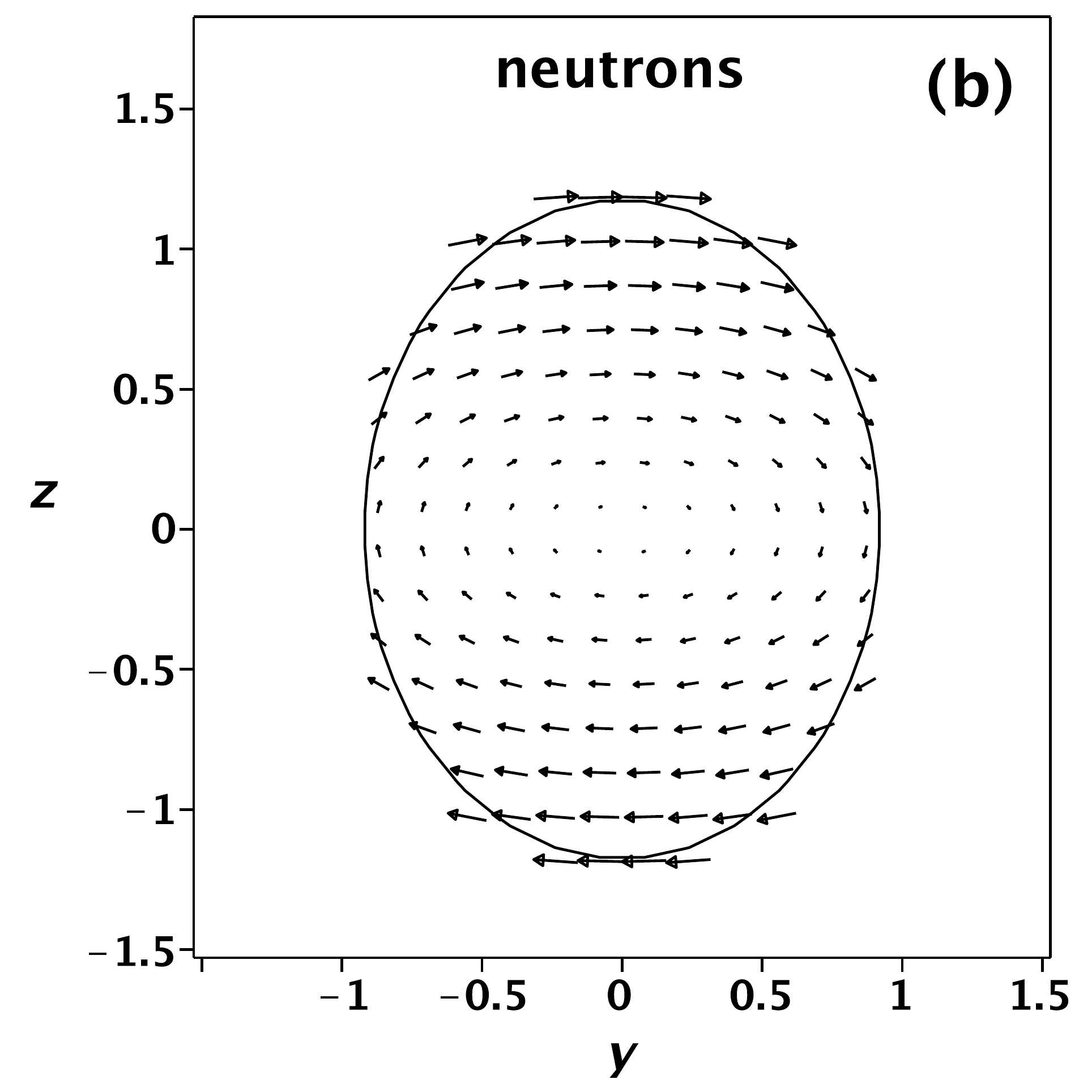}\\
\includegraphics[width=0.5\columnwidth]{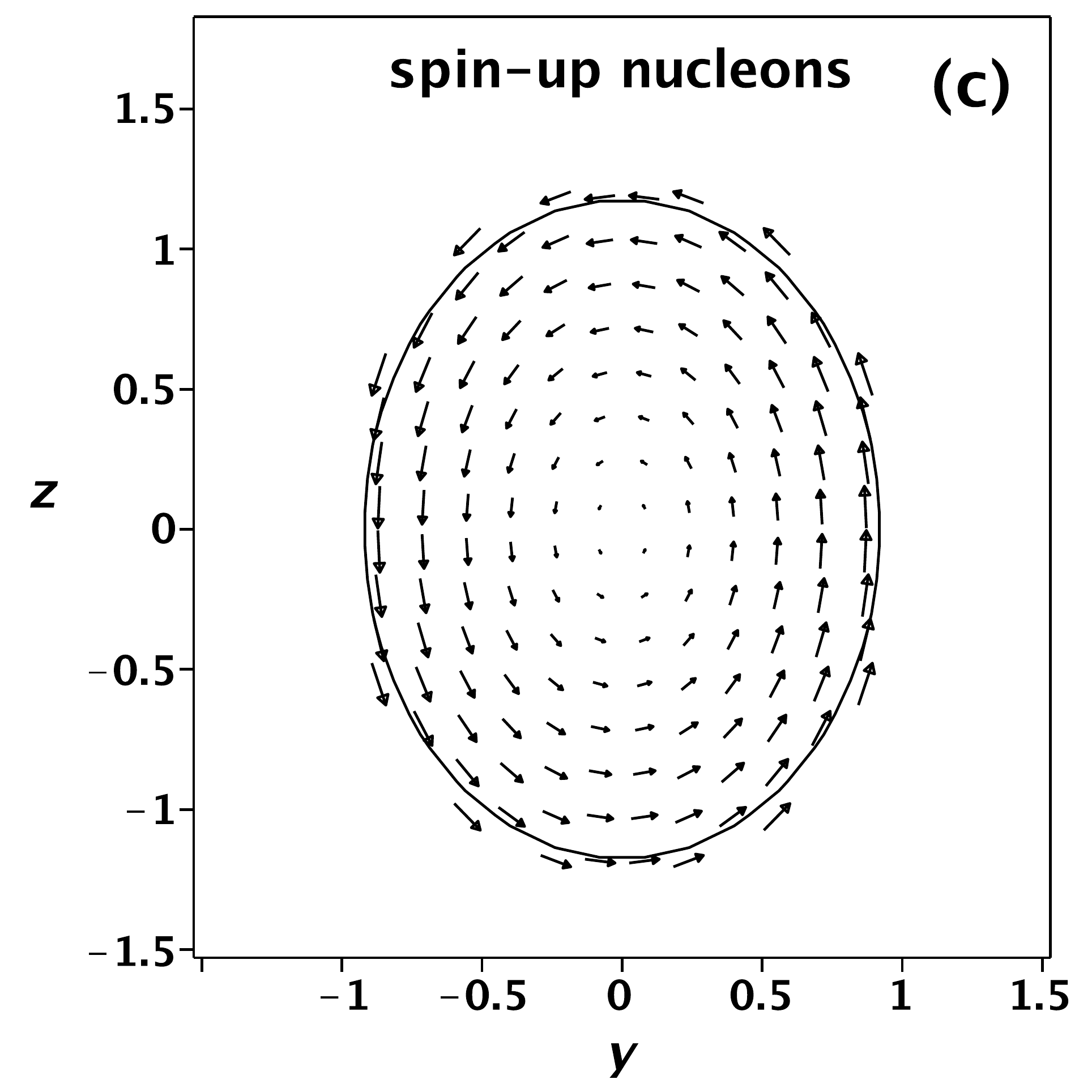}\includegraphics[width=0.5\columnwidth]{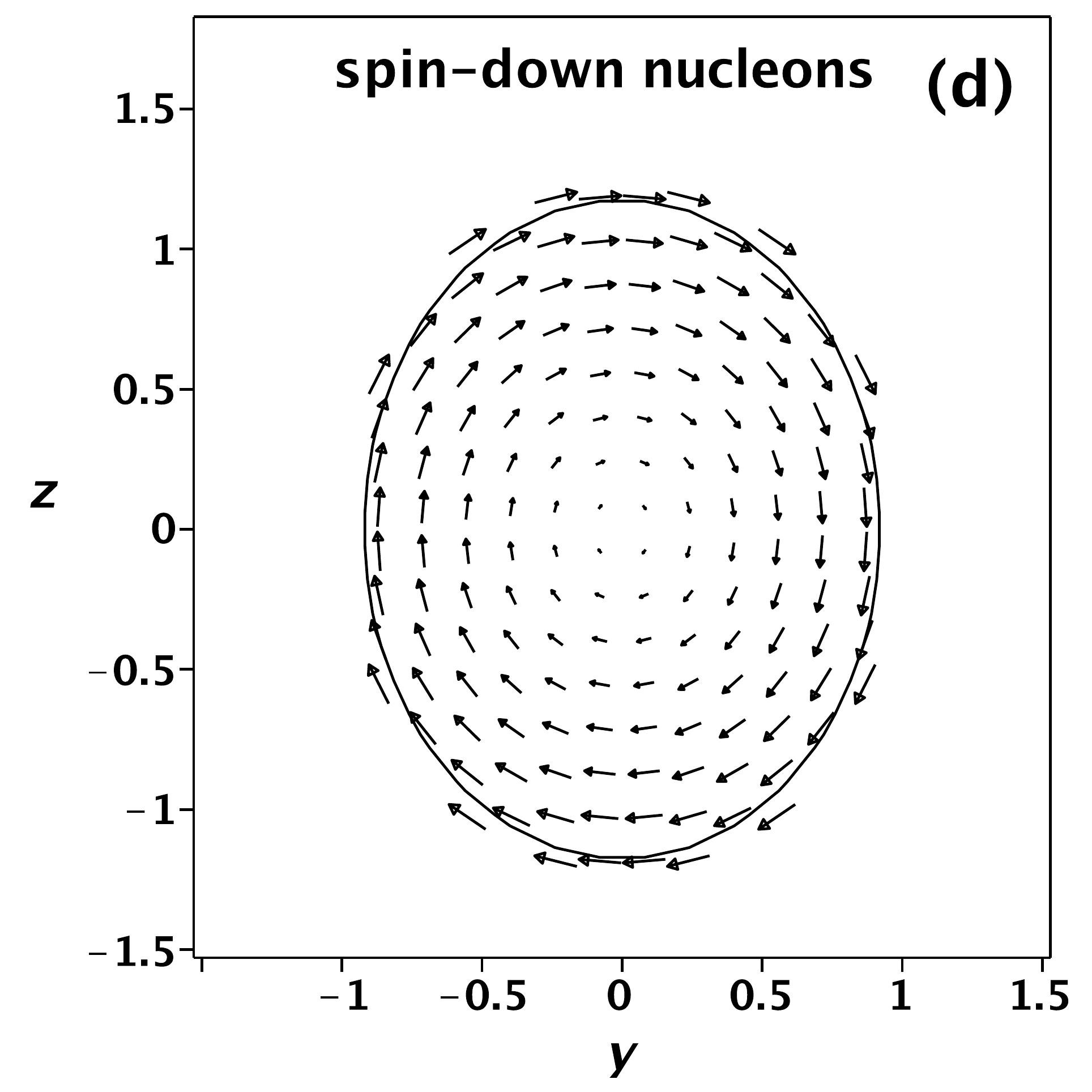}\\
\includegraphics[width=0.5\columnwidth]{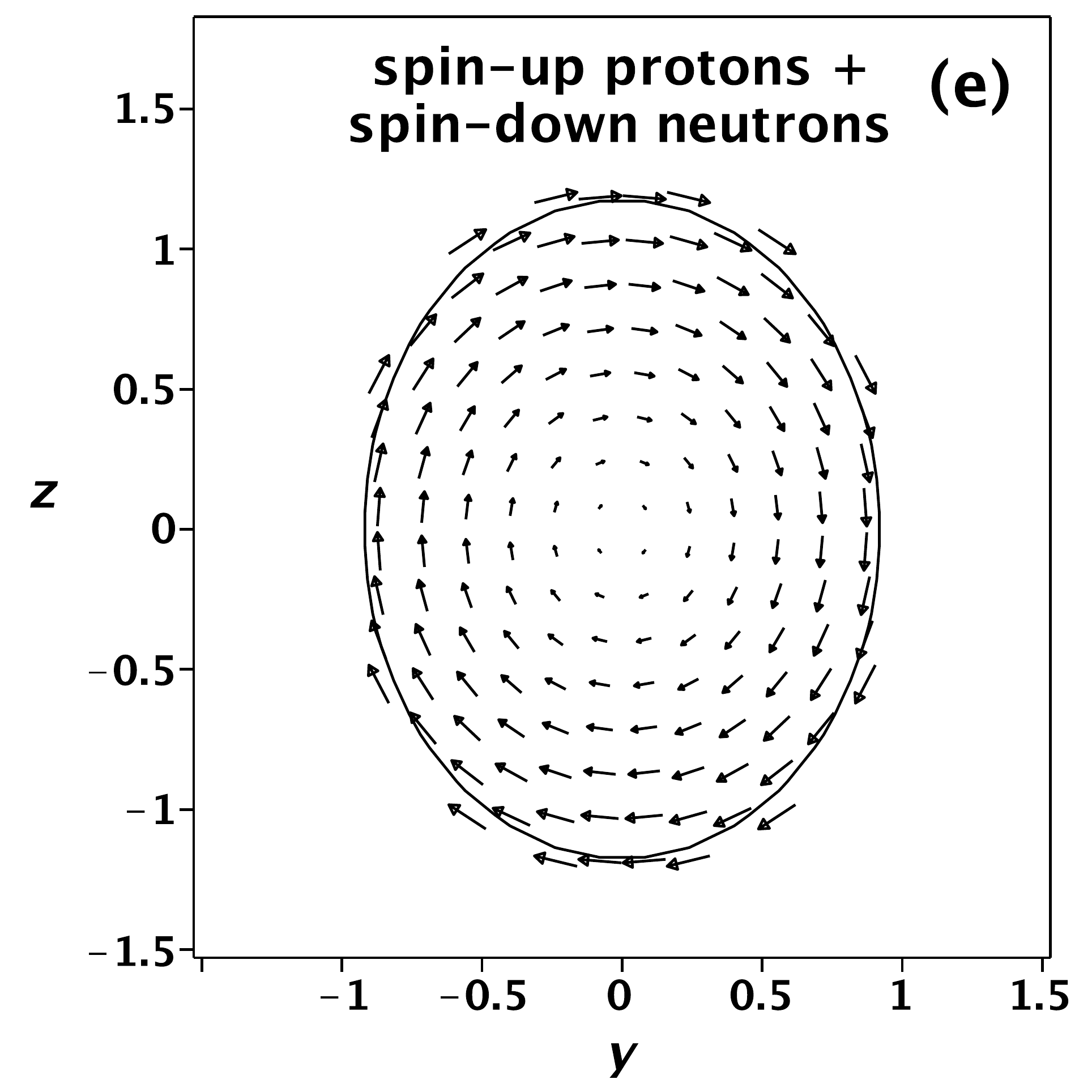}\includegraphics[width=0.5\columnwidth]{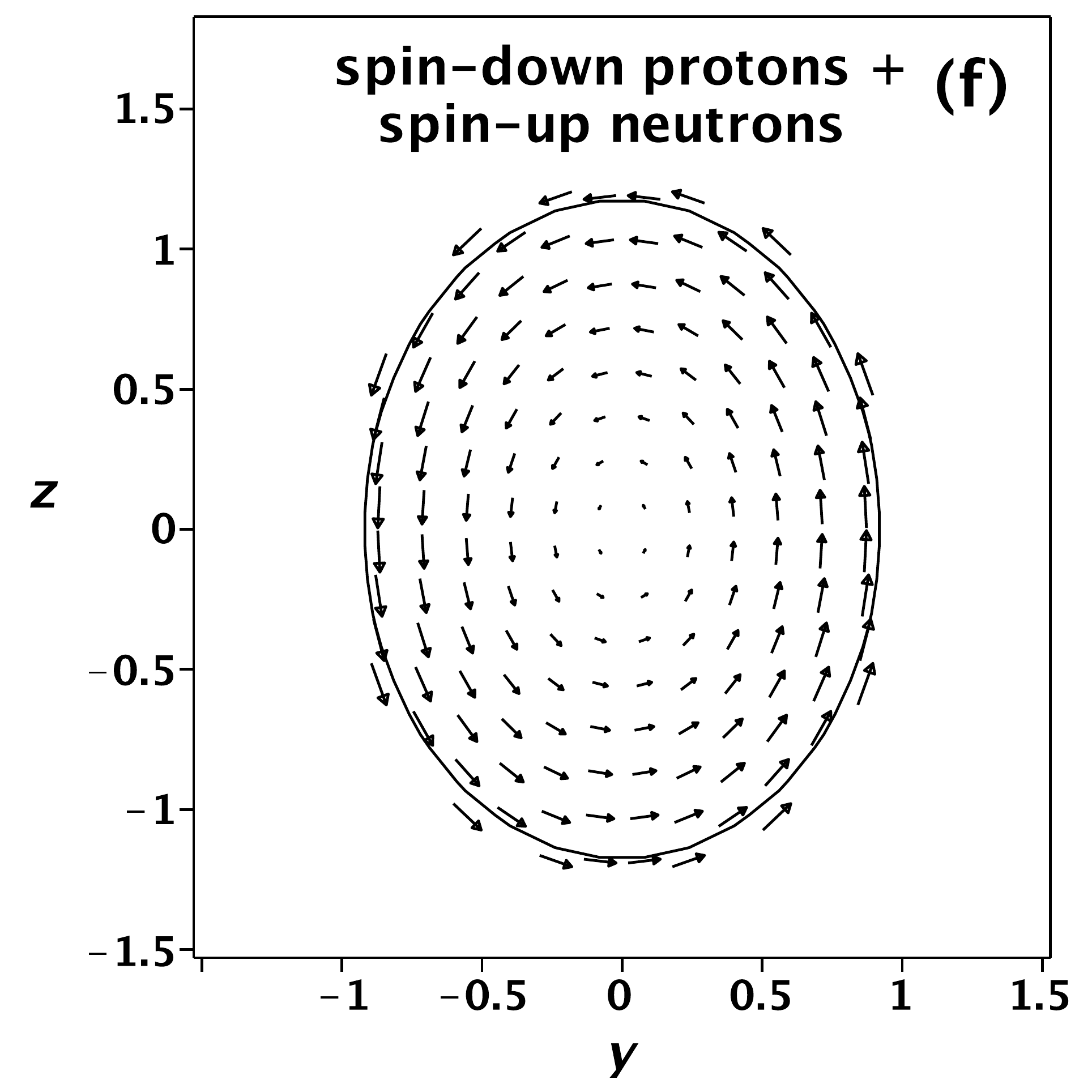}
\caption{The currents in $^{164}$Dy for $E=2.20$ MeV:
$\delta J^{+}_{\rm p}$~(a), $\delta J^{+}_{\rm n}$~(b), 
$\delta J^{\uparrow\uparrow}$~(c), $\delta J^{\downarrow\downarrow}$~(d),
$\delta J^{\uparrow\uparrow}_{\rm p}+\delta J^{\downarrow\downarrow}_{\rm n}$~(e), 
$\delta J^{\downarrow\downarrow}_{\rm p}+\delta J^{\uparrow\uparrow}_{\rm n}$~(f).
\mbox{{\textsf y} $=y/R$, {\textsf z} $=z/R$.}}
\label{E1}
\end{figure}
The comparison of the set of equations for \mbox{$\mu=1$~(\ref{iv})} with the 
analogous set of equations for $\mu=-1$ allows one to find that $\L^{\varsigma}_{2-1}=\L^{\varsigma}_{21}$ and 
$\L^{\varsigma}_{1-1}=-\L^{\varsigma}_{11}$ (with $\varsigma=+,-$). Therefore we have:
\begin{eqnarray}
\label{JCart}
&&\delta J^{\varsigma}_x=0,
\nonumber\\
&&\delta J^{\varsigma}_y=-i\frac{\sqrt{3}}{A_1}n^+\!\left(\L^{\varsigma}_{11}-\L^{\varsigma}_{21}\right)z,
\nonumber\\
&&\delta J^{\varsigma}_z=
-i\frac{\sqrt{3}}{A_2}n^+\!\left(\L^{\varsigma}_{11}+\L^{\varsigma}_{21}\right)y.
\end{eqnarray}
This result is quite remarkable. The first equation $\delta J^{\varsigma}_x=0$ says that 
all motions take place only in two dimensions, i.e. in one plane. Obviously 
it is one of the properties to be satisfied by the scissors mode. Another obvious and 
necessary property of the scissors mode is the rotational out of phase motion of 
its subentities. 
This property is demonstrated by the pictures of currents
(see Figs.~\ref{E1},~\ref{E2},~\ref{E3}) constructed 
with the help of second and third equations of (\ref{JCart}).


\begin{figure}[t!]
\includegraphics[width=0.5\columnwidth]{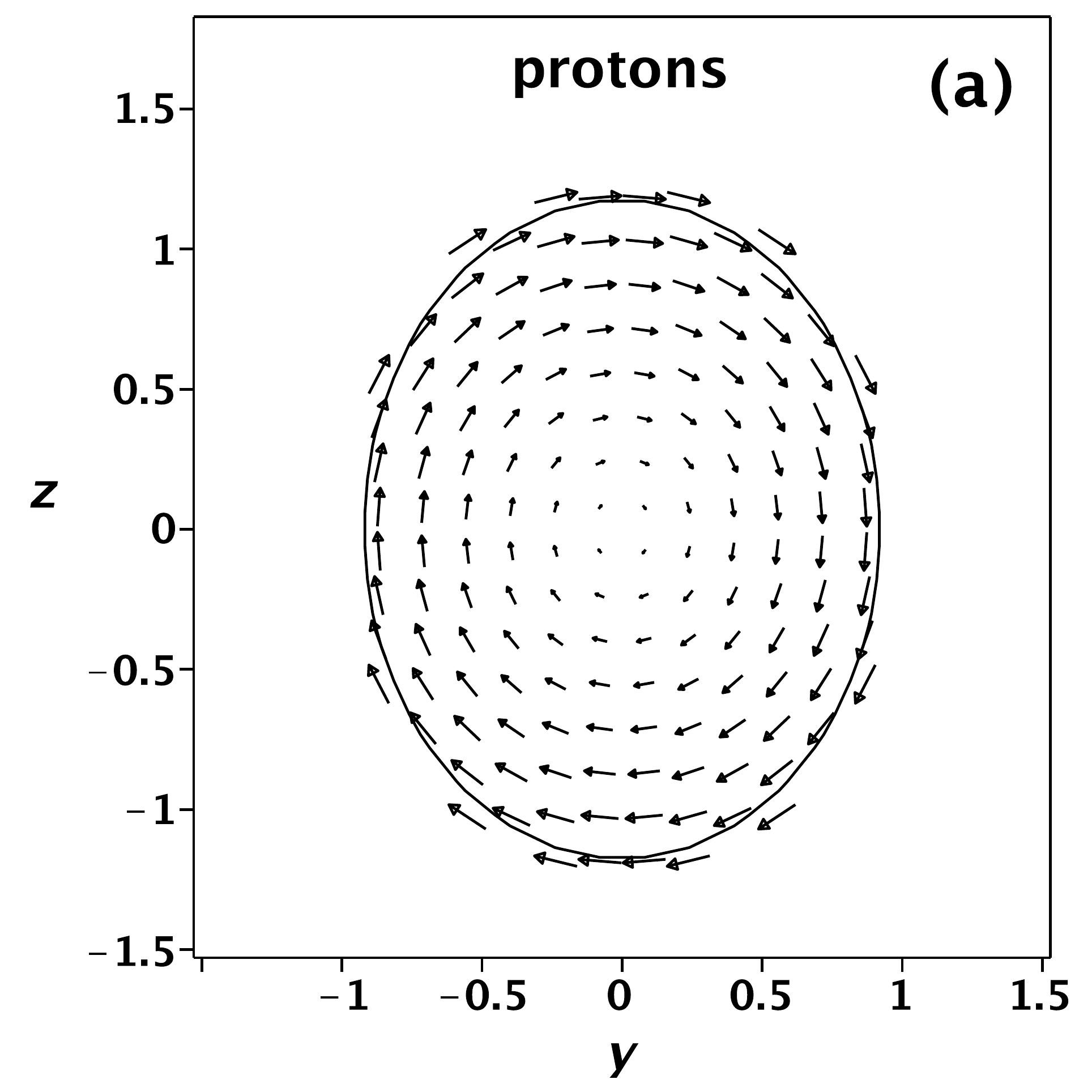}\includegraphics[width=0.5\columnwidth]{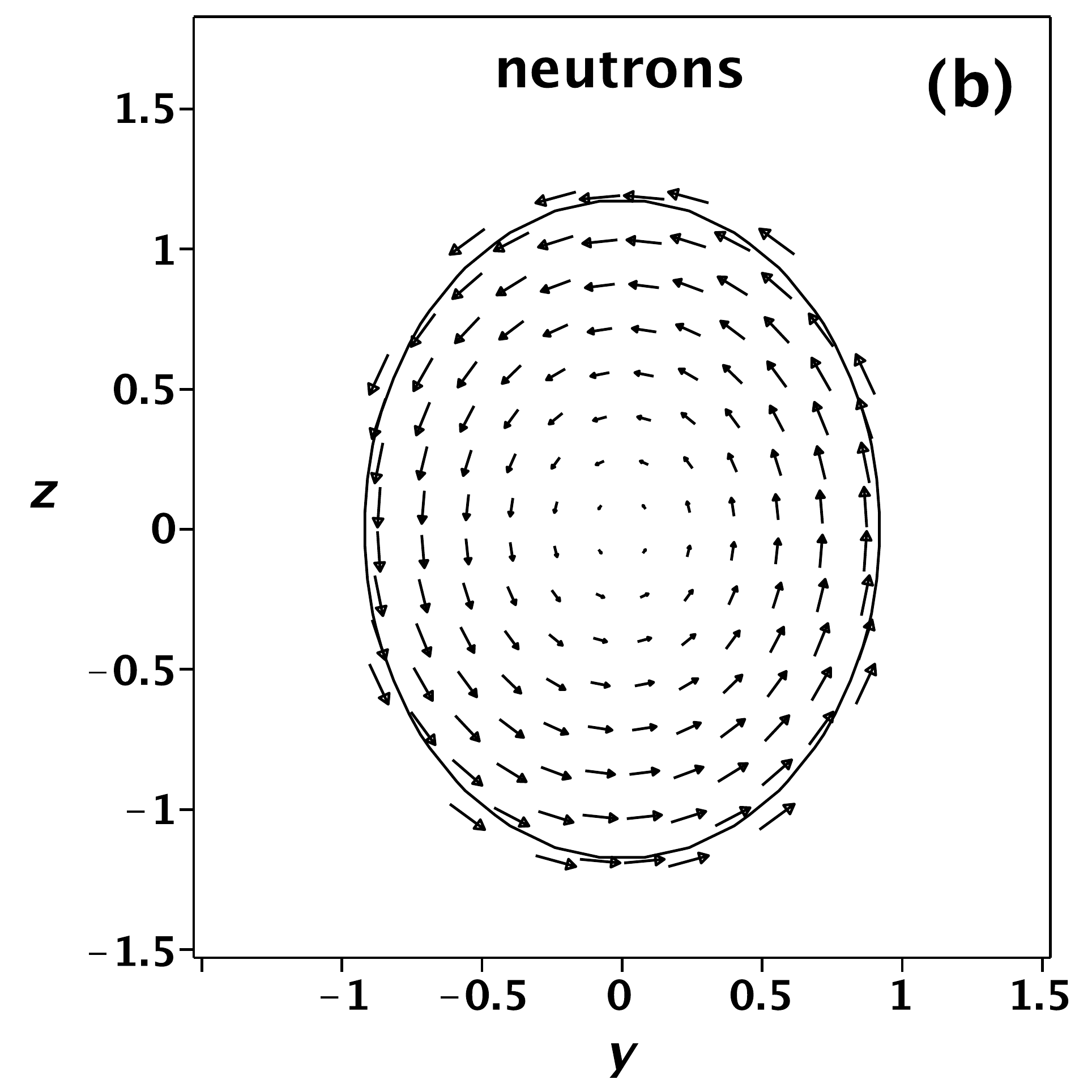}\\
\includegraphics[width=0.5\columnwidth]{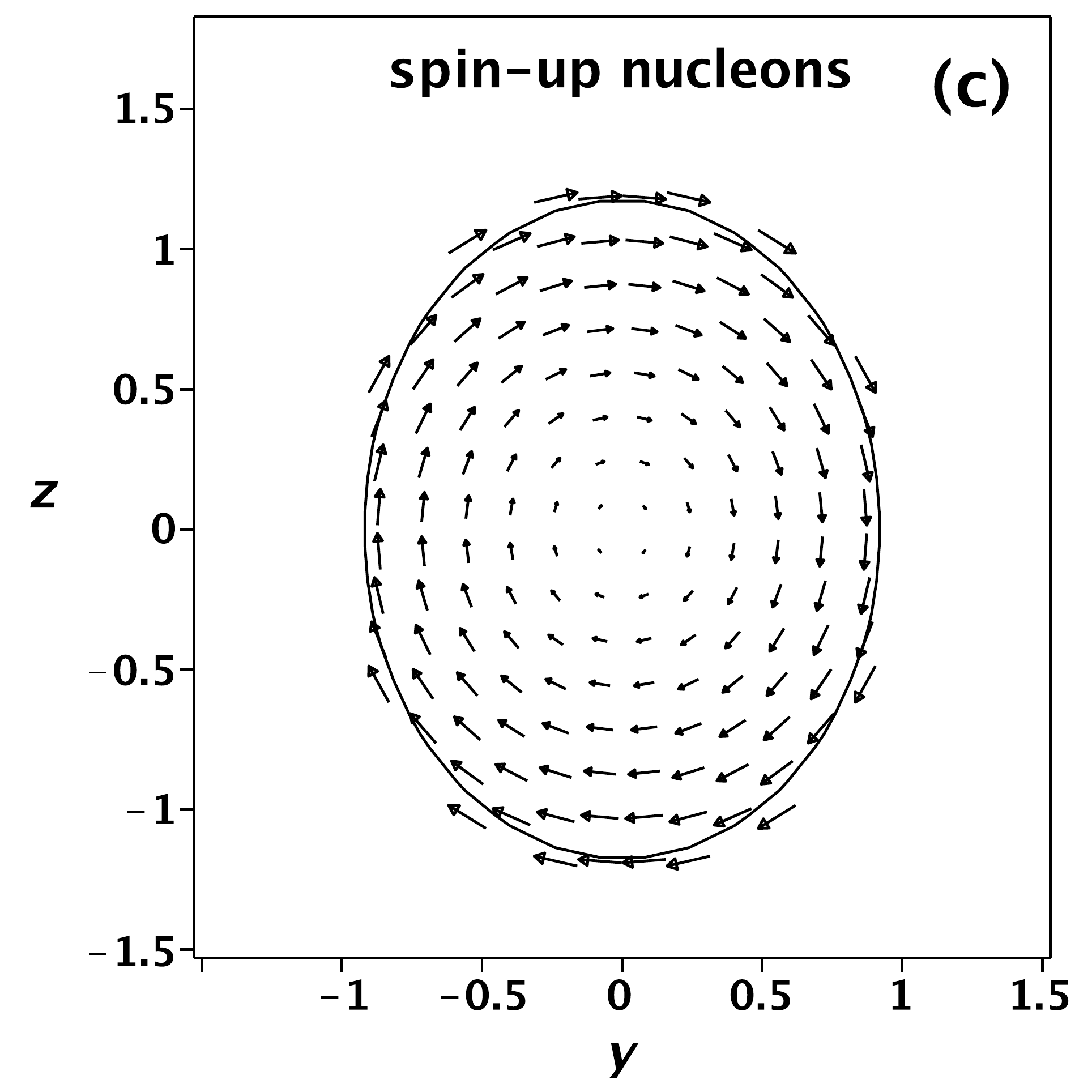}\includegraphics[width=0.5\columnwidth]{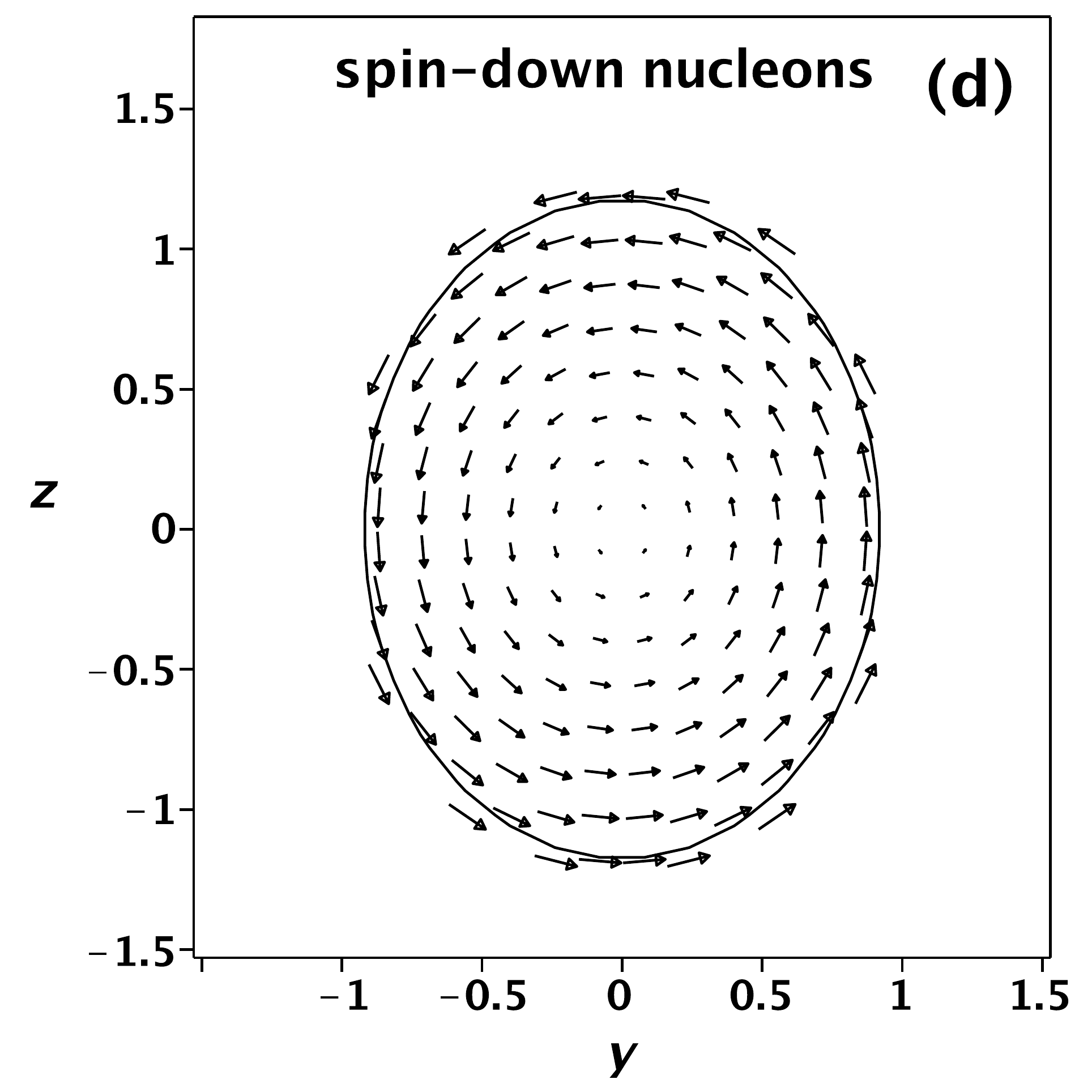}\\
\includegraphics[width=0.5\columnwidth]{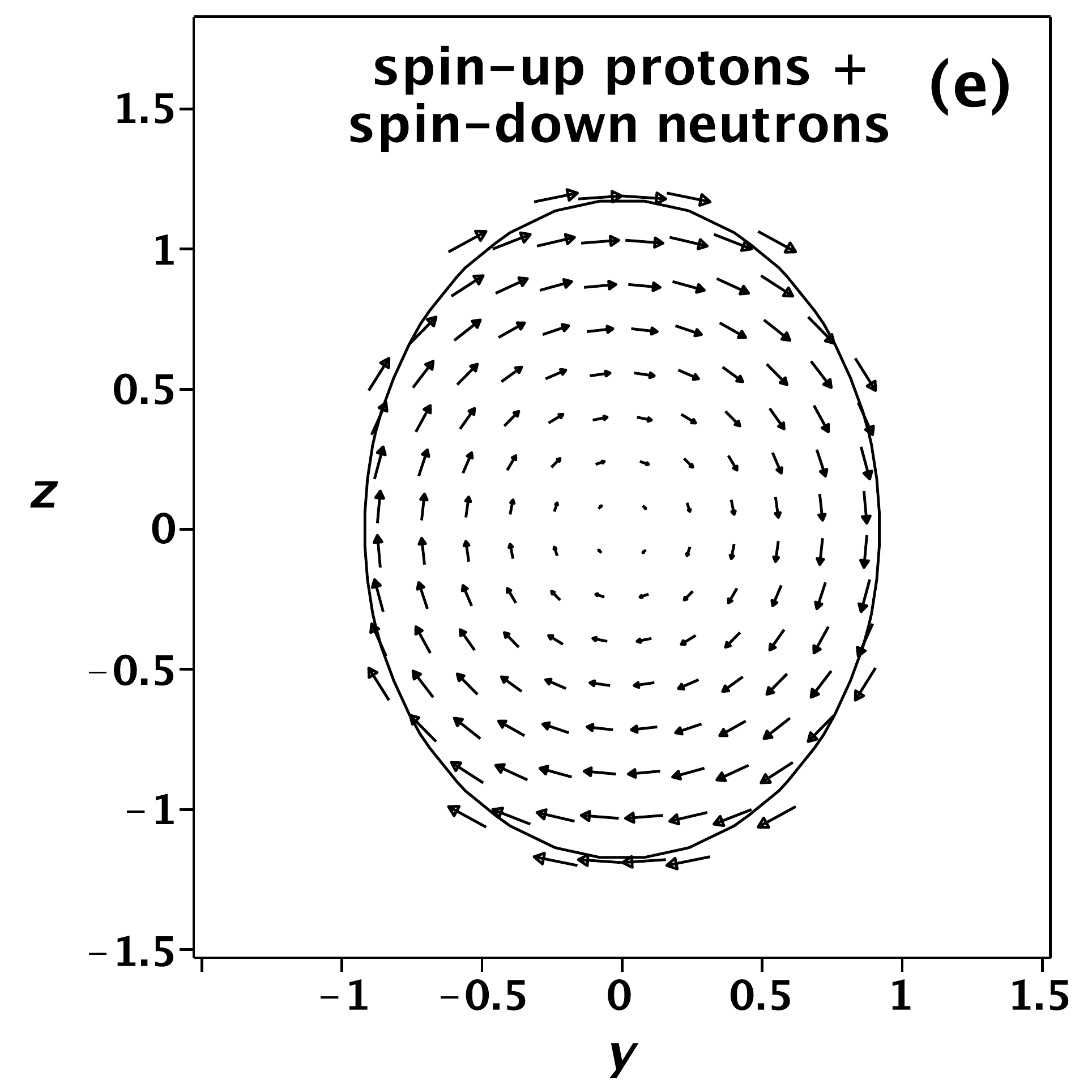}\includegraphics[width=0.5\columnwidth]{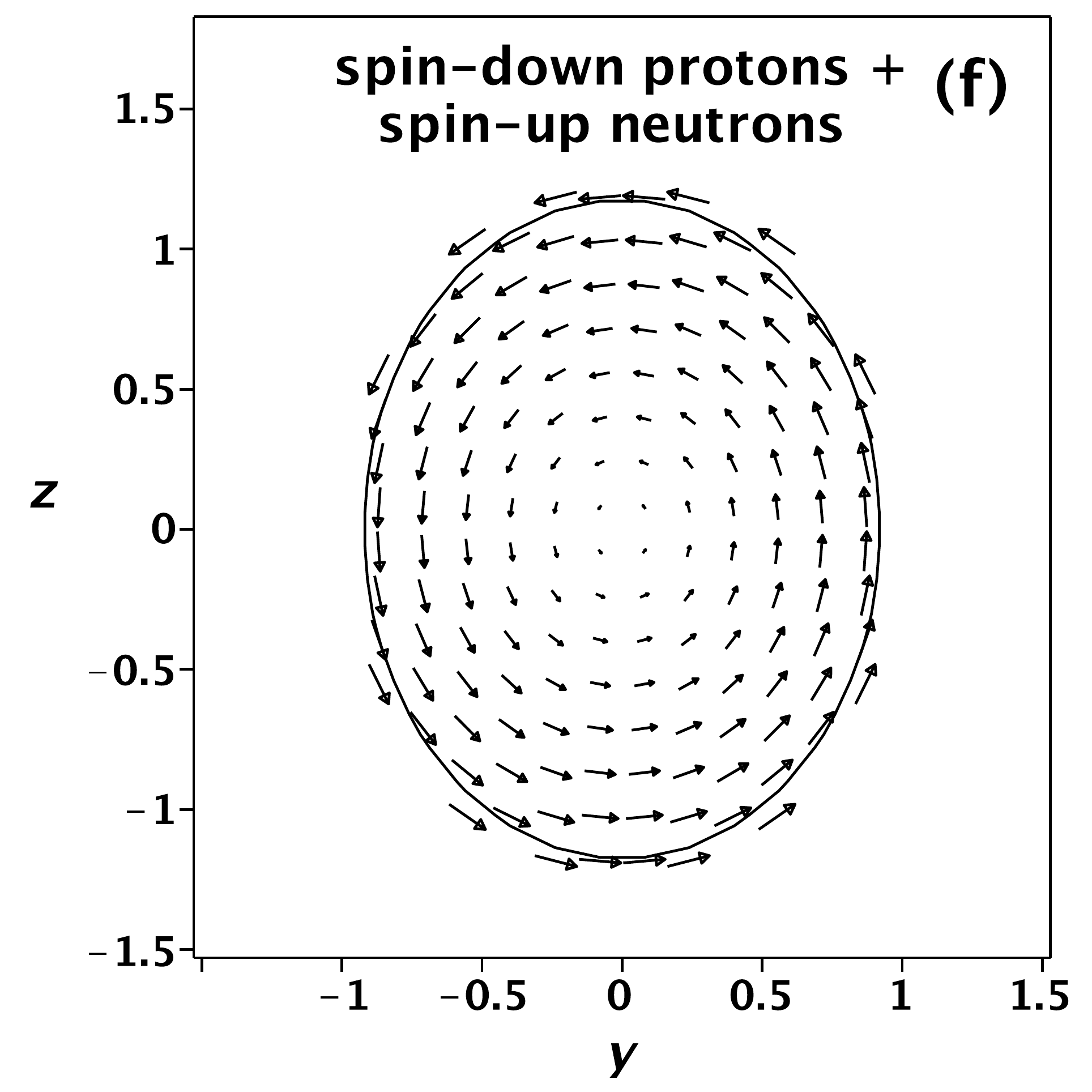}
\caption{The currents in $^{164}$Dy for $E=2.87$ MeV: $\delta J^{+}_{\rm p}$~(a), $\delta J^{+}_{\rm n}$~(b), 
$\delta J^{\uparrow\uparrow}$~(c), $\delta J^{\downarrow\downarrow}$~(d),
$\delta J^{\uparrow\uparrow}_{\rm p}+\delta J^{\downarrow\downarrow}_{\rm n}$~(e), 
$\delta J^{\downarrow\downarrow}_{\rm p}+\delta J^{\uparrow\uparrow}_{\rm 
n}$~(f).
\mbox{{\textsf y} $=y/R$, {\textsf z} $=z/R$.}}
\label{E2}
\end{figure}
\begin{figure}[t!]
\includegraphics[width=0.5\columnwidth]{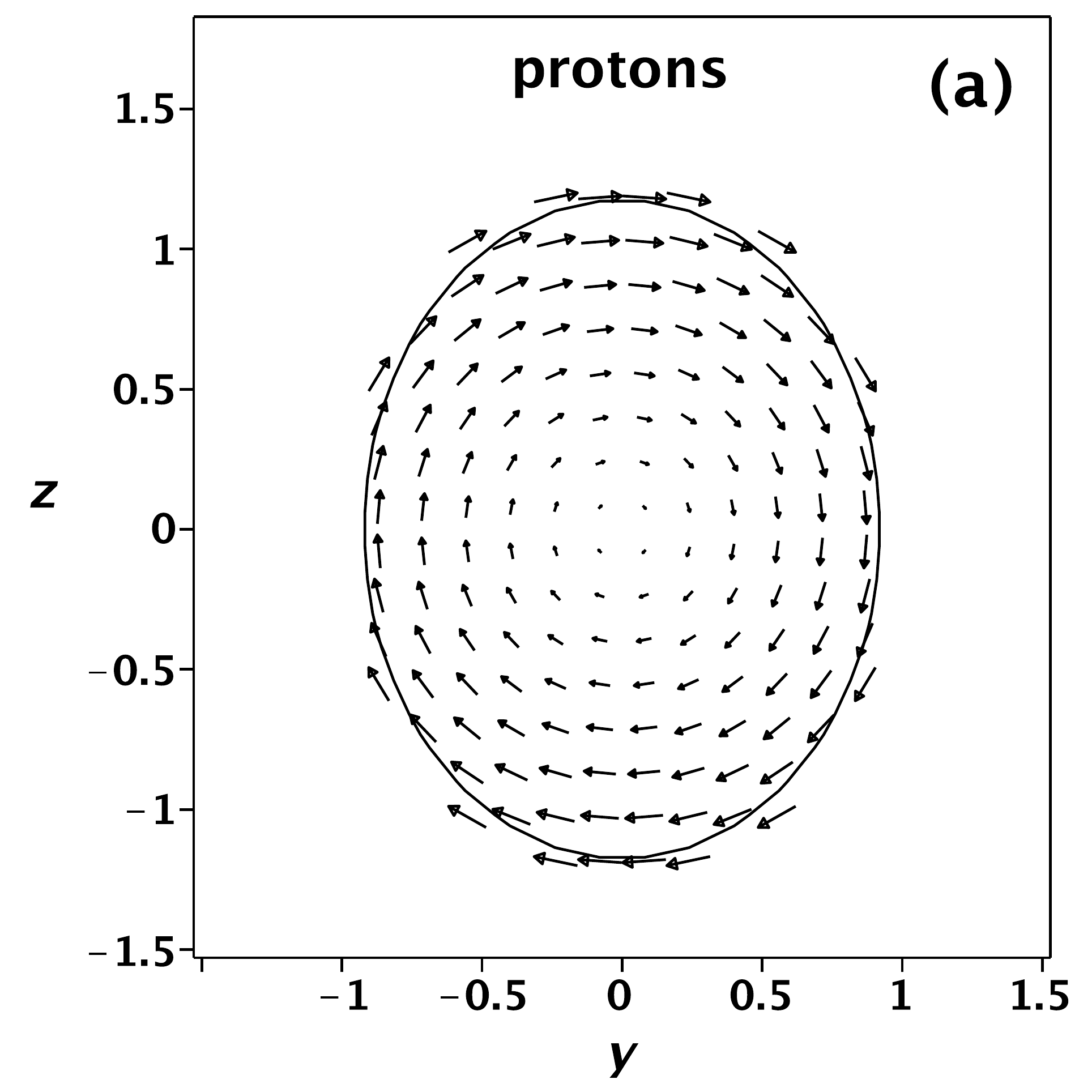}\includegraphics[width=0.5\columnwidth]{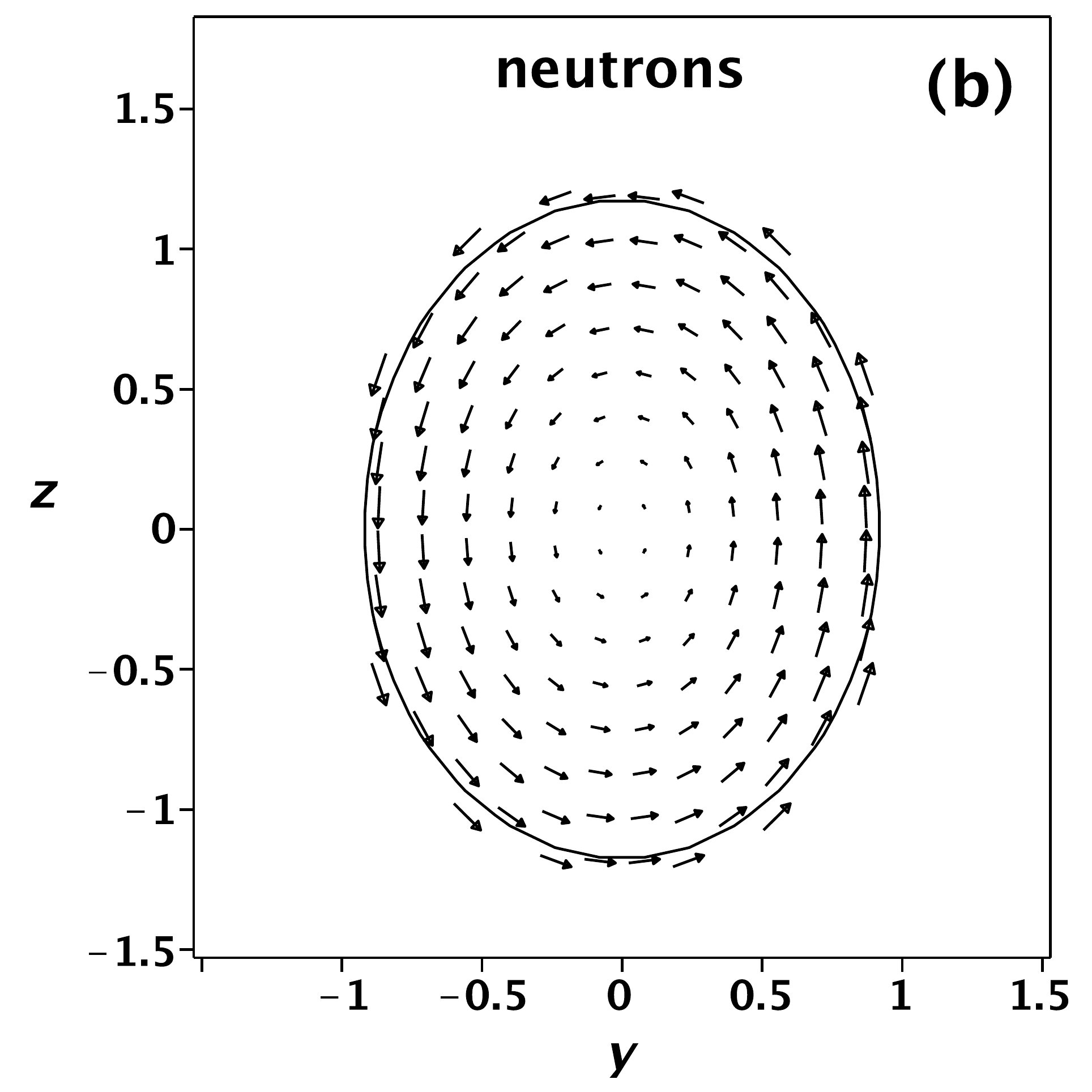}\\
\includegraphics[width=0.5\columnwidth]{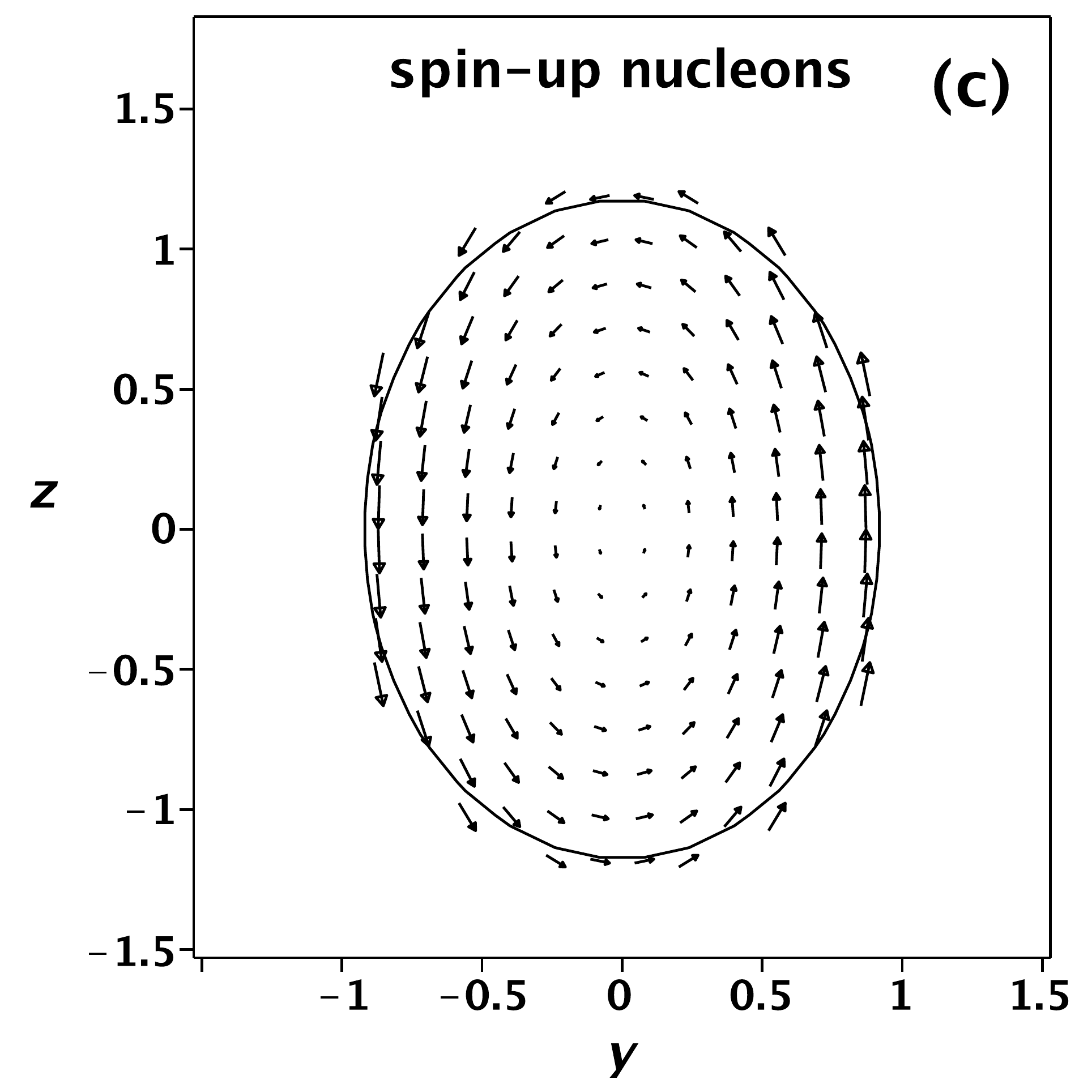}\includegraphics[width=0.5\columnwidth]{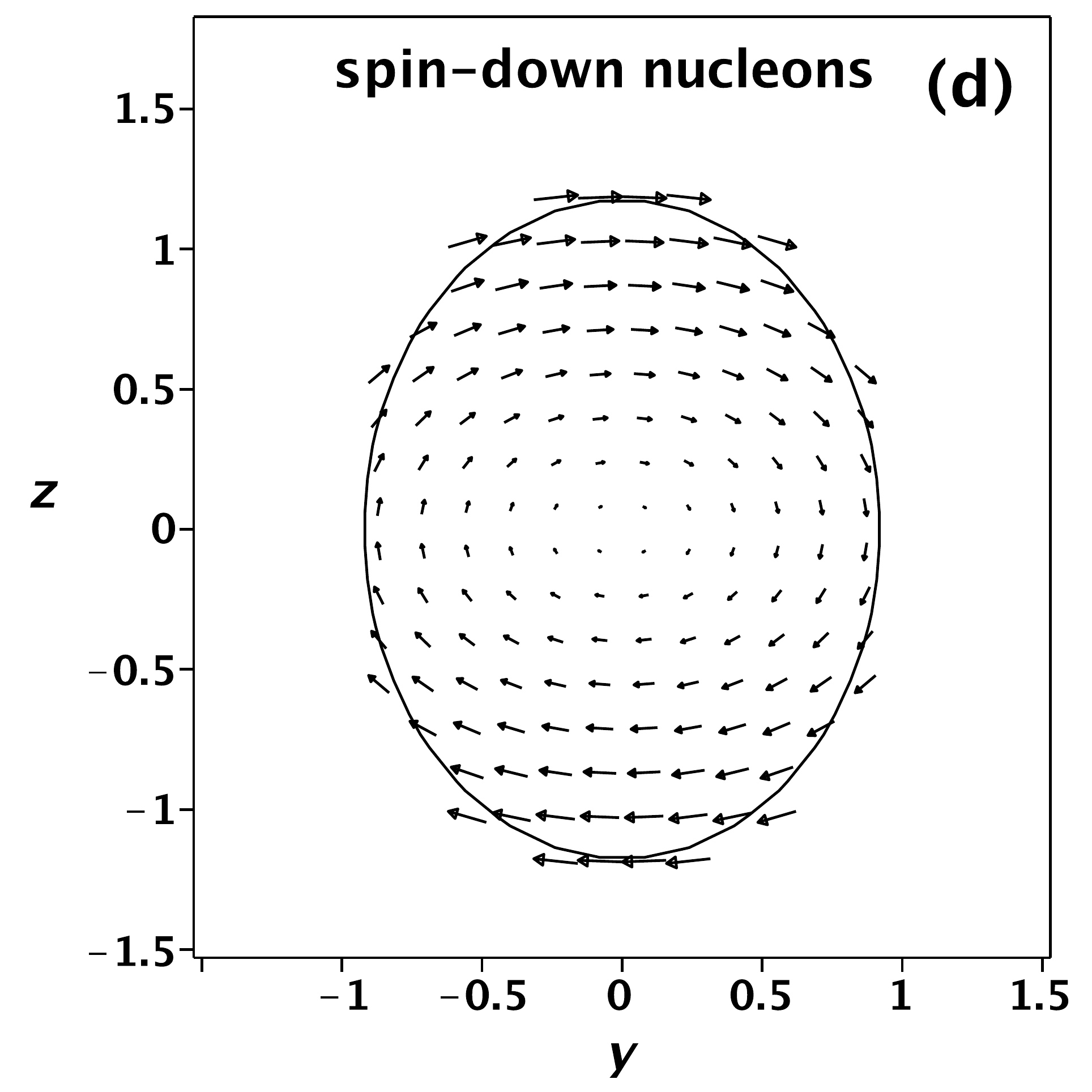}\\
\includegraphics[width=0.5\columnwidth]{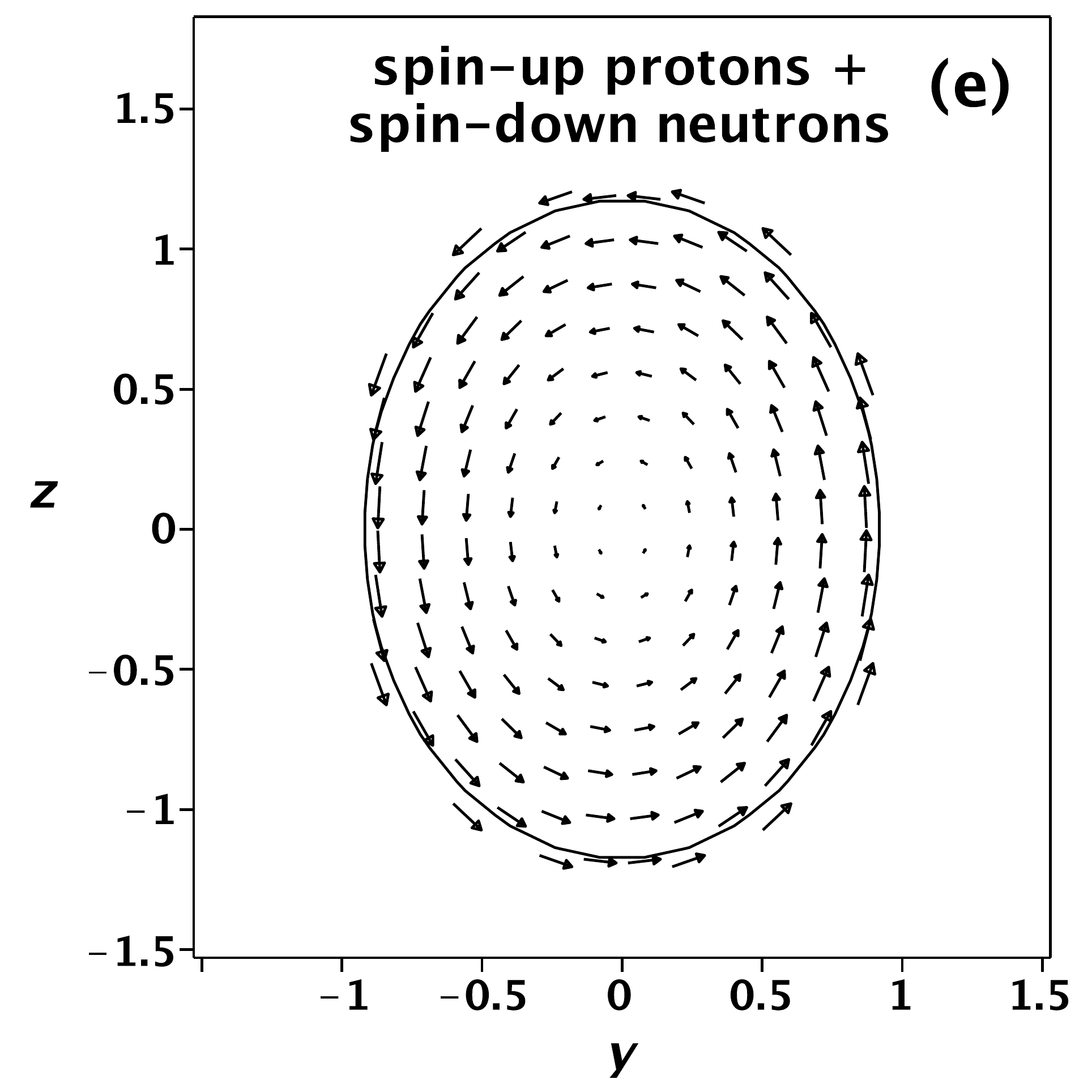}\includegraphics[width=0.5\columnwidth]{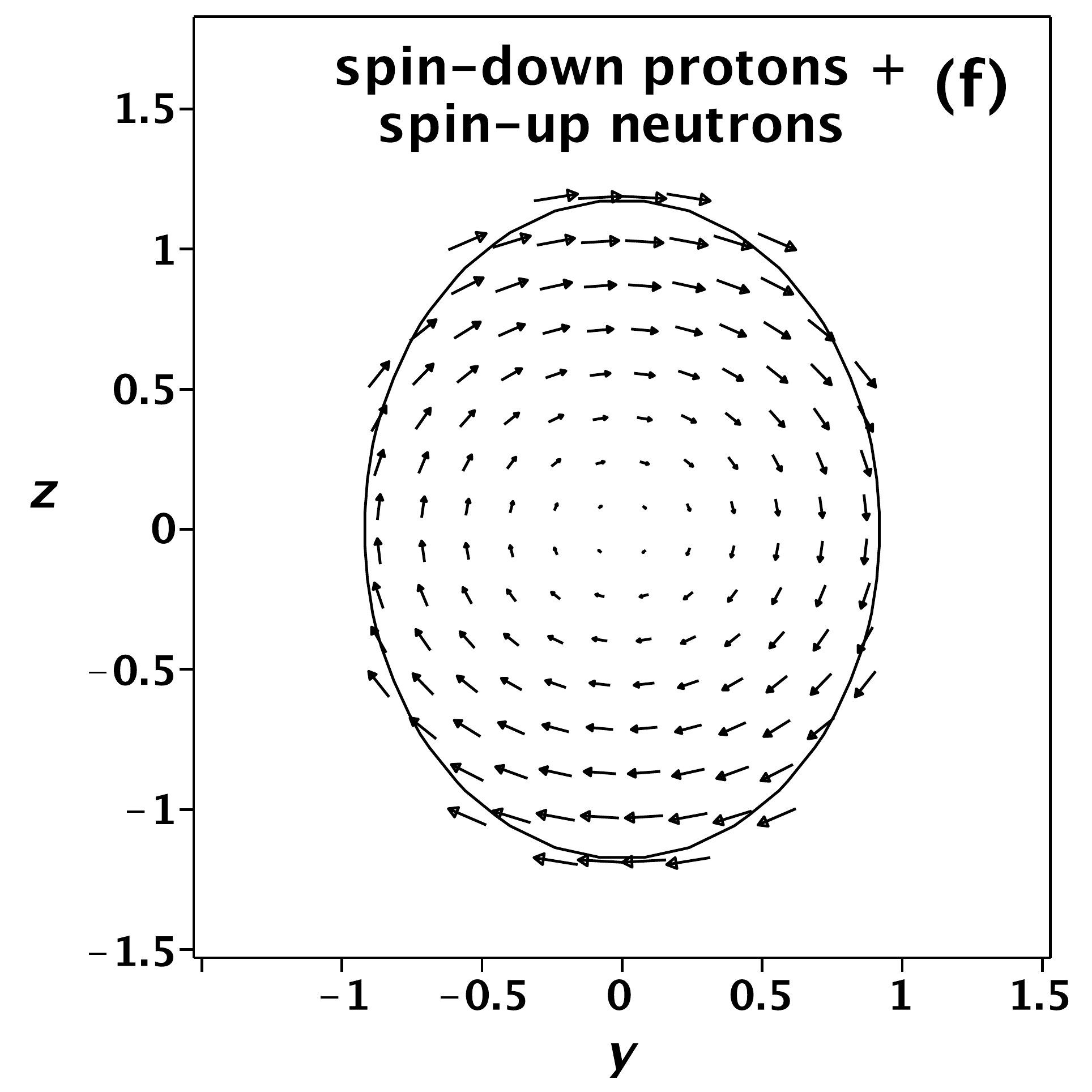}
\caption{The currents in $^{164}$Dy for $E=3.59$ MeV: $\delta J^{+}_{\rm p}$~(a), $\delta J^{+}_{\rm n}$~(b), 
$\delta J^{\uparrow\uparrow}$~(c), $\delta J^{\downarrow\downarrow}$~(d),
$\delta J^{\uparrow\uparrow}_{\rm p}+\delta J^{\downarrow\downarrow}_{\rm n}$~(e), 
$\delta J^{\downarrow\downarrow}_{\rm p}+\delta J^{\uparrow\uparrow}_{\rm 
n}$~(f).
\mbox{{\textsf y} $=y/R$, {\textsf z} $=z/R$.}}
\label{E3}
\end{figure}


Let us analyze these figures. 
First of all it is seen that one can not 
identify any of three $M1$ excitations with only one type of motions shown on
Fig.~\ref{fig1} -- it turns out that every excitation is a 
 mixture of all three possible scissors. Nevertheless an
approximate identification can be made. 
It is necessary to introduce some numerical measure of the contribution of 
every type of scissors into the particular excitation. Introducing the 
notations (see (\ref{JCart}))
\begin{eqnarray}
\label{AB}
&&A=
-i\frac{\sqrt{3}}{A_2}\!\left(\L^{\varsigma}_{11}+\L^{\varsigma}_{21}\right),
\nonumber\\
&&B=-i\frac{\sqrt{3}}{A_1}\!\left(\L^{\varsigma}_{11}-\L^{\varsigma}_{21}\right)
\nonumber
\end{eqnarray}
we can construct the following indicator characterizing the 
definite scissors, for example, conventional one:
$$AB_{(ab)}=[A^2+B^2]_{(a)}+[A^2+B^2]_{(b)}.$$
Analogous values $AB_{(cd)}$ and $AB_{(ef)}$ are defined also for spin 
scissors.
After normalization all three values are transformed in percents, which are 
shown in Table~\ref{tab-cur} together with the respective values of $A$ and 
$B$. The simple analysis of this Table allows one to conclude that:
\begin{enumerate}
\item  excitation with $E=2.20$ MeV represents
predominantly ($51\% $) the "complicate" spin scissors (Fig.~\ref{E1}~(e),~(f)) with rather strong 
admixture ($47\% $) of the "simple" spin scissors (Fig.~\ref{E1}~(c),~(d)),
\item  excitation with $E=2.87$ MeV represents
predominantly ($54\% $) the "simple" spin scissors (Fig.~\ref{E2}~(c),~(d)) with rather big 
admixture ($32\% $) of the conventional scissors (Fig.~\ref{E2}~(a),~(b)),
 \item  excitation with $E=3.59$ MeV represents
predominantly ($62\% $) the conventional scissors (Fig.~\ref{E3}~(a),~(b)) with a rather strong admixture ($31\% $) of the "complicate" spin scissors 
(Fig.~\ref{E3}~(e),~(f)).
\end{enumerate}
It is worth to note, that introduced in~\cite{BaSc} indicator 
$\beta=-B/A$ works here too: if $\beta$ is positive or negative the lines of current produce ellipse or hyperbola.

The situation with currents in Actinides is exactly the same as in Rare 
Earths. The picture of currents in $^{232}$Th is indistinguishable from that
of $^{164}$Dy.


\section{Conclusion}

We have solved the dynamical equations describing the nuclear collective motion
without the artificial division into isovector and isoscalar parts, an approximation we had 
applied in our previous work. As a result a new, third, type of nuclear scissors is 
found. The three types  of scissors modes can be approximately classified as isovector spin-scalar (conventional), isovector spin-vector and
isoscalar spin-vector, see Fig.~\ref{fig1}.
The analysis of currents has shown that three low-lying $1^+$ magnetic
excitations, predicted by the theory (see Table~\ref{tab1}), represent 
quite strong mixture of all three scissors.
The calculated energy centroids and summarized
transition probabilities of even-even Dy isotopes are in very good agreement
with the experimental results of the Oslo group. The experimental NRF data
for $^{164}$Dy are also in excellent agreement with our calculations,
whereas the data for $^{160,162}$Dy are in  good agreement only 
with the calculated centroids of the two higher lying scissors.
So we agree with the conclusion of the authors of~\cite{Renstrom}:
{\it ``It is  highly  desirable  to  remeasure  the  Dy  isotopes  by  
performing NRF experiments using quasi-monochromatic beams
in the interesting energy region between 2 and 4 MeV as done
for $^{232}$Th.''} According to our latest findings it is necessary to extend 
their proposal to all Rare Earth and Actinide nuclei.

More precisely, a satisfactory agreement is achieved for well deformed 
nuclei of the rare earth region with standard values of all possible parameters.   
The accuracy of the description of the scissors mode by the WFM method is comparable with that of QRPA, what is demonstrated by the comparison  of our 
results of calculations in the frame of WFM and QPNM approaches.
A satisfactory agreement is also achieved for weakly deformed (transitional)
nuclei of the same region by a very modest re-fit of the spin-orbit strength.
We suppose that fourth order moments and more realistic interactions are required for the 
adequate description of transitional nuclei. This will be the subject of future work.

\begin{acknowledgments}
Valuable discussions with M. Urban are gratefully acknowledged.
The work was supported by the \mbox{IN2P3/CNRS-JINR 03-57} Collaboration agreement.
\end{acknowledgments}

\begin{widetext}

\appendix

\section{Pairing }
\label{AppA}

\begin{eqnarray}
&&I^{\kappa\Delta}_{pp}(\br,p)=
\frac{r_p^3}{\sqrt{\pi}\hbar^3}{\rm e}^{-\alpha p^2}
\int\!\kappa^r(\br,p')\left[\phi_0(x)
-4\alpha^2p'^4\phi_2(x)\right]
{\rm e}^{-\alpha p'^2}p'^2dp',
\\
&&I^{\kappa\Delta}_{rp}(\br,p)=
\frac{r_p^3}{\sqrt{\pi}\hbar^3}{\rm e}^{-\alpha p^2}
\int\!\kappa^r(\br,p')[\phi_0(x)
-2\alpha p'^2\phi_1(x)]{\rm e}^{-\alpha p'^2}p'^2dp',
\end{eqnarray}
where $x=2\alpha pp'$,

\begin{eqnarray}
&&\phi_0(x)=\frac{1}{x}\sinh(x),
\qquad\phi_1(x)=\frac{1}{x^2}\left[\cosh(x)-\frac{1}{x}\sinh(x)\right],
\nonumber \\
&&\phi_2(x)=\frac{1}{x^3}\left[\left(1
+\frac{3}{x^2}\right)\sinh(x)-\frac{3}{x}\cosh(x)\right].
\end{eqnarray}

Anomalous density and semiclassical gap equation \cite{Ring}:
\begin{eqnarray}
&&\kappa(\br,\bp)=\frac{1}{2}
\frac{\Delta(\br,\bp)}{\sqrt{h^2(\br,\bp)+\Delta^2(\br,\bp)}},
\\
&&\Delta(\br,\bp)=-\frac{1}{2}\int\!\frac{d^3\!p'}{(2\pi\hbar)^3}
v(|\bp-\bp'|)
\frac{\Delta(\br,\bp')}{\sqrt{h^2(\br,\bp')+\Delta^2(\br,\bp')}},
\end{eqnarray}
where $v(|\bp-\bp'|)=\beta {\rm e}^{-\alpha |\bp-\bp'|^2}\!$
with $\beta=-|V_0|(r_p\sqrt{\pi})^3$ and $\alpha=r_p^2/4\hbar^2$.

The functions $\Delta(r')\equiv\Delta_{eq}(r',p_F(r'))$, 
$I_{rp}^{\kappa\Delta}(r')\equiv I_{rp}^{\kappa\Delta}(r',p_F(r'))$, 
$I_{pp}^{\kappa\Delta}(r')\equiv I_{pp}^{\kappa\Delta}(r',p_F(r'))$ 
depend on the radius $r'$  and the local Fermi momentum $p_F(r')$. 
The value of~$r'$ is not fixed by the theory and can be used 
as the fitting parameter.
 Nevertheless, to get rid off the fitting 
parameter, we use the averaged values of these functions:
$\bar\Delta=\int d\br\, n(\br)\Delta(r,p_F(r))/A$, etc.

\section{Currents}
\label{AppB}

\begin{eqnarray}
\L_{\lambda,\mu}^\varsigma
=\int\! d^3r \{r\otimes \delta J^\varsigma\}_{\lambda\mu}
=\frac{1}{\sqrt3}(-1)^\lambda 
\left[{A}_1 C_{1\mu,10}^{\lambda\mu}K_{\mu,0}^\varsigma-
{A}_2 \left(C_{1\mu+1,1-1}^{\lambda\mu}K_{\mu+1,-1}^\varsigma+
C_{1\mu-1,11}^{\lambda\mu}K_{\mu-1,1}^\varsigma\right)\right].\qquad
\end{eqnarray}
\begin{eqnarray}
\label{K} 
\nonumber&&K_{-1,-1}^\varsigma=-{\frac{\sqrt3\,\L_{2-2}^\varsigma}{{A}_2}},
\quad
K_{-1,0}^\varsigma={\frac{\sqrt3\,(\L_{1-1}^\varsigma+\L_{2-1}^\varsigma)}{\sqrt2\,{A}_1}},
\quad
K_{-1,1}^\varsigma=-{\frac{\sqrt3\, \L_{10}^\varsigma+\L_{20}^\varsigma+\sqrt2\, \L_{00}^\varsigma}
{\sqrt2\, {A}_2}},\\
\nonumber&&K_{0,-1}^\varsigma={\frac{\sqrt3\,(\L_{1-1}^\varsigma-\L_{2-1}^\varsigma)}{\sqrt2\,{A}_2}},
\quad
K_{0,0}^\varsigma={\frac {\sqrt {2}\L_{2,0}^\varsigma-\L_{0,0}^\varsigma}{{A}_1}},
\quad
K_{0,1}^\varsigma=-{\frac{\sqrt3\,(\L_{11}^\varsigma+\L_{21}^\varsigma)}{\sqrt2\,{A}_2}},\\
&&K_{1,-1}^\varsigma={\frac{\sqrt3\, \L_{10}^\varsigma-\L_{20}^\varsigma-\sqrt2\, \L_{00}^\varsigma}
{\sqrt2\, {A}_2}},
\quad
K_{1,0}^\varsigma={\frac{\sqrt3\,(\L_{21}^\varsigma-\L_{11}^\varsigma)}{\sqrt2\,{A}_1}},
\quad
K_{1,1}^\varsigma=-{\frac{\sqrt3\,\L_{22}^\varsigma}{{A}_2}},
\end{eqnarray}
where
\begin{eqnarray}
{A}_1=\frac{Q_{00}}{\sqrt3}\left(1+\frac{4}{3}\delta\right),\quad
{A}_2=-\frac{Q_{00}}{\sqrt3}\left(1-\frac{2}{3}\delta\right),\quad
Q_{00}=A<r^2>=\frac35 AR^2.
\nonumber\\
\end{eqnarray}


\section{Excitation probabilities }
\label{AppC}
   
Excitation
probabilities are calculated with the help of the theory of linear 
response of the system to a weak external field
\begin{equation}
\label{extf}
\hat O(t)=\hat O \,\e^{-i\Omega t}+\hat O^{\dagger}\,e^{i\Omega t}.
\end{equation}
A detailed explanation can be found in \cite{BaMo,BaSc}. 
We recall only the main points.
The matrix elements of the operator $\hat O$ obey the relationship \cite{Lane}
\begin{equation}
\label{matel}
|\langle\psi_a|\hat O|\psi_0\rangle|^2=
\hbar\lim_{\Omega\to\Omega_a}(\Omega-\Omega_a)
\overline{\langle\psi '|\hat O|\psi '\rangle\e^{-i\Omega t}},
\end{equation}
where $\psi_0$ and $\psi_a$ are the stationary wave functions of the
unperturbed ground and excited states; $\psi'$ is the wave function
of the perturbed ground state, $\Omega_a=(E_a-E_0)/\hbar$ are the
normal frequencies, the bar means averaging over a time interval much
larger than $1/\Omega$.

To calculate the magnetic transition probability, it is necessary
to excite the system by the following external field:
\begin{equation}
\label{Magnet}
\hat O_{\lambda\mu}=\mu_N\left(g_s^{\tau}\hat\bS/\hbar-ig_l^{\tau}\frac{2}{\lambda+1}[\br\times\nabla]\right)
\nabla(r^{\lambda}Y_{\lambda\mu}), \quad 
\mu_N=\frac{e\hbar}{2mc}.
\end{equation}
Here $g_l^{\rm p}=1,$ $g_s^{\rm p}=5.5856$ for protons and $g_l^{\rm n}=0,$ $g_s^{\rm n}=-3.8263$ for neutrons.
The dipole operator \mbox{($\lambda=1,\ \mu=1$)} 
in cyclic coordinates looks like
\begin{equation}
\label{Magnet_1}
\hat O_{11}=
\sqrt{\frac{3}{4\pi}}\left[g_s^{\tau}\hat S_{1}/\hbar
-g_l^{\tau}\sqrt2\sum_{\nu,\sigma}
C_{1\nu,1\sigma}^{11}r_{\nu}\nabla_{\sigma}\right]\mu_N.
\end{equation}
Its Wigner transform is
\begin{equation}
\label{MagWig}
(\hat O_{11})_W=
\sqrt{\frac{3}{4\pi}}\left[g_s^{\tau}\hat S_{1}
-ig_l^{\tau}\sqrt2\sum_{\nu,\sigma}
C_{1\nu,1\sigma}^{11}r_{\nu}p_{\sigma}\right]\frac{\mu_N}{\hbar}.
\end{equation}
 For the matrix element we have
\begin{eqnarray}
\label{psiO}
\langle\psi'|\hat O_{11}|\psi'\rangle &=&
\sqrt{\frac{3}{2\pi}}\left[-\frac{\hbar}{2}
(g_s^{\rm n}\F^{\rm n\downarrow\uparrow}
+g_s^{\rm p}\F^{\rm p\downarrow\uparrow})
-ig_l^{\rm p}\L_{11}^{\rm p+}\right]\frac{\mu_N}{\hbar}
\nonumber\\
&=&\sqrt{\frac{3}{8\pi}}\left[-\frac{1}{2}
[(g_s^{\rm n}-g_s^{\rm p})\bar{\F}^{\downarrow\uparrow}
+(g_s^{\rm n}+g_s^{\rm p}) \F^{\downarrow\uparrow}]
+\frac{i}{\hbar}g_l^{\rm p}(\bar\L_{11}^{+}- \L_{11}^{+})\right]\mu_N
\nonumber\\
&=&\sqrt{\frac{3}{8\pi}}\left[
\frac{1}{2}(g_s^{\rm p}-g_s^{\rm n})\bar{\F}^{\downarrow\uparrow}
+\frac{i}{\hbar}g_l^{\rm p} \bar{\L}_{11}^{+}
+\frac{i}{\hbar}[g_s^{\rm n}+g_s^{\rm p}-g_l^{\rm p}]\L_{11}^{+}
\right]\mu_N.
\end{eqnarray}
 Deriving (\ref{psiO}) we have used the relation $2i\L^+_{11}=-\hbar \F^{\d}$,
which follows from the angular momentum conservation~\cite{BaMo}.

One has to add the external field (\ref{Magnet_1}) to the Hamiltonian (\ref{Ham}). 
Due to the external field some dynamical equations of 
(\ref{iv}) become inhomogeneous:
\begin{eqnarray}
     \dot {\bar\R}^{+}_{21}
&=&\ldots\; +i\frac{3}{\sqrt{\pi}}\frac{\mu_N}{2\hbar} g^{\rm p}_l R^{{\rm p}+}_{20}({\rm eq})\,\e^{i\Omega t},
\nonumber\\
     \dot {\bar\L}^{-}_{11}
&=&\ldots\; + i\sqrt{\frac{3}{\pi}}\frac{\mu_N}{2\hbar} g^{\rm p}_l L^{{\rm p}-}_{10}({\rm eq})\,\e^{i\Omega t},
\nonumber\\
     \dot {\bar\L}^{\d}_{10}
&=&\ldots\; 
+i\sqrt{\frac{3}{2\pi}}\frac{\mu_N}{2\hbar}\left[ g^{\rm n}_s L^{{\rm n}-}_{10}({\rm eq})-g^{\rm p}_s L^{{\rm p}-}_{10}({\rm eq})\right]  \e^{i\Omega t}.
\end{eqnarray}
For the isoscalar set of equations, respectively, we obtain:
\begin{eqnarray}
     \dot {\R}^{+}_{21}
&=&\ldots\; - i\frac{3}{\sqrt{\pi}}\frac{\mu_N}{2\hbar} g^{\rm p}_l R^{{\rm p}+}_{20}({\rm eq})\,\e^{i\Omega t},
\nonumber\\
     \dot {\L}^{-}_{11}
&=&\ldots\; - i\sqrt{\frac{3}{\pi}}\frac{\mu_N}{2\hbar} g^{\rm p}_l L^{{\rm p}-}_{10}({\rm eq})\,\e^{i\Omega t},
\nonumber\\
     \dot {\L}^{\d}_{10}
&=&\ldots\; 
+i\sqrt{\frac{3}{2\pi}}\frac{\mu_N}{2\hbar}\left[ g^{\rm n}_s L^{{\rm n}-}_{10}({\rm eq})+g^{\rm p}_s L^{{\rm p}-}_{10}({\rm eq})\right] \e^{i\Omega t}.
\end{eqnarray}\end{widetext}
Solving the inhomogeneous set of equations 
one can find the required in (\ref{psiO}) values of
$\L_{11}^{+}$ , $\bar\L_{11}^{+}$ and $\bar\F^{\d}$ and  calculate 
$B(M1)$ factors for all excitations with the help of 
relationship (\ref{matel}).

One also should be aware of the fact that straightforward application of 
Lane's formula to the present WFM approach which leads to non-symmetric 
eigenvalue problems may yield negative transition 
probabilities violating the starting relation~(\ref{matel}). However, with the 
parameters employed here and also in our previous works~\cite{BaMo,BaMoPRC,BaMoPRC2,Malov,Urban,BaSc}, 
this never happened.

\end{document}